\documentclass[useAMS,usenatbib]{mnras}
\usepackage[utf8]{inputenc}
\usepackage{longtable}
\usepackage{graphicx}
\usepackage{adjustbox}
\usepackage{pdflscape}
\usepackage{rotating}
\usepackage{amsmath}
\usepackage{epstopdf}
\usepackage{afterpage}
\usepackage{float}
\usepackage{bookmark}

\newcommand{\Teff}{\mbox{$T_{\mathrm{eff}}$}}

\title[Kinematics of  white dwarfs]{The kinematics of  the white dwarf
  population from the SDSS DR12}

\author[B. Anguiano et al.]{B.~Anguiano$^{1,2}$\thanks{E-mail: astrobaj@gmail.com},
  A.~Rebassa-Mansergas$^{3,4}$,
  E.~Garc\'\i a--Berro$^{3,4}$,
  S.~Torres$^{3,4}$,\newauthor
  K.C.~Freeman$^{5}$,
  T.~Zwitter$^{6}$\\
  $^{1}$ Department of Astronomy, University of Virginia, Charlottesville, VA 22904-4325, USA\\ 
$^{2}$ Department of Physics $\&$ Astronomy, Macquarie University, Balaclava Rd, NSW 2109, Australia\\
$^{3}$   Departament  de   F\'isica,   Universitat  Polit\`ecnica   de
  Catalunya, c/Esteve Terrades 5, 08860 Castelldefels, Spain\\
$^{4}$ Institute for Space Studies of Catalonia, c/Gran Capit\`a 2--4,
  Edif. Nexus 201, 08034 Barcelona, Spain\\
$^{5}$  Research  School  of Astronomy  and  Astrophysics,  Australian
  National University, Cotter Rd., Weston, ACT 2611, Australia\\
$^{6}$ University  of Ljubljana,  Faculty of Mathematics  and Physics,
  Ljubljana, Slovenia
}

\begin{document}
\date{Accepted Mar 2017. Received Jan 2017}
\pagerange{\pageref{firstpage}--\pageref{lastpage}}
\pubyear{2016}
\maketitle

\begin{abstract}
We use  the Sloan  Digital Sky  Survey Data Release  12, which  is the
largest available white dwarf catalog  to date, to study the evolution
of the kinematical properties of the population of white dwarfs in the
Galactic  disc.  We  derive  masses, ages,  photometric distances  and
radial velocities for all white dwarfs with hydrogen-rich atmospheres.
For those stars for which proper  motions from the USNO-B1 catalog are
available the  true three-dimensional components of  the stellar space
velocity are obtained.   This subset of the  original sample comprises
20,247  objects, making  it the  largest sample  of white  dwarfs with
measured three-dimensional velocities.  Furthermore, the volume probed
by our  sample is  large, allowing us  to obtain  relevant kinematical
information. In  particular, our sample extends  from a Galactocentric
radial distance $R_{\rm G}=7.8$~kpc to 9.3~kpc, and vertical distances
from  the Galactic  plane ranging  from $Z=-0.5$~kpc  to 0.5~kpc.   We
examine  the   mean  components   of  the   stellar  three-dimensional
velocities,  as  well  as  their   dispersions  with  respect  to  the
Galactocentric and vertical distances.  We  confirm the existence of a
mean   Galactocentric  radial   velocity  gradient,   $\partial\langle
V_{\rm R}\rangle/\partial  R_{\rm G}=-3\pm5$~km~s$^{-1}$~kpc$^{-1}$.   We
also  confirm  North-South   differences  in  $\langle  V_{\rm z}\rangle$.
Specifically,  we find  that white  dwarfs  with $Z>0$  (in the  North
Galactic hemisphere) have $\langle  V_{\rm z}\rangle<0$, while the reverse
is  true for  white dwarfs  with $Z<0$. The age-velocity dispersion relation 
derived from the present sample indicates that the Galactic population 
of white dwarfs may have experienced an additional source of heating, which 
adds to the secular evolution of the Galactic disc.
\end{abstract}

\begin{keywords}
(stars:)  white dwarfs;  Galaxy: general;  Galaxy: evolution;  Galaxy:
  kinematics  and  dynamics;  (Galaxy:)  solar  neighborhood;  Galaxy:
  stellar content
\end{keywords}

\label{firstpage}

\section{Introduction}

White dwarfs  are the  most usual  stellar evolutionary  endpoint, and
account  for about  97 per  cent  of all  evolved stars  --- see,  for
instance,  the comprehensive  review  of  \citet{Alt2010a}.  Once  the
progenitor main-sequence star has  exhausted all its available nuclear
energy sources, it evolves to the white dwarf cooling phase, which for
the coolest and fainter stars last for ages comparable to the age of
the Galactic  disc.  Hence, the  white dwarf population  constitutes a
fossil record  of our  Galaxy, thus allowing  to trace  its evolution.
Moreover, the population of Galactic white dwarfs also conveys crucial
information about the evolution of  most stars. Furthermore, since the
cooling process  itself, as  well as  the structural properties   of  white   
dwarfs,   are   reasonably  well   understood \citep{Alt2010a,Renedo2010}, 
white  dwarfs can  be used  to retrieve information about our Galaxy that 
would complement that obtained using other stars, like main sequence stars. 

The ensemble properties of the  white dwarf population are recorded in
the white  dwarf luminosity function, which  therefore carries crucial
information  about  the  star  formation  history,  the  initial  mass
function, or the nature and history of the different components of our
Galaxy --- see the recent review  of \cite{Garcia2016} for an extensive list
of possible  applications, as well  as for updated references  on this
topic.  Among these  applications perhaps the most well  known of them
is   that    white   dwarfs   are   frequently    used   as   reliable
cosmochronometers.  Specifically,  since the  bulk of the  white dwarf
population has  not had time  enough to cool down  to undetectability,
white dwarfs provide an independent estimate  of the age of the Galaxy
\citep{Winget,Nature}.   But   this  is   not  the   only  interesting
application of white dwarfs. In particular, the observed properties of
the population of white dwarfs can  also be employed to study the mass
lost during  the previous stellar evolutionary  phases.  Additionally,
on  the asymptotic  giant branch  white dwarf  progenitors eject  mass
which has been  enriched in carbon, nitrogen, and  oxygen during the 
evolutionary  stages.   Hence, the  progenitors  of  white dwarfs  are
significant contributors to the chemical evolution of the Galaxy. 
Finally, binary systems made of a main sequence star
and a white dwarf can also be used to probe the evolution of
the metal content of the Galaxy \citep{2011AJ....141..107Z,rebassa-mansergasetal16-2}. 

All these studies are done analyzing  the mass distribution of field white
dwarfs.   Several   studies  have   focused  on  obtaining   the  mass
distribution  of  Galactic  white  dwarfs.  In  particular,  the  mass
distribution  of white  dwarfs of  the most  common spectral  type ---
namely those  with hydrogen-rich atmospheres,  also known as  DA white
dwarfs --- has been extensively  investigated in recent years --- see,
for example  \cite{Kepleretal07} and \cite{rebassa-mansergasetal15-2},
and references therein.

The  first  study of  the  kinematic  properties  of the  white  dwarf
population was that  of \cite{Sion1988}, where a sample  of over 1,300
degenerate  stars  from  the   catalog  of  \cite{McCookSion1987}  was
employed.  Since  then, several  studies have  been done,  focusing in
different important aspects that can  be obtained from the kinematical
distributions,  such  as  the  identification  of  halo  white  dwarfs
candidates \citep{Liebert1989,Torres1998}, the  dark matter content of
a               potential                halo               population
\citep{Oppenheimer2001,Reid2001,Torres2002}, or the possible existence
of a  major merger  episode in  the Galactic  disc \citep{Torres2001}.
However,  the  small sample  sizes  used  in these  studies  prevented
obtaining conclusive  results. But this  is not the only  problem that
these studies  faced.  Specifically,  in addition to  poor statistics,
the  major  drawback   of  these  works  was  the   lack  of  reliable
determinations of  true three-dimensional velocities.  The  reason for
this is that the surface gravity of  white dwarf stars is so high that
gravitational  broadening of  the  Balmer lines  is important.   Thus,
disentangling the  gravitational redshift from the  true Doppler shift
is, in most of the cases, a difficult task.  Consequently, determining
the true radial  velocity of white dwarfs require a model that predicts 
stellar masses and radii and hence, the value of the gravitational redshift. 
Even  more, a  sizable fraction of  cool white  dwarfs has
featureless spectra, and hence for these stars only proper motions can
be determined.  All this precluded accurate measurements of the radial
component of the velocity, and the assumption of zero radial velocity,
or the method proposed by \cite{DehnenBinney1998} were adopted in most
studies    ---     see,    for    instance,     \cite{Sion2009}    and
\cite{WeggPhinney2012},  and   references  therein.    More  recently,
\cite{Silvestri2001} presented a way to overcome this drawback.  The 
authors  studied a  sample of  white  dwarfs in  common proper  motion
binary systems. Among these pairs  they selected a sub-sample in which
the secondary star  was an M dwarf,  for which the sharp  lines in its
spectrum  allowed   to  derive  easily  reliable   radial  velocities.
Nevertheless, it  was not until  the ESO SNIa Progenitor  surveY (SPY)
project  --- see  \citet{Pauli2006}  and references  therein ---  that
radial velocities  were measured  for the  first time  with reasonable
precisions. This was  made because within this  survey high resolution
UVES VLT spectra  were obtained for a sample  of stars.  Nevertheless,
the  sample of  \citet{Pauli2006}  only contained  $\sim400$ DA  white
dwarfs  with  radial  velocities  measurements  better  than  $2\,{\rm
km\,s^{-1}}$, of  which they estimated that  2 per cent of  stars were
members of  the Galactic  halo and  7 per cent  belonged to  the thick
disc.

In  summary,  the  need  of  a complete sample,  or  at  least  statistically
significant, with  accurate   measurements  of   true  space
velocities, distances, masses and ages, is crucial for studying the 
evolution of our  Galaxy.  In this sense, it is worth
emphasizing that  little progress  has been done  to use  the Galactic
white dwarf population  to unravel the evolution of  the Galactic disc
studying the  age-velocity relationship (AVR).   Age-velocity diagrams
reflect the slow increase of the random velocities with age due to the
heating of the  disc by massive objects. The term  ``disc heating'' is
often applied to the sum of the effects that may cause larger velocity
dispersions in  the population of  disc stars.  In  principle, heating
injects kinetic energy into the random component of the stellar motion
over  time.   In  order  to  understand  the  origin  of  the  present
assemblage of  disc stars, it  is necessary to quantify  the kinematic
properties  of  the  populations  in the  disc  and  characterize  the
properties of their stars as  accurately as possible.  In other words,
the space motions of the stars as a function of age allows us to probe
the dynamical evolution of the Galactic disc.

Several  heating mechanisms  have  been proposed  in  the last  years.
Among   them    we   explicitly   mention   transient    spiral   arms
\citep{DeSimone2004, Minchev2006},    giant    molecular    clouds,
\citep{Lacey1984,  Hanninen2002},  massive  black holes  in  the  halo
\citep{Lacey1985},  repeated  disc  impact of  the  original  Galactic
globular cluster  population \citep{VandePutte2009},  satellite galaxy
mergers  \citep{Quinn1993, Walker1996,  Velazquez1999, Villalobos2008,
Moster2010,  House2011}.   Recently,  \citet{Martig2014}  studied  the
impact  of different  merger  activity  in the  shape  of  the AVR  in
simulated disc galaxies  and found that the shape of  the AVR strongly
depends on the  merger history at low redshift for  stars younger than
9~Gyr.   A mechanism  called  radial  mixing \citep{Sellwood2002}  was
suggested as  a possible source of  disc heating \citep{Schonrich2009,
Loebman2011}.  However,  \citet{Minchev2012} and  \citet{VeraCiro2014}  found that
the contribution  of radial  mixing to disc  heating is  negligible in
their simulations.  Recently,  \citet{Grand2016} used state-of-the-art
cosmological magnetohydrodynamical simulations and  found that the 
dominant  heating mechanism is the bar,  whereas spiral  arms and  radial
migration are all subdominant in  Milky Way-sized galaxies.  They also
found  that the  strongest source,  though less  prevalent than  bars,
originates  from external  perturbations from  satellites/subhaloes of
masses $\log_{\rm 10}(M/M_{\sun}) \ge 10$.

\begin{table}
\rotatebox{90}{
\begin{minipage}{\textheight}
\centering
\caption{\label{tab_cat}  The  stellar parameters,  distances,  proper
  motions, radial  velocities, gravitational  redshifts, plate-mjd-fib
  identifiers,  spectral  signal-to-noise  ratios  and  ages  for  the
  complete sample  of 20,247 SDSS  hydrogen-rich white dwarfs  used in
  this work.  Ages 1, 2 and  3 are obtained using the initial-to-final
  mass  relations  of   \citet{Catalan2008},  \citet{Gesicki2014}  and
  \citet{ferrarioetal05-1}, respectively.   The table is  published in
  its entirety in the online version of the paper.}
\setlength{\tabcolsep}{0.35ex}
\begin{small}
\begin{tabular}{cccccccccccccccc}
\hline
\hline
               Star    &  $T_\mathrm{eff}$  &    $M_\mathrm{WD}$   &       $\log\,g$    &     $d$            &  $\mu_\mathrm{\alpha}$ & $\mu_\mathrm{\delta}$ &    $V_\mathrm{r}$   & $V_\mathrm{grav}$  &   mjd  &    plt   &  fib  &    SN   &     Age 1            &    Age 2             &  Age 3        \\
             (SDSS)    &  (K)             &    $(M_{\odot})$      &         (dex)     & (pc)               &     (mas)            & (mas)                &      (km/s)        &  (km/s)           &       &          &       &         &    (Gyr)              &    (Gyr)            & (Gyr)  \\
\hline
J000006.75$-$004653.9  &   10793$\pm$ 244 &    0.680$\pm$0.130  &   8.137$\pm$0.205 &   234.28$\pm$ 35.94 &    $-$1.70$\pm$3.08  &    $-$1.31$\pm$3.08 &    40.81$\pm$20.27   &   37.15         & 52203  &     685  &   225  &   13.23 &  1.20$^{+0.08}_{-3.56}$ &  1.16$^{+0.07}_{-1.92}$ &  1.13$^{+0.04}_{-2.87}$  \\                   
J000012.32$-$005042.5  &    7570$\pm$ 361 &    0.603$\pm$0.426  &   8.016$\pm$0.678 &   233.72$\pm$120.07 &    $-$2.15$\pm$3.70  &    $-$5.49$\pm$3.70 & $-$98.36$\pm$55.59   &   30.42         & 52902  &    1091  &   117  &    5.24 &  2.61$^{+0.90}_{-2.74}$ &  2.27$^{+1.23}_{-3.08}$ &  2.19$^{+1.29}_{-3.16}$  \\                   
J000022.87$-$000635.6  &   22550$\pm$ 658 &    0.417$\pm$0.033  &   7.430$\pm$0.103 &   641.58$\pm$ 57.94 &      15.53$\pm$3.01  &       4.67$\pm$3.01 &  $-$1.15$\pm$14.22   &   12.91         & 51791  &     387  &   166  &   19.32 &     ---              &    ---               &    ---          \\                 
J000034.06$-$052922.3  &   19640$\pm$ 167 &    0.505$\pm$0.017  &   7.730$\pm$0.037 &   221.78$\pm$  7.17 &      85.34$\pm$2.48  &   $-$14.64$\pm$2.48 & $-$10.97$\pm$ 6.04   &   20.06         & 54380  &    2624  &   261  &   55.48 &     ---              &    ---               &    ---          \\                    
J000051.84+272405.2    &   21045$\pm$ 402 &    0.606$\pm$0.042  &   7.960$\pm$0.075 &   476.17$\pm$ 27.28 &      11.55$\pm$3.10  &    $-$6.73$\pm$3.10 & $-$25.33$\pm$11.74   &   28.59         & 54452  &    2824  &   272  &   23.33 &  1.32$^{+0.50}_{-1.73}$ &  1.01$^{+0.28}_{-0.83}$ &  0.93$^{+0.25}_{-0.98}$ \\                    
J000104.05+000355.8    &   13405$\pm$ 583 &    0.505$\pm$0.081  &   7.784$\pm$0.147 &   383.54$\pm$ 45.28 &       4.07$\pm$3.13  &    $-$9.67$\pm$3.13 & $-$26.96$\pm$19.30   &   21.37         & 52203  &     685  &   490  &   14.46 &     ---              &    ---               &    ---         \\                     
J000110.09+273520.4    &   12681$\pm$ 470 &    0.488$\pm$0.106  &   7.750$\pm$0.200 &   607.84$\pm$ 97.89 &    $-$4.69$\pm$3.76  &    $-$8.66$\pm$3.76 & $-$33.21$\pm$29.55   &   20.20         & 54452  &    2824  &   207  &    9.74 &     ---              &    ---               &    ---         \\                    
J000115.76+285647.2    &   13417$\pm$1244 &    0.509$\pm$0.113  &   7.792$\pm$0.201 &   588.26$\pm$ 95.54 &       9.29$\pm$3.62  &   $-$15.57$\pm$3.62 &   102.01$\pm$21.49   &   21.64         & 54452  &    2824  &   432  &   12.63 &     ---              &    ---               &    ---         \\                     
J000120.42$-$052140.8  &    7924$\pm$ 285 &    0.707$\pm$0.318  &   8.190$\pm$0.522 &   291.09$\pm$115.47 &      13.86$\pm$3.40  &   $-$60.69$\pm$3.40 & $-$68.69$\pm$36.23   &   40.27         & 54327  &    2630  &   265  &    6.13 &  2.03$^{+1.18}_{-3.41}$ &  1.99$^{+1.21}_{-3.45}$ &  1.98$^{+1.20}_{-3.46}$  \\                     
J000126.85+272000.1    &   10023$\pm$  46 &    0.566$\pm$0.044  &   7.933$\pm$0.067 &   119.80$\pm$  6.21 &     113.18$\pm$2.72  &   $-$26.20$\pm$2.72 &     56.72$\pm$9.12   &   26.80         & 54368  &    2803  &   211  &   32.12 &  3.47$^{+1.66}_{-1.41}$ &  2.31$^{+0.78}_{-2.57}$ &  2.36$^{+0.90}_{-2.52}$  \\                           
J000135.06+244958.5    &   13668$\pm$2054 &    0.541$\pm$0.267  &   7.861$\pm$0.434 &   814.08$\pm$287.12 &       1.06$\pm$3.68  &      -5.11$\pm$3.68 &     61.06$\pm$ 44.00 &   24.13         & 54389  &    2822  &   273  &    7.70 &     ---              &    ---               &    ---          \\  
J000208.73$-$050855.1  &   17505$\pm$ 521 &    0.501$\pm$0.056  &   7.740$\pm$0.114 &   659.22$\pm$ 63.46 &       ---            &       ---           &     75.28$\pm$ 27.01 &   20.21         & 54327  &    2630  &   273  &   14.07 &     ---              &    ---               &    ---          \\ 
J000216.02+120309.3    &    9913$\pm$ 933 &    0.907$\pm$0.426  &   8.512$\pm$0.839 &   678.84$\pm$487.93 &       ---            &       ---           &    102.98$\pm$110.92 &   65.72         & 55912  &    5649  &   467  &    2.05 &  1.69$^{0.50}_{-1.03}$  &  1.66$^{0.53}_{-1.00}$  &  1.64$^{0.55}_{-0.98}$ \\
J000228.15+235547.9    &   26801$\pm$ 428 &    0.559$\pm$0.026  &   7.820$\pm$0.070 &   731.77$\pm$ 39.05 &      13.01$\pm$3.05  &       0.07$\pm$3.05 &     12.98$\pm$ 12.39 &   23.42         & 54389  &    2822  &   251  &   29.81 &  3.43$^{1.55}_{-0.86}$  &  2.04$^{0.79}_{-2.25}$  &  2.27$^{1.11}_{-2.02}$ \\
J000236.49+254143.2    &   18330$\pm$ 886 &    0.585$\pm$0.116  &   7.930$\pm$0.200 &   874.14$\pm$134.61 &      -0.93$\pm$4.93  &      -0.89$\pm$4.93 &     70.85$\pm$ 43.43 &   27.14         & 54389  &    2822  &   371  &    7.55 &  1.94$^{1.33}_{-2.42}$  &  1.31$^{0.73}_{-3.05}$  &  1.23$^{0.67}_{-3.13}$ \\
\hline
\end{tabular}
\end{small}
\end{minipage}
}
\end{table}

From the observational point of view  it is worth mentioning that some
of the properties  of the AVR are still  controversial.  For instance,
\citet{Wielen1977}, \citet{Holmberg2007},  and \citet{Aumer2009} found
that the stellar velocity dispersion increases steadily for all times,
following a power law. In the case of the vertical velocity dispersion
they  found that  $\sigma_{\rm z}\propto  t^{\alpha}$,  where $\alpha$  is
close to 0.5. On the other hand, \citet{Dawson1984} found that the AVR
rises fairly steeply for stars younger than 6~Gyr, thereafter becoming
nearly constant with  age.  Another observational finding  is that 
heating takes place for the first  2 to 3~Gyr, but then saturates when
$\sigma_{\rm z}$ reaches $\sim 20$~km~s$^{-1}$.   This suggests that stars
of higher  velocity dispersion spend  most of their orbital  time away
from   the  Galactic   plane  where   the  sources   of  heating   lie
\citep[e.g.][]{Stromgren1987,        Quillen2001,       Soubiran2008}.
\citet{Seabroke2007}  found that  vertical  disc  heating models  that
saturate after 4.5~Gyr are  equally consistent with observations.  The
difficulty    of   obtaining    precise   ages    for   field    stars
\citep{Soderblom2010}, and selection effects  in the different samples
used by different  authors might be responsible  for the discrepancies 
mentioned above.


\begin{figure}
  \includegraphics[width=1.01\columnwidth]{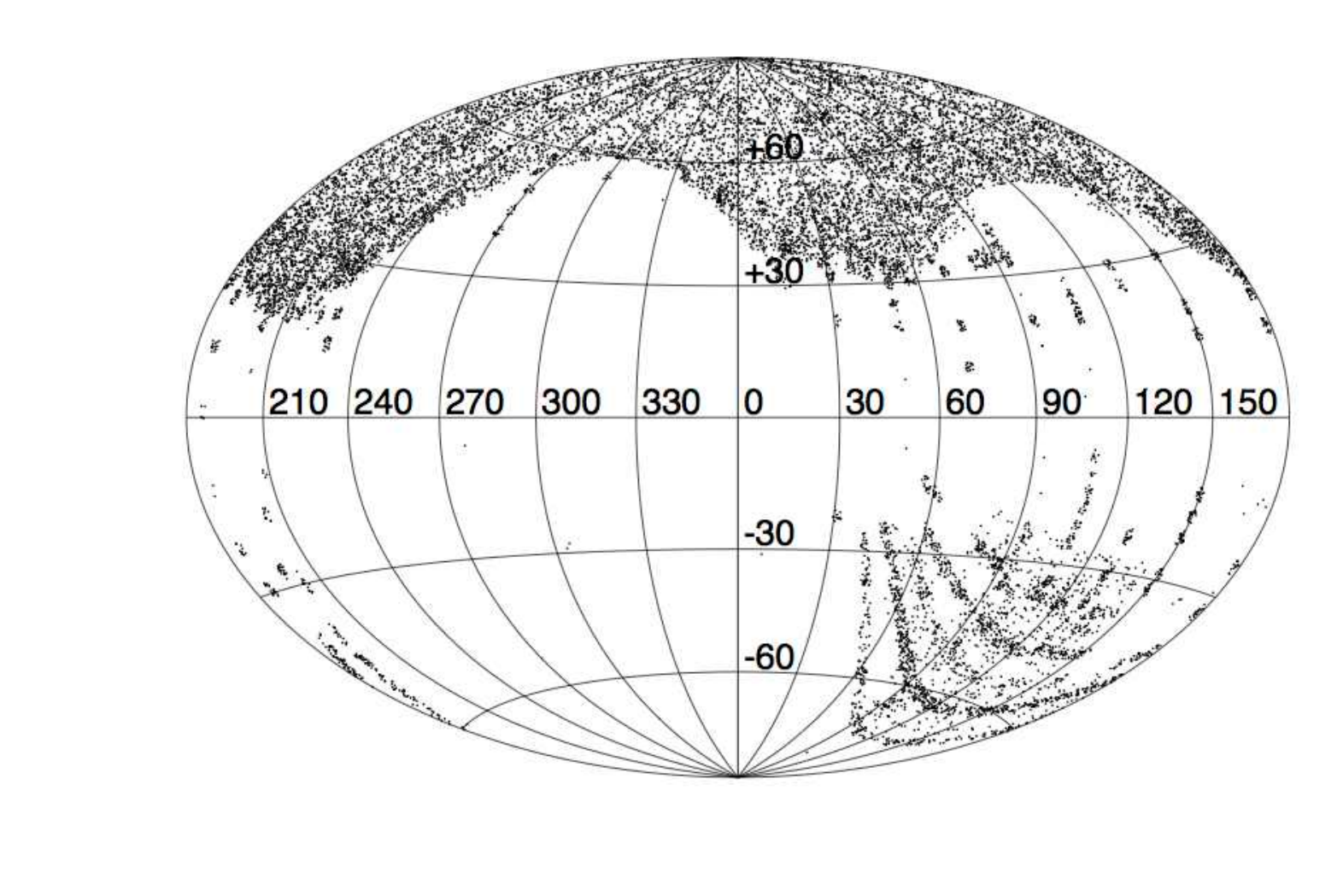}
  \caption{A Hammer-Aitoff  projection in Galactic coordinates  of the
    SDSS DR 12 spectroscopic plate distribution used in this work.}
  \label{fig:aitoff}
\end{figure}

In this  paper we use  the sample  of white dwarfs  with hydrogen-rich
atmospheres from the Sloan Digital Sky Survey (SDSS) data release (DR)
12,  which is  the largest  catalog  available to  date, to  determine
velocity gradients  in the  $(R_{\rm G},\,Z)$ space.   Moreover, since
white  dwarfs are  excellent natural  clocks  we use  them to  compute
accurate  ages and  in this  way  we determine  the AVR  in the  solar
vicinity.  All this  allows us to investigate  the kinematic evolution
of  the  Galactic disc.   This  paper  is  organized as  follows.   In
Sect.~\ref{catalog} we describe the white dwarf catalog and we explain
how     we      assess     the     quality     of      white     dwarf
spectra. Sect.~\ref{distributions} is devoted to explain how we derive
effective  temperatures  and  surface  gravities,  as  well  as  their
respective errors, and to  discuss the corresponding distributions. We
also  present  the set  of  derived  masses, ages,  distances,  radial
velocities   and   proper   motions.     This   is   later   used   in
Sect.~\ref{Velo_maps}  where  we present  the  velocity  maps and  age
gradients.   The  AVR  is   discussed  in  Sect.~\ref{AVR},  while  in
Sect.~\ref{conclu} we summarize our most important results and we draw
our conclusions.

\section{The white dwarf catalog}
\label{catalog}

Modern large scale  surveys have been very  profficient at identifying
white  dwarfs.   Among  them  we  mention  the  Palomar  Green  survey
\citep{Liebert2005}, the Kiso  survey \citep{LimogesBergeron2010}, and
the   LAMOST  spectroscopic   survey  of   the  Galactic   Anti-Center
\citep{rebassa-mansergasetal15-1}.   However,  it  has been  the  SDSS
\citep{York2000,  Alam2015}  the  survey  that  in  recent  years  has
significantly increased the number of known white dwarfs.  Indeed, the
SDSS has  produced the largest  spectroscopic sample of  white dwarfs,
and its  latest release,  the DR~12, contains  more than  30,000 stars
\citep{Kepler2016}.   Because  of  its  considerably  larger  size  as
compared to  other available white  dwarf catalogs, we adopt  the SDSS
catalog as  the sample  of study  in this work.   From this  sample we
select only white dwarfs with  hydrogen dominated atmospheres, that is
of the DA spectral type. For  these white dwarfs radial velocities and
the relevant  stellar parameters  can be  derived using  the technique
described  below.  The  Galactic coordinates  of the  20,247 DA  white
dwarfs  selected  in  this  way are  shown  in  Fig.~\ref{fig:aitoff},
whereas  Table~\ref{tab_cat}  lists  the  stellar  parameters,  proper
motions and radial velocities of these stars.  This table is published
in  its integrity  in machine-readable  format.  However,  for obvious
reasons, only a portion is shown  here for guidance regarding its form
and content.

\subsection{Signal-to-noise ratio of the SDSS white dwarf spectra}
\label{SNR}

Before providing details on how  we measure the stellar parameters and
radial velocities from the SDSS white dwarf spectra it is important to
mention that these spectra are in several cases rather noisy.  This is
because white dwarfs are generally  faint objects.  Hence, the stellar
parameters and radial velocities derived from their spectra often have
large uncertainties. In order to select  a clean DA white dwarf sample
for the AVR of the  Galactic disc we hence  need to exclude
all bad  quality data.  We will  do this based on  the signal-to-noise
ratio (SNR) of the SDSS spectra, which is calculated for each spectrum
in this section.

The featureless spectral  region of the continuum in a  white dwarf DA
spectrum allows us to derive a statistical estimate of the SNR using a
rather simple approach. This is done comparing the flux level (signal)
within a  particular wavelength  range to the  intrinsic noise  of the
spectrum in the same wavelength region.  That is, we compute the ratio
of the average signal to its  standard deviation --- see, for example,
\citet{Rosales-Ortega2012}.

\begin{figure}
\includegraphics[width=1.05\columnwidth]{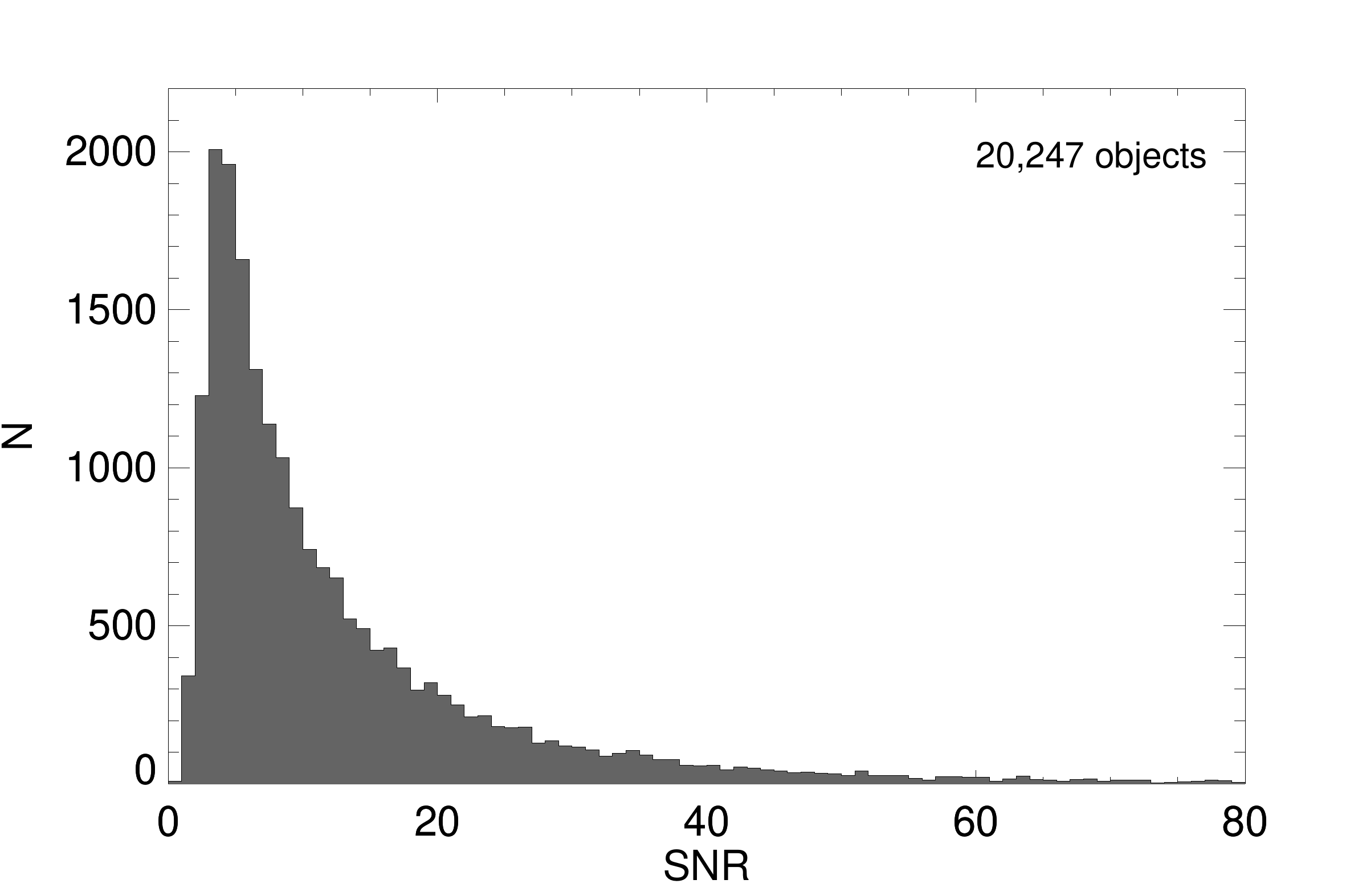}
\caption{Distribution of the signal-to-noise ratios of the white dwarf
  spectra in the sample analyzed in this work.  The distribution peaks
  at SNR$\,\sim\,$5, while the mean value is 13.2.}
\label{fig:SNR}
\end{figure}

\begin{table}
\caption{The  four  selected  central wavelength  ($\lambda_{\rm  C}$)
  regions, C1, C2, C3 and C4 respectively, and their wavelength widths
  ($\Delta \lambda$) for a statistical estimate of the signal-to-noise
  ratio in the SDSS spectra.}
\label{tab:cen_lambda}
\begin{center}
\begin{tabular}{ccc}
\hline
\hline
& $\lambda_{\rm C}$ (\rm \AA) & $\Delta \lambda$ (\rm \AA) \\
\hline
C1 & 4600 & 200 \\
C2 & 5300 & 400 \\
C3 & 6100 & 600 \\
C4 & 6900 & 200 \\
\hline
\end{tabular}
\end{center}
\end{table}

We defined four continuum bands centered at $\lambda_{\rm C}$ of width
$\Delta  \lambda$ (see  Table~\ref{tab:cen_lambda}) from  the observed
spectrum, $f(\lambda$). We applied a  correction for the presence of a
slope  within  the  continuum  band, where  $\sigma_{\rm  C}$  is  the
standard deviation in the difference between $f(\lambda$) and a linear
fit to  $f(\lambda$). We then  calculated the statistical SNR  for the
four continuum  bands employing  the expression (SNR)$_{\rm  C_{i}}$ =
$\mu_{\rm C_{i}}/\sigma_{\rm  C_{i}}$, where $\mu_{\rm C_{i}}$  is the
mean flux in the given continuum  band and $\sigma_{\rm C_{i}}$ is the
standard  deviation  of the  flux  within  the given  continuum  band,
dominated by noise instead of  real features.  We found that SNR$_{\rm
C1}>$SNR$_{\rm  C2}>$SNR$_{\rm   C3}>$SNR$_{\rm  C4}$,  for   all  the
spectra, as  expected for these  stars, where emergent flux  appears in
the  blue wavelengths.   We finally  derived the  continuum SNR  of an
observed spectrum as the average of the individual (SNR)$_{\rm C_{i}}$
calculated for  each of the pseudocontinuum  bands, C1, C2, C3  and C4
respectively.    Fig.~\ref{fig:SNR}  shows   the   histogram  of   the
distribution  of the  statistical SNR  for the  total of  20,247 white
dwarf spectra  selected for  this study.   We found  that half  of the
sample has  a statistical SNR for  the continuum larger than  12 while
most of the spectra a SNR around 5.

\begin{figure*}
\includegraphics[width=\columnwidth]{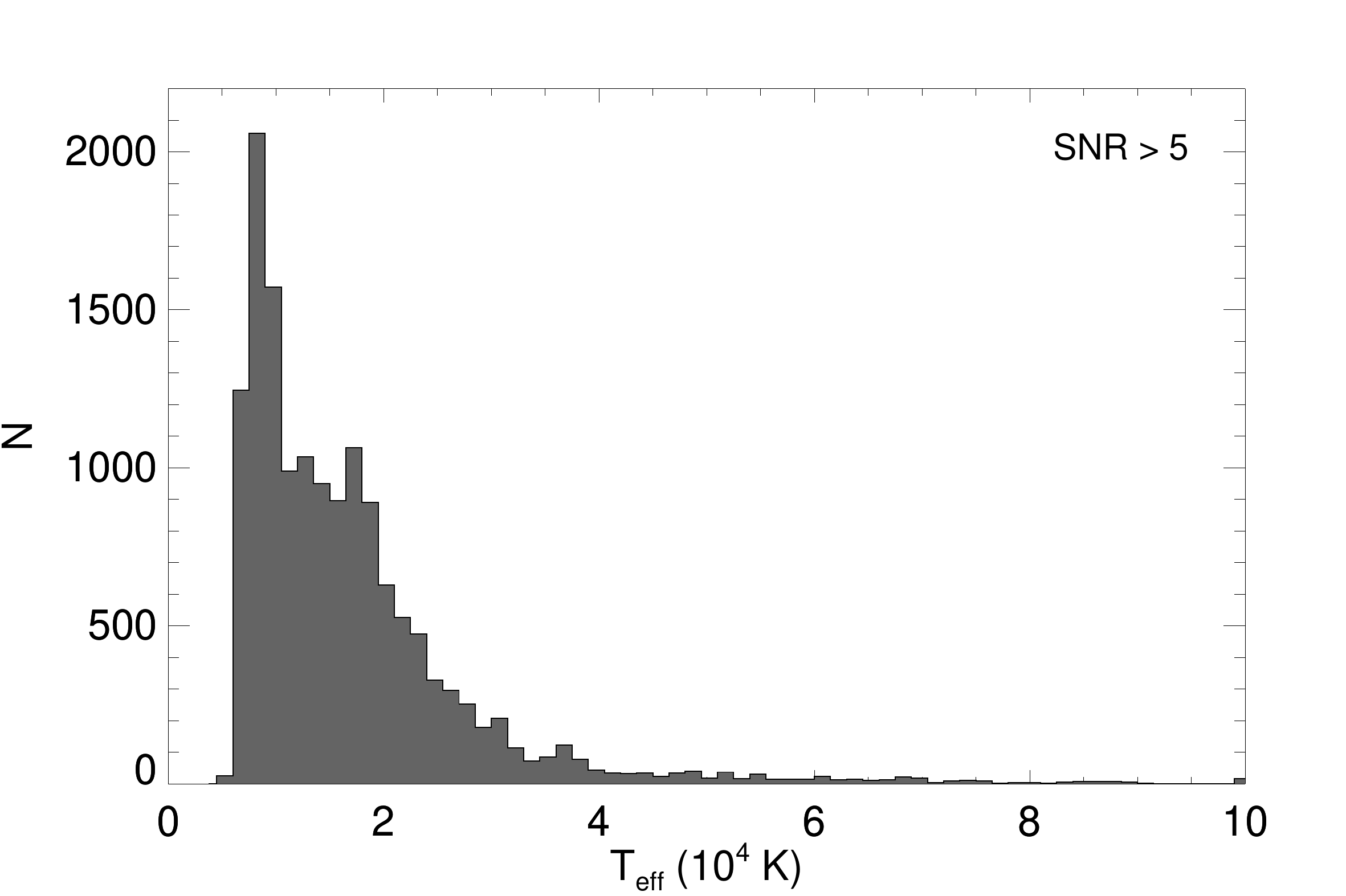}
\includegraphics[width=\columnwidth]{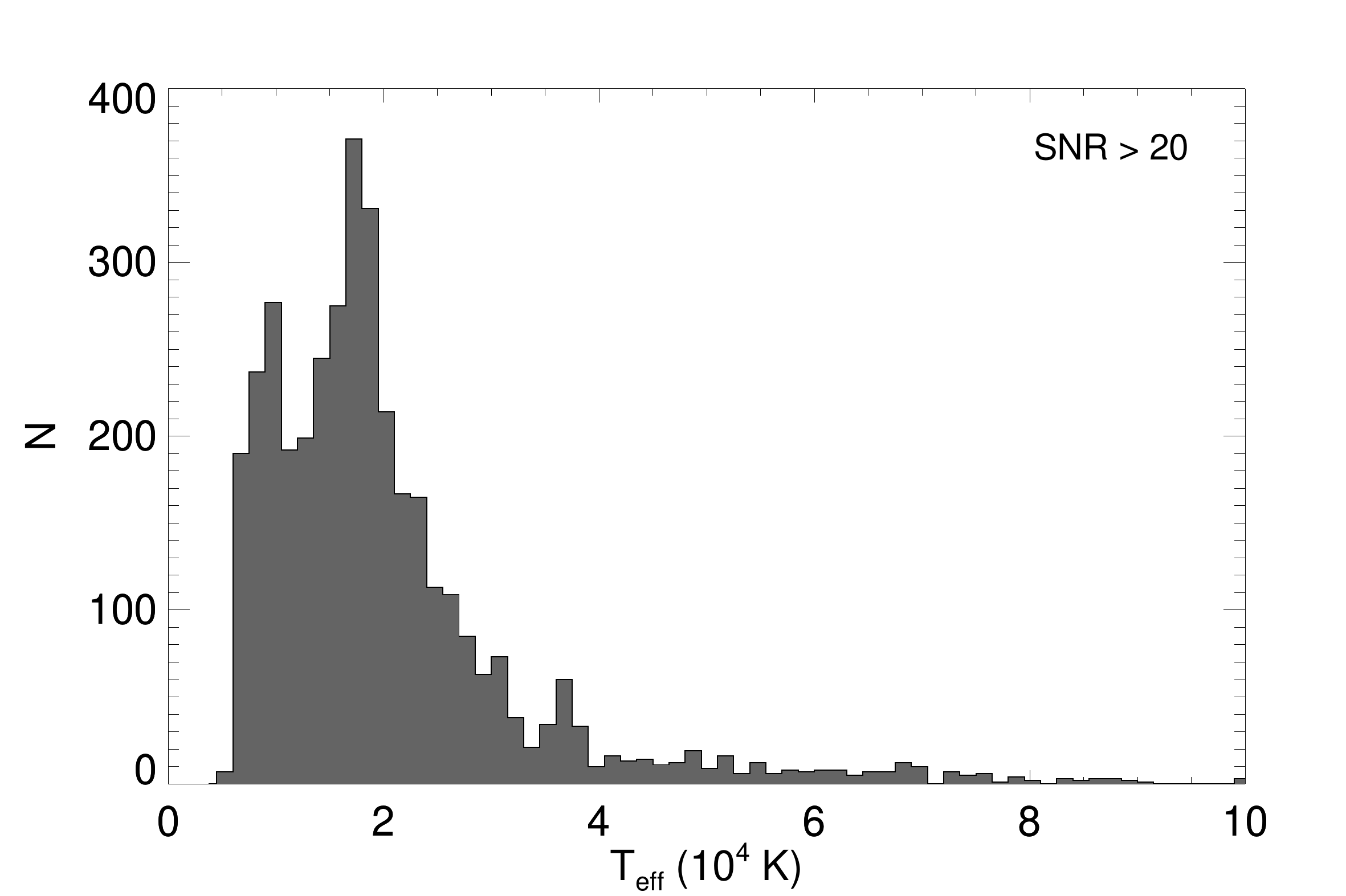}  
\caption{Distribution of effective temperatures  for the SDSS DA white
  dwarf sample  employed in this  study.  Most  of the objects  in the
  sample   have   effective   temperatures  between   $7,000\,$K   and
  $30,000\,$K.  The left panel shows the distribution for white dwarfs
  with spectra with  SNR$>$5, while the right panel  displays the same
  distribution for stars with spectra with SNR$>$20.}
  \label{fig:teff}
\end{figure*}

\begin{figure*}
\includegraphics[width=\columnwidth]{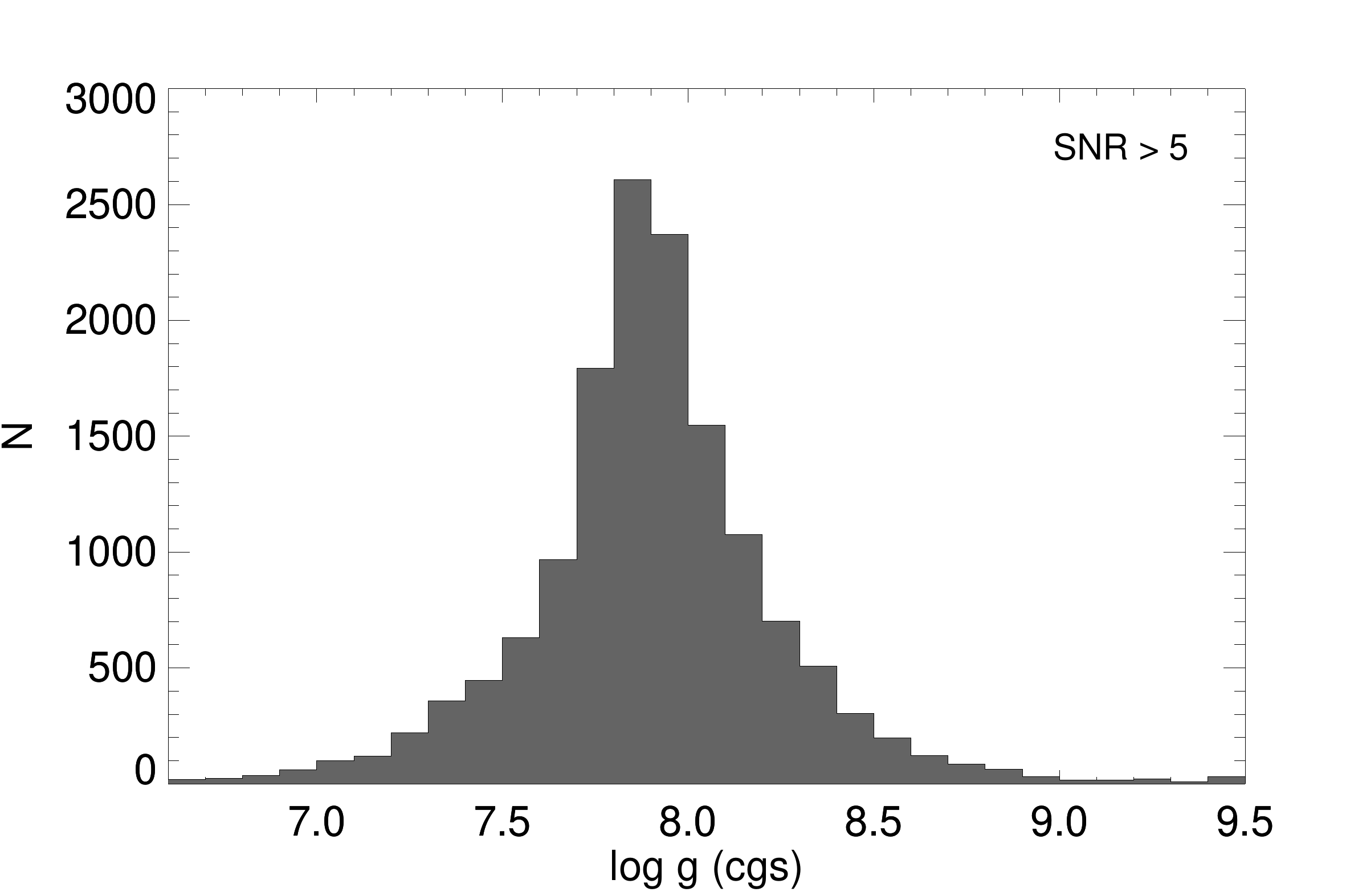}
\includegraphics[width=\columnwidth]{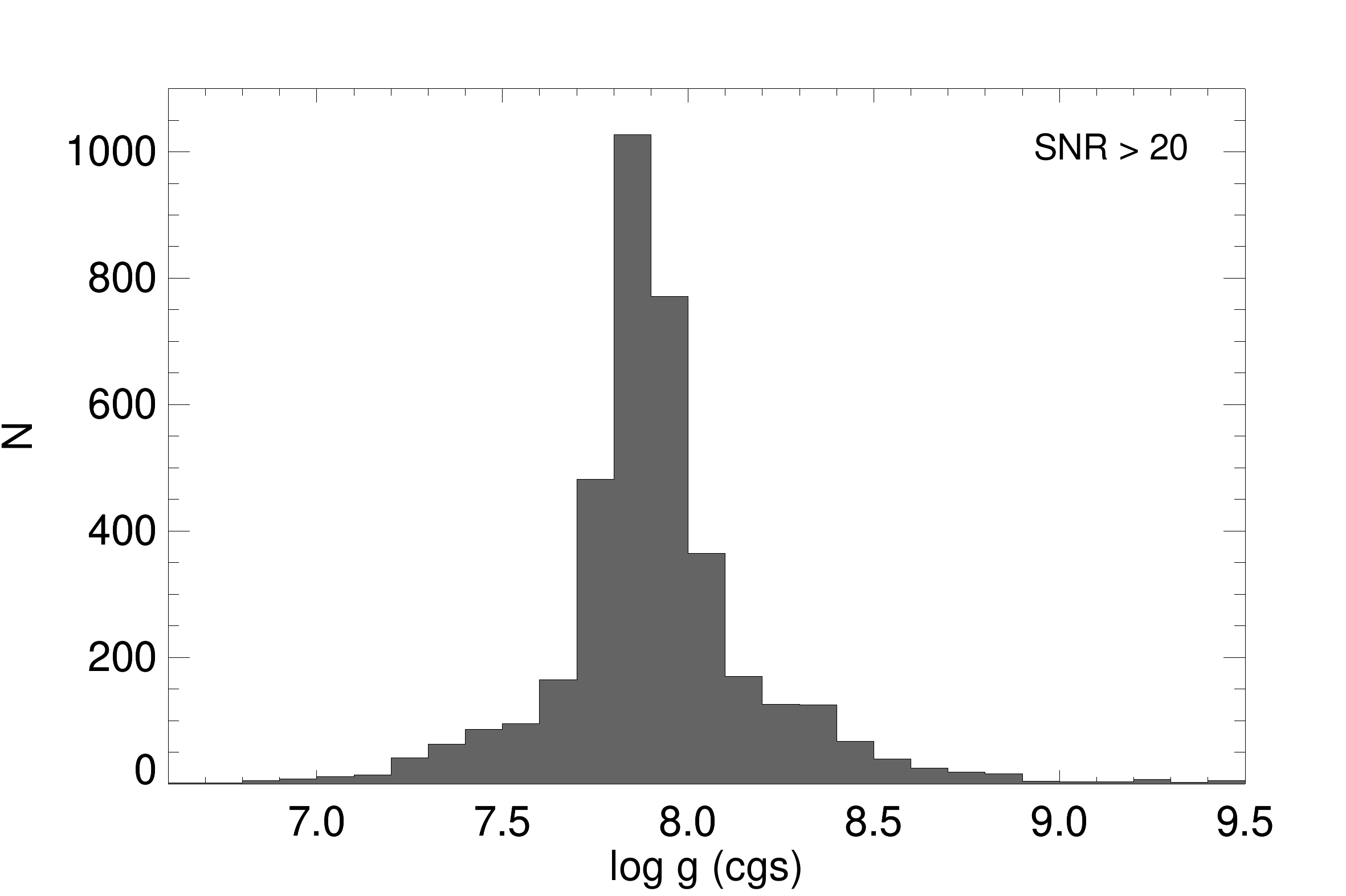}  
\caption{Distribution of surface gravities for the SDSS DA white dwarf
  sample employed in this study.  The surface gravity distribution for
  the sample displays a narrow peak  at around 7.8 dex. The left panel
  shows the distribution for white  dwarfs with spectra SNR$>$5, while
  the  right  panel displays  the  same  distribution for  stars  with
  spectra with SNR$>$20.}
  \label{fig:logg}
\end{figure*}

\section{Distributions of stellar parameters}
\label{distributions}

White dwarfs are classified into several different sub-types depending
on their atmospheric composition.  The  most common of these sub-types
are  DA  white dwarfs,  which  have  hydrogen-rich atmospheres.   They
comprise  $\sim$85  per  cent  of  all white  dwarfs  ---  see,  e.g.,
\citet{kleinmanetal13-1}.  The most distinctive spectral feature of DA
white dwarfs is the Balmer series.   These lines are sensitive to both
the  effective temperature  and  the surface  gravity.  We fitted  the
Balmer  lines sampled  by the  SDSS spectra  with the  one-dimensional
model  atmosphere  spectra  of   \citet{koester10-1},  for  which  the
parameterization  of convection  follows the  mixing length  formalism
ML2/$\alpha =  0.8$.  In order  to account for  the higher-dimensional
dependence of convection, which is important for cool white dwarfs, we
applied the three-dimensional corrections of \citet{tremblayetal13-1}.

\subsection{Effective temperatures and surface gravities}
\label{s-param}

Fig.~\ref{fig:teff} shows the  distributions of effective temperatures
for two sets of data. In  the left panel the distribution of effective
temperatures  for a  sub-sample of  white dwarfs  with spectra  having
SNR$>$5  is  displayed,  whereas  the   right  panel  shows  the  same
distribution  for those  white  dwarfs with  spectra  of an  excellent
quality,  namely those  with  SNR$>$20. For  those  white dwarfs  with
spectra   with  SNR$>$5   the   effective   temperatures  are   within
$6,000\,{\rm  K}\,<$\Teff$<\,100,000\,$K, but  most white  dwarfs have
effective temperatures  between 7,000~K  and 30,000~K,  while although
for  the  sub-sample with  SNR$>$20  the  same  is  true, there  is  a
secondary peak at $T_{\rm eff}\sim10,000$~K.   The origin of this peak
remains unclear, but it is related to the longer cooling timescales of
faint white dwarfs.  However, a  precise explanation of this peak must
be  addressed with  detailed population  synthesis studies,  which are
beyond  the scope  of  this paper.   Fig.~\ref{fig:logg} displays  the
distribution of surface gravities for both the sub-sample with SNR$>$5
(left panel) and that with SNR$>20$  (right panel). As can be seen, in
both  cases surface gravities  ranges from  6.5~dex~$<\log
g<$~9.5~dex with  a narrow peak around  7.85~dex.  These distributions
are very similar to those presented by other authors using SDSS DA
white  dwarfs   \citep[e.g.][]{kleinmanetal13-1,  kepleretal15-1}.  In
summary, we conclude  that the sub-sample of white  dwarfs with SNR$>$20
is  totally  representative  of  the  sample  of  white  dwarfs.  This
sub-sample  will  be used  below  to  derive  in  a reliable  way  the
kinematical properties of the white dwarf population.

\begin{figure}
\includegraphics[width=1.1\columnwidth]{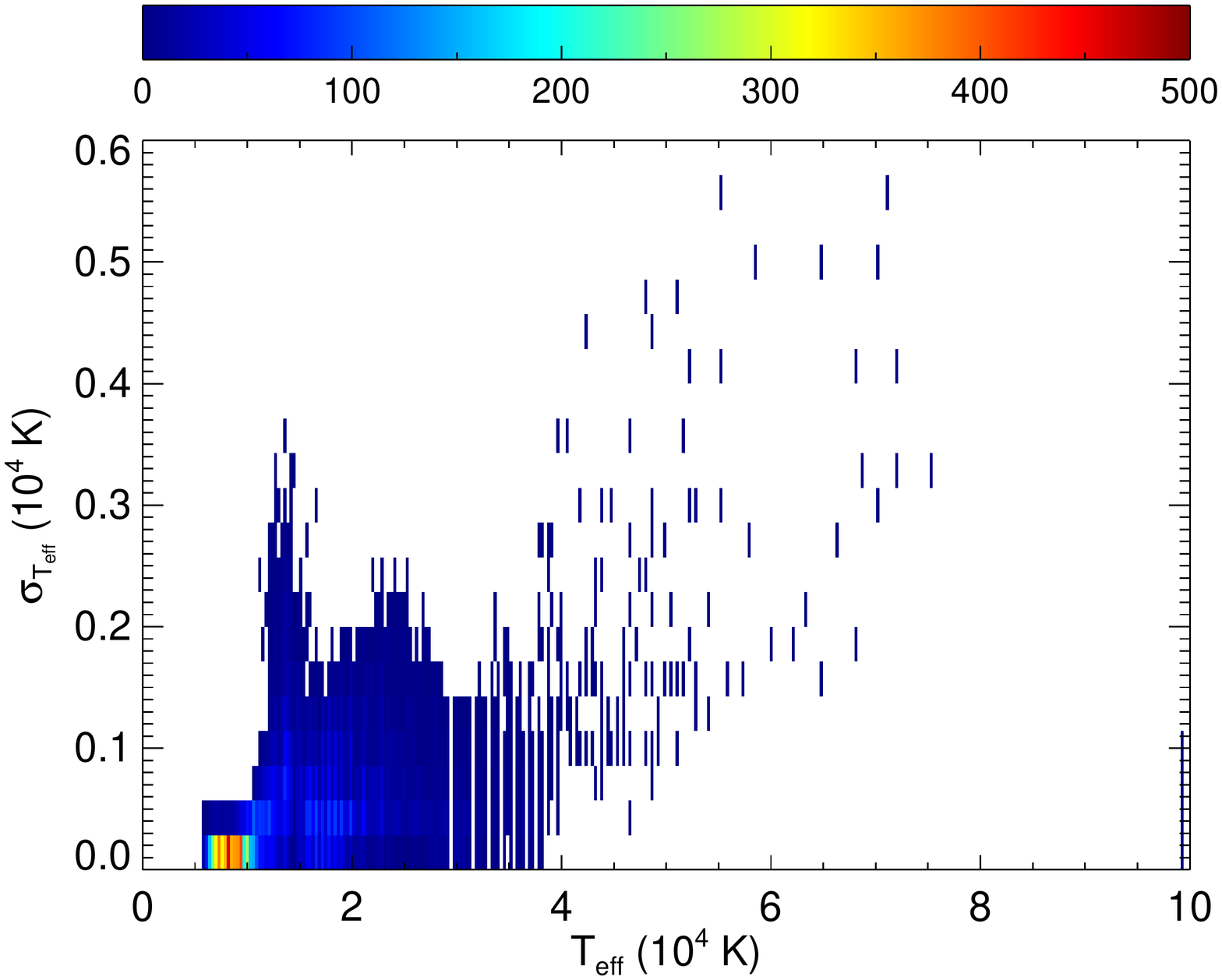}
\includegraphics[width=1.1\columnwidth]{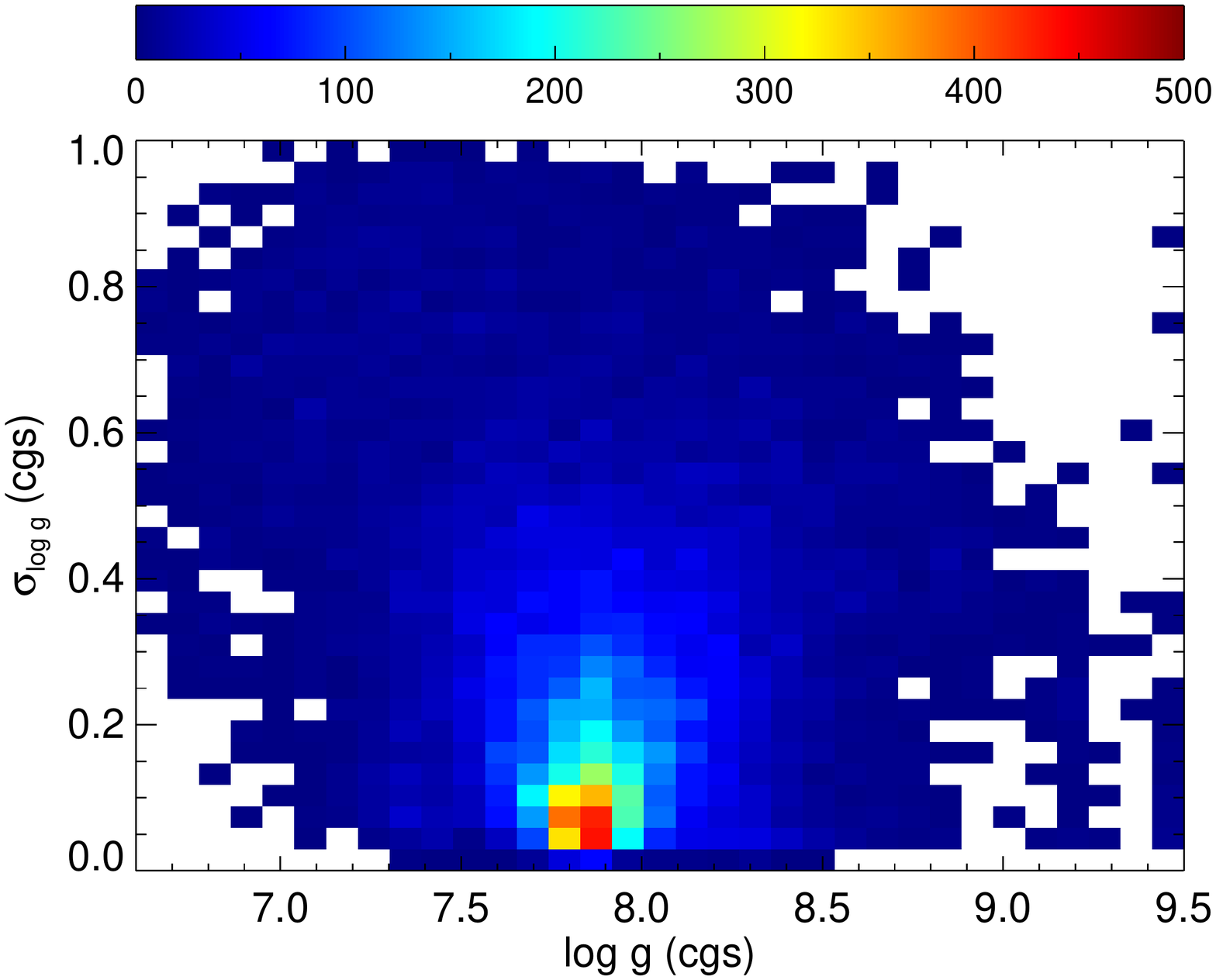}  
\caption{Distribution   of  uncertainties   for  both   the  effective
  temperatures (upper  panel) and  surface gravities (lower panel) for
  the  SDSS  DA  white  dwarf  sample employed  in  this  study.  This
  distribution of errors corresponds to  all white dwarfs with spectra
  having SNR$>$5.}
  \label{fig:errors}
\end{figure}

\begin{figure*}
\includegraphics[width=\columnwidth]{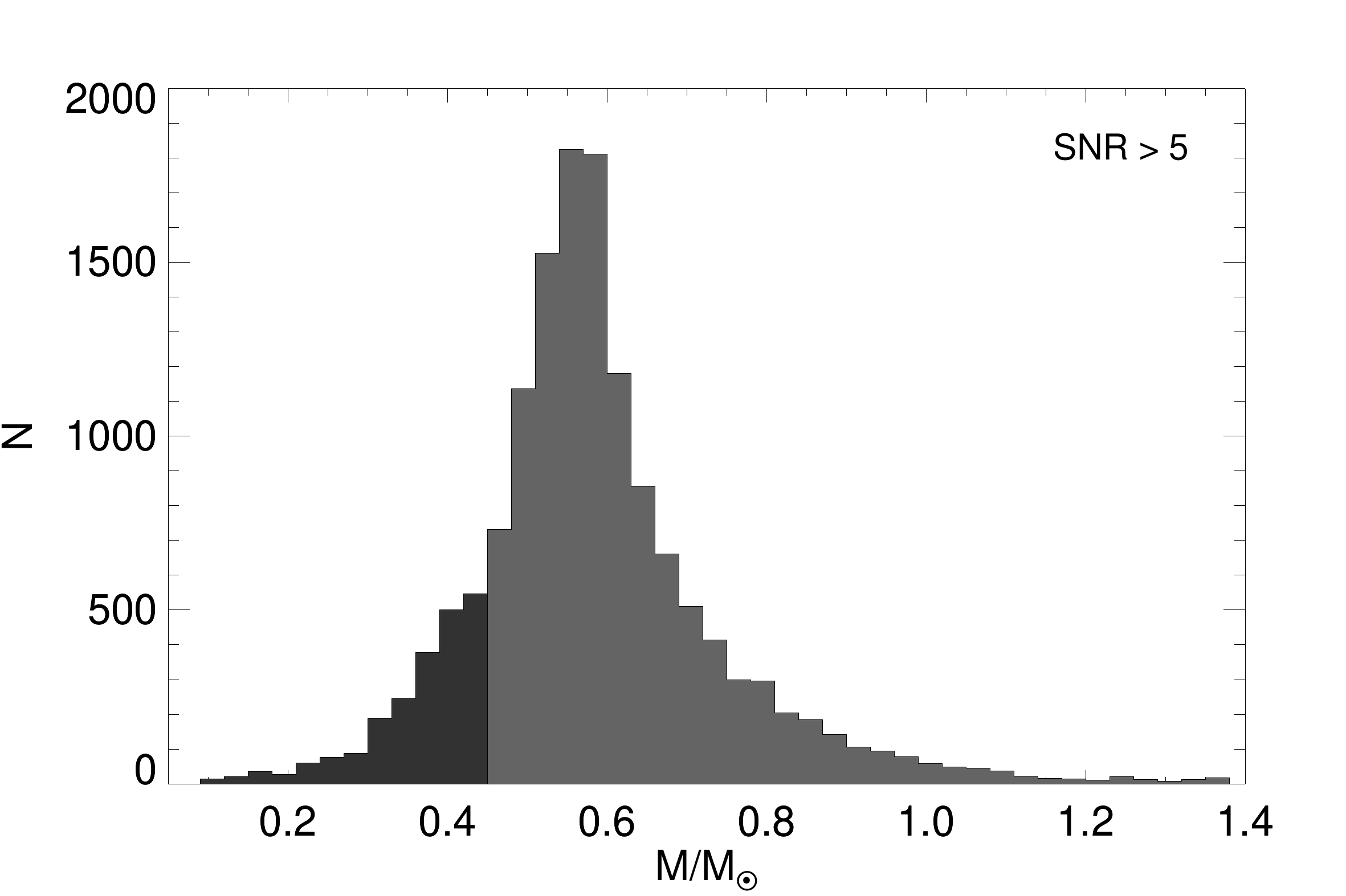}
\includegraphics[width=\columnwidth]{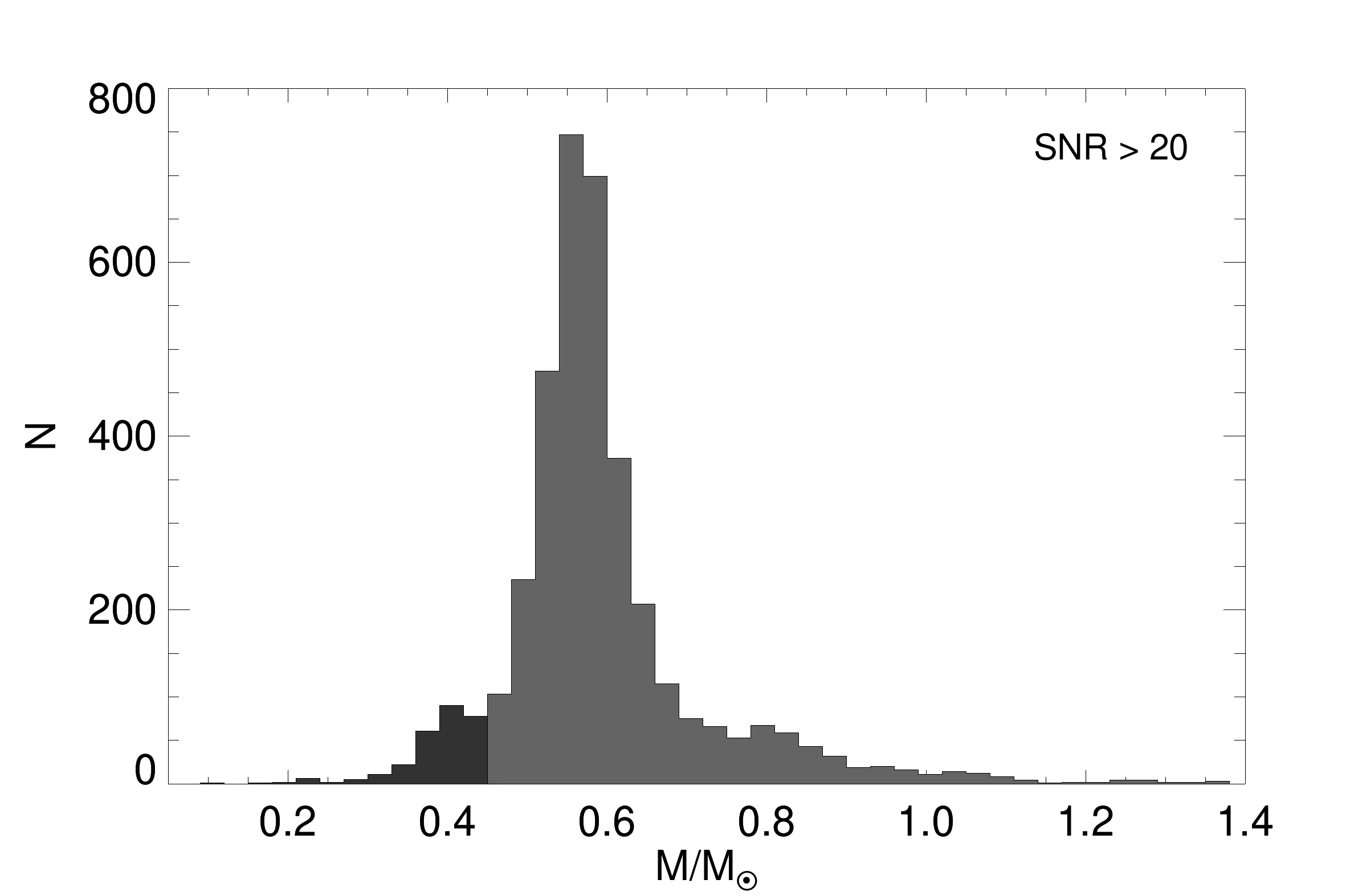}  
\caption{Mass distribution for the SDSS DA white dwarf sample employed
  in  this study.   The left  panel shows  the distribution  for white
  dwarfs with spectra SNR$>$5, while the right panel displays the same
  distribution for stars with spectra with SNR$>$20. In both cases the
  distributions  have a  narrow  peak at  $\sim 0.55\,M_{\sun}$.   The
  black   shaded  area   shows  white   dwarfs  with   masses  $M_{\rm
  WD}<0.45\,M_{\sun}$, which  are thought to be  formed through binary
  evolution (see text for more details).}
  \label{fig:masses_wd}
\end{figure*}

Fig.~\ref{fig:errors}  shows   the  distribution  of  errors   of  the
effective temperatures as a function  of the effective temperature for
all 20,427 DA  white dwarfs with spectra having  SNR$>$5 (upper panel),
and  the  corresponding  distribution   of  uncertainties in surface
gravities as a function of surface gravity (lower panel). It is worth
noting that both  distributions have very narrow  peaks. Actually, the
typical uncertainties peak around $\sigma_{T_{\rm eff}}\sim200\,$K and
$\sigma_{\log {\rm g}} \sim0.15\,$dex.  These values are comparable to
those    obtained    in    equivalent    studies    of    this    kind
\citep{kleinmanetal13-1, kepleretal15-1}.

We  interpolated  the  measured  effective  temperatures  and  surface
gravities   on   the   cooling    tracks   of   \citet{Alt2010a}   and
\citet{Renedo2010} to  derive the  white dwarf masses,  radii, cooling
ages and absolute $UBVRI$ magnitudes. The absolute magnitudes in the
UBVRI  system  were converted  into  the  $ugriz$ system  using  the
equations of  \citet{jordietal06-1}.  

In passing we  note that although white dwarfs  cool down at  almost constant radii, the  radius (hence,
the   surface   gravity)  slightly   evolves   as   time  passes   by.
Consequently,  mass  determinations,   along  with  the  corresponding
uncertainties, depend on both T$_{\rm eff}$ and $\log g$.

\subsection{Mass distribution}
\label{sec:masses}

In Fig.~\ref{fig:masses_wd} we  show the mass distribution  for the white
dwarf  sub-sample  with spectra  with  SNR$>$5  (left panel)  and  the
sub-sample with SNR$>$20  (right panel).  For the  sample with spectra
with SNR$>$5  the observed  mass distribution  exhibits a  narrow peak
with broad  and flat  tails which  extend both  to larger  and smaller
masses.  The mean  mass of the distribution  is near $0.55\,M_{\sun}$.
We found that most white dwarfs in this sample have mass uncertainties
$\sigma_{M_{\rm  WD}}<0.15\,M_{\sun}$.  The  sub-sample with  SNR$>$20
shows  a  similar  behavior,  but   although  the  mean  mass  of  the
distribution  is  essentially  the  same,  the  mass  distribution  is
narrower, and the tails at low  and high masses are much less evident.
This is  particularly true for  white dwarfs with masses  smaller than
$0.45\,M_{\sun}$ --- those that populate the black shaded area in this
figure.  This is  a direct consequence of the  better determination of
white dwarf masses for the sub-sample with spectra with larger SNR.

Theoretical models predict  that white dwarfs of  masses below $0.45\,
M_{\sun}$  have   helium  cores.   The  existence   of  such  low-mass
helium-core white dwarfs cannot be explained by single stellar 	evolution,
as the main sequence lifetimes  of the corresponding progenitors would
be  larger than  the  Hubble time.   It is  then  expected that  these
low-mass white  dwarfs are  formed as a  consequence of  mass transfer
episodes in binary systems, which lead to a common envelope phase, and
hence to a  dramatic decrease of the  binary orbit \citep{Liebert2005,2011MNRAS.413.1121R}. 
We  therefore expect the vast  majority of all low-mass (those with  masses $\leq0.45\,M_{\sun}$) white dwarfs in
our sample to be members of close binaries.

Additionally,  there  is  another  potential source  of  close  binary
contamination in the  white dwarf mass distribution,  for masses above
$0.45\,M_{\sun}$.  Post-common-envelope  binaries, typically including
a carbon-oxygen core  white dwarf, represent a  noticeable fraction of
the    Galactic    white    dwarf   population    ---    see,    e.g.,
\citet{rebassa-mansergasetal12-1}  and  \citet{Camacho2014}.  Some  of
these systems are  likely present in our sample when  the companion is
an unseen  low-mass star or a  second (less luminous) white  dwarf. In
those cases, the brighter and lower  mass white dwarf will likely have
experienced   a   mass    loss   episode.    \citet{Badenes2012}   and
\citet{Maoz2016}  estimated that  the number  of close  binary systems
represent $\sim10$ per cent of the population.

\begin{figure}
  \includegraphics[width=\columnwidth]{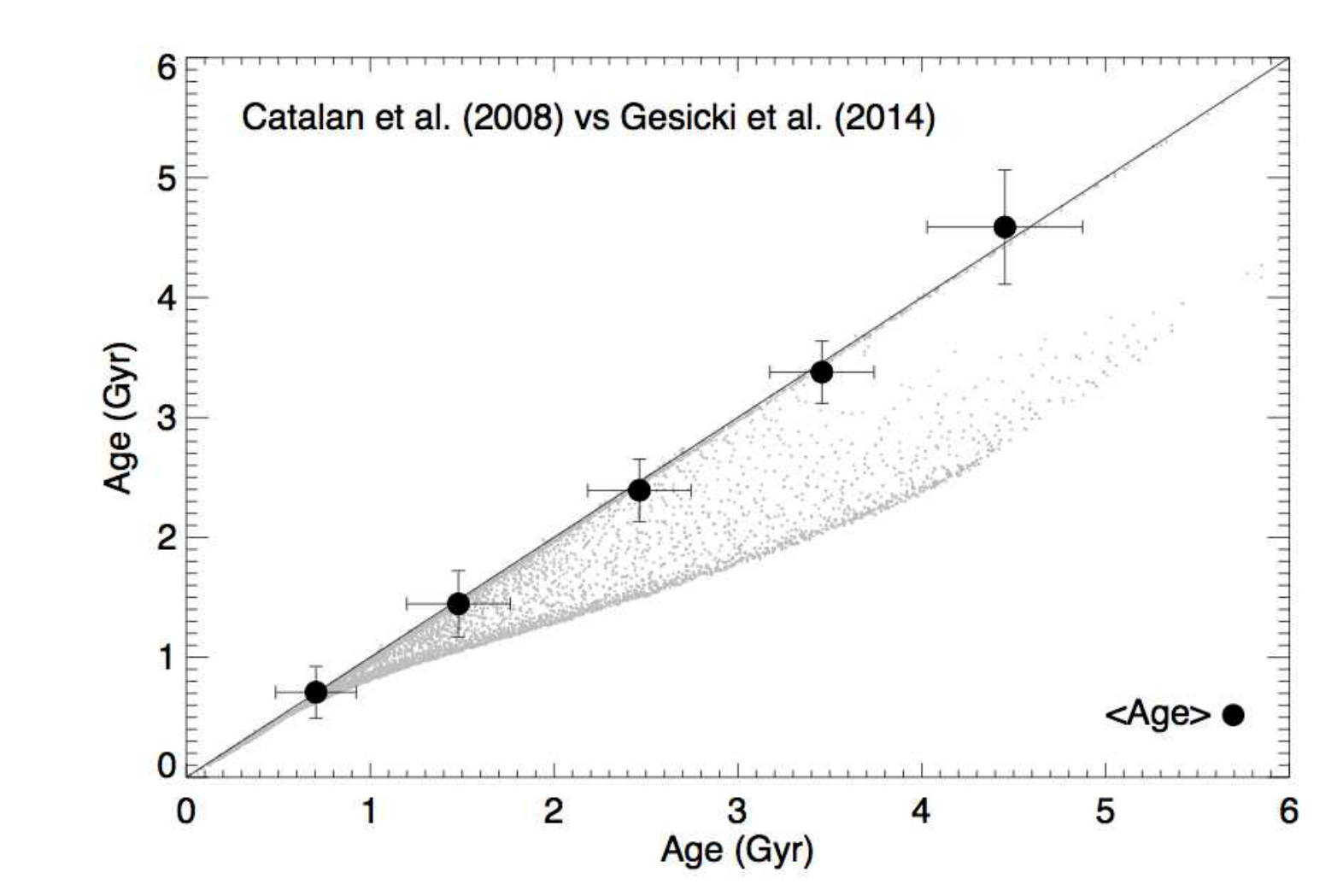}
  \includegraphics[width=\columnwidth]{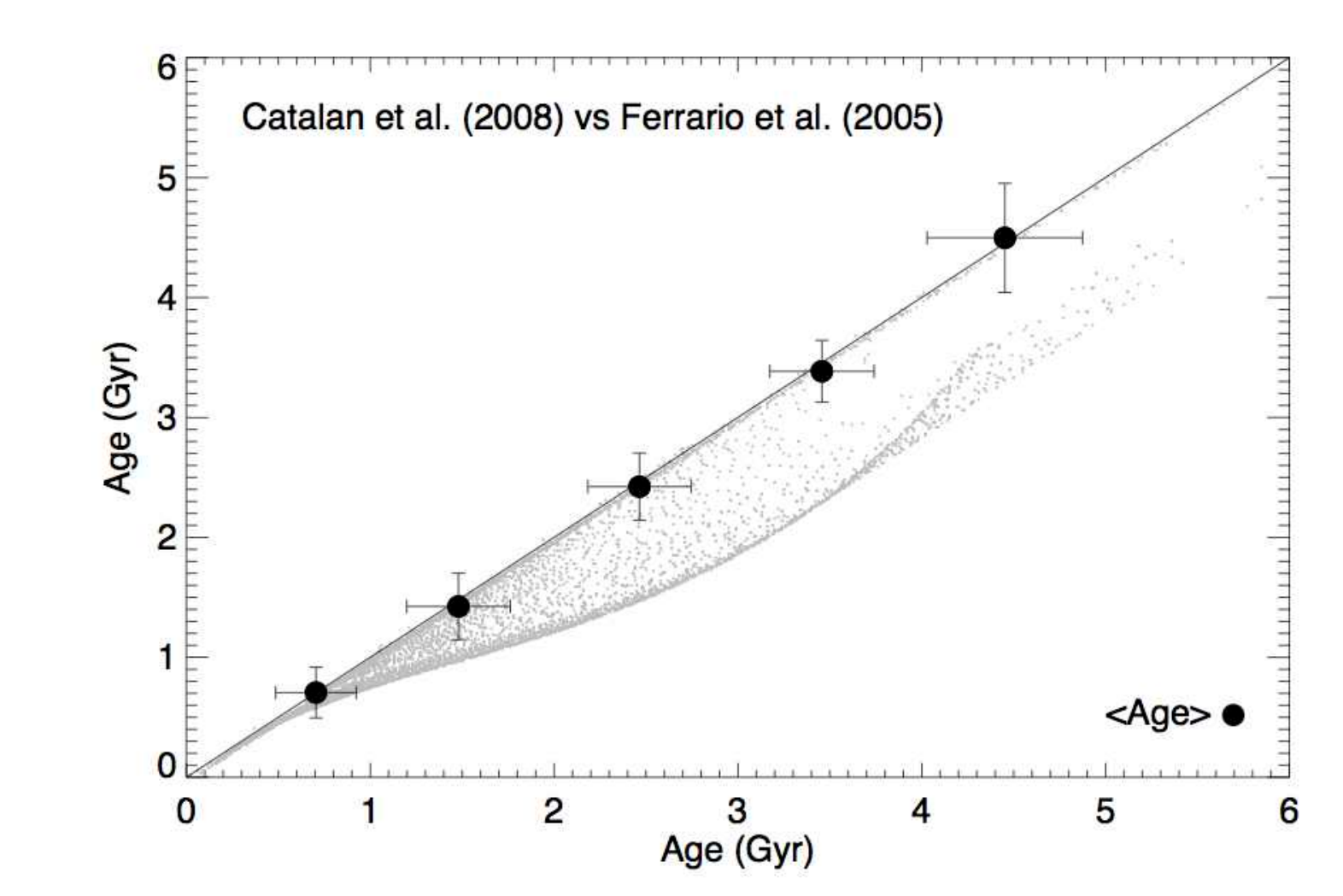}
  \includegraphics[width=\columnwidth]{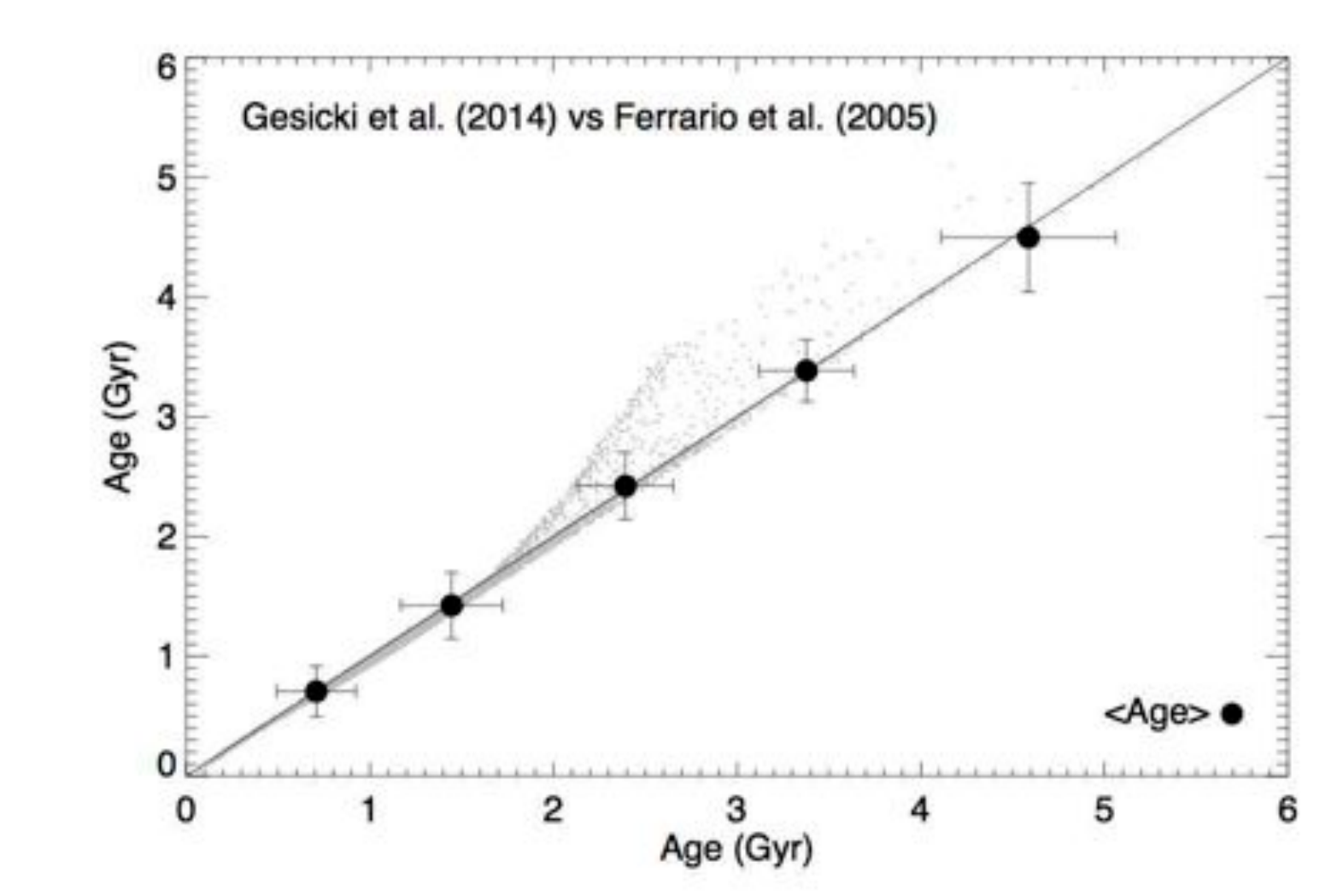}
  \caption{Comparison between the total  ages of white dwarfs obtained
    using the  IMFRs of  \citet{Catalan2008}, \citet{ferrarioetal05-1}
    and \citet{Gesicki2014}.   Note the impact of  the different IMFRs
    in the final estimated individual ages.}
\label{fig:ages_comp}
\end{figure}

\subsection{Ages}
\label{ages_WD}

The age  of a white dwarf  is computed as  the sum of its  cooling age
plus the  main sequence lifetime  of its progenitor.  Using  the white
dwarf  effective  temperatures  and   surface  gravities  obtained  in
Sect.~\ref{s-param} we interpolated these values on the cooling tracks
of \citet{Renedo2010}  to obtain  the corresponding cooling  ages.  To
obtain their  main sequence  progenitor lifetimes  an initial-to-final
mass  relation (IFMR)  --- i.e.,  the relationship  between the  white
dwarf mass  and the mass of  its main sequence progenitor  --- must be
adopted.   Given  that  the  currently  available  IFMRs  suffer  from
relatively  large observational  uncertainties,  especially for  white
dwarf  with  masses  below  $0.55\,M_{\sun}$, here  we  adopted  three
different relationships. More specifically, we used the semi-empirical
IFMR of \citet{Catalan2008} as our reference case.  However, to assess
the uncertainties  in the cooling ages  we also employed the  IFMRs of
\citet{ferrarioetal05-1}  and  \citet{Gesicki2014}.  All  three  IFMRs
cover the range  of white dwarf masses with  carbon-oxygen cores. Even
more,  we only  calculated  ages  for stars  with  masses larger  than
$0.55\,M_{\sun}$.   Using these  IFMRs  we  derived three  independent
values  of  the progenitor  masses  for  each  white dwarf.   We  then
interpolated  the progenitor  masses  in  the solar-metallicity  BASTI
isochrones of \citet{pietrinfernietal04-1} and  computed the time that
the white dwarf  progenitors spent on the main sequence,  and thus the
total age.

\begin{figure}
  \includegraphics[width=\columnwidth]{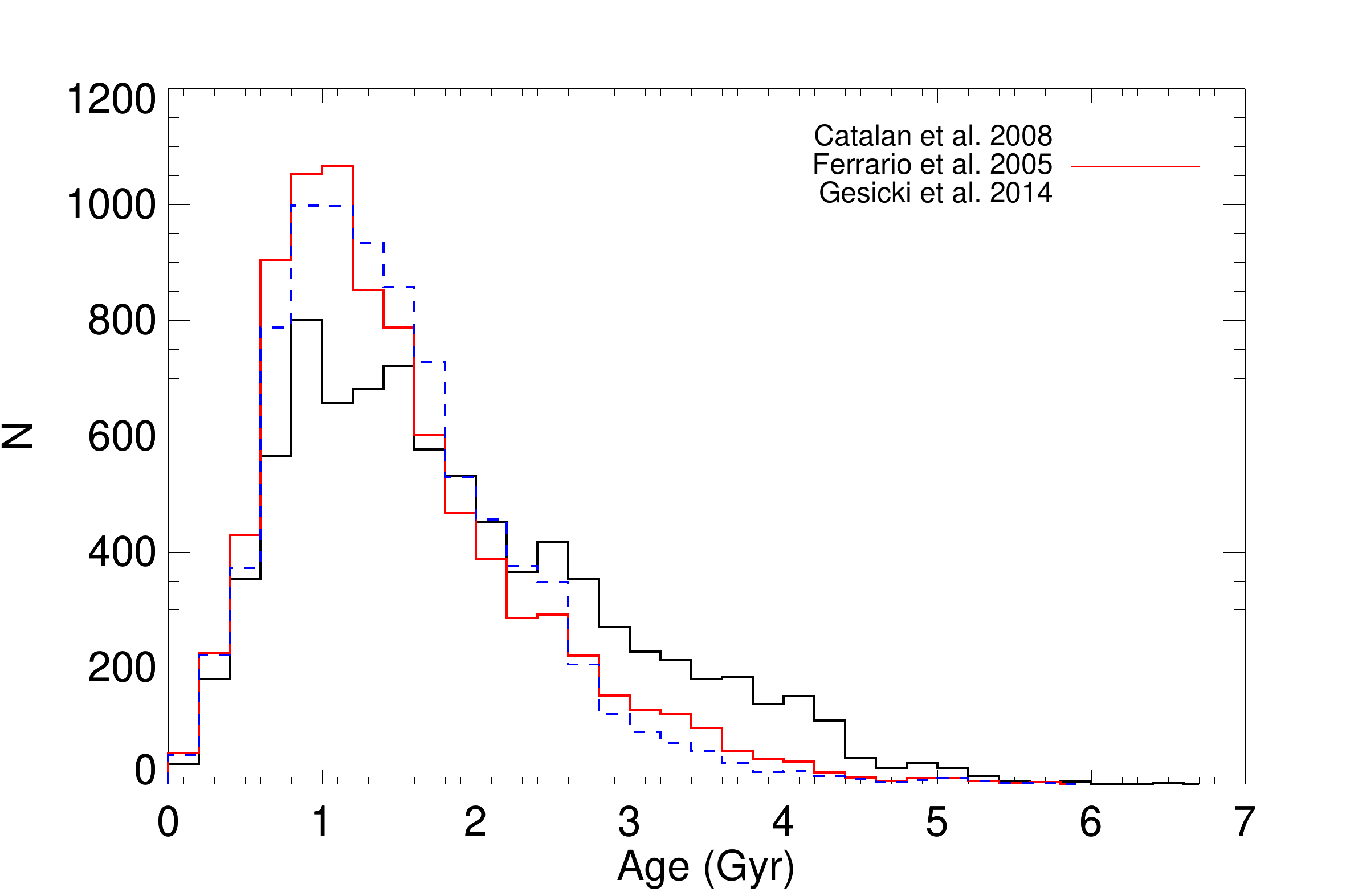}  
  \caption{Age   distributions   using   our   three   adopted   IFMRs
    \citep{Catalan2008,  ferrarioetal05-1,  Gesicki2014}.   The  three
    distributions show a clear peak at around 1~Gyr.}
\label{fig:age_hist}
\end{figure}

In Fig.~\ref{fig:ages_comp}  we compare the total  ages obtained using
different IFMRs for white dwarfs with spectra with SNR$>$5.  As can be
seen, although there  is a sizable fraction of stars  that have nearly
equal ages, systematic  effects arise when using the  IFMRs derived by
different authors.  Thus, we turn our attention to quantify the impact
of  the different  IFMRs  on the  ages  of stars  in  our sample.   In
Fig.~\ref{fig:ages_comp} we  also show  the average ages  and standard
deviations for  our sample of white  dwarfs when stars are  grouped in
bins  of 1~Gyr  duration.  As  can  be seen  the average  ages are  in
excellent agreement with  each other, and the  standard deviations are
substantially  smaller than  the size  of  the bin.   This means  that
although  the  ages  of  individual  white dwarfs  may  differ 
depending on  the adopted IFMR, the average  ages in each 1~Gyr bin are in
good  agreement. Thus,  these ages  can be  safely employed  to derive
average structural properties of our Galaxy.

We also  study the distribution of ages of white  dwarfs  in our
sample. Fig.~\ref{fig:age_hist} shows the  distribution of white dwarf
ages for the  three IFMRs already mentioned.   The three distributions
show  a peak  around 1~Gyr.   However,  for the  distribution of  ages
obtained   using    the   IFMRs   of    \citet{ferrarioetal05-1}   and
\citet{Gesicki2014} the peak is  narrower, whereas the distribution of
ages obtained using the IFMR of \citet{Catalan2008} has a smaller peak
and shows an  extended tail with a significant number  of white dwarfs
with ages ranging  from 2 to 5~Gyr.  This is because the  slope of the
IFMR of \citet{Catalan2008}  is steeper, and  thus results in
an extended age  distribution. Finally, in all three cases  there is a
clear paucity of white dwarfs with ages larger than $\sim 6$~Gyr. This
is because relatively massive white dwarfs older than $\sim  6$~Gyr are
typically rather  cool (\Teff  $<$ 7,000  K) and  hence too  faint and
difficult to detect by the SDSS.  Hence,  we will not be able to probe
the very first stages of the Galactic disc.

Finally, we emphasize that the age uncertainty for  individual   white  
dwarfs   depends  on   both  its   mass  (or, equivalently,  the surface  
gravity) and  its luminosity  (namely, the effective  temperature).   
We  found that  typical  age  uncertainties cluster around 
$\sigma_{\rm age}=0.5$~Gyr and that most of the SDSS DA white 
dwarfs in this study have $\sigma_{\rm age}< 2.5$~Gyr. This final 
error budget may be slightly underestimated as we employed solar 
metallicity isochrones for all the objects, however this is a reasonable 
assumption for our local sample as most of the objects belong to the 
thin disc where a metallicity close to solar is the most common 
value for a sample between $R_{\rm G}=7$~kpc to 9~kpc, and 
vertical distances from  the Galactic  plane ranging from $Z=-0.5$~kpc  
to 0.5~kpc \citep{2015ApJ...808..132H}.

\subsection{Rest-frame radial velocities}
\label{sec:rvs}

We  determined the  radial  velocities (RVs)  from  the observed  SDSS
spectra using  a cross-correlation  procedure originally  described by
\citet{TonryDavis1979}. We used  128 DA white dwarf model 
spectra \citep{koester10-1}  at  the  SDSS  nominal
spectral resolution.   The models cover effective  temperatures in the
range  $6,000\,$K to  $80,000\,$K and  surface gravities  from 7.5  to
9~dex.  We used the IRAF\footnote{IRAF  is distributed by the National
Optical Astronomy Observatories, which are operated by the Association
of  Universities for  Research in  Astronomy, Inc.,  under cooperative
agreement with  the National  Science Foundation} package  {\tt rvsao}
\citep{KurtzMink1998}.

\begin{figure}
  \includegraphics[width=\columnwidth]{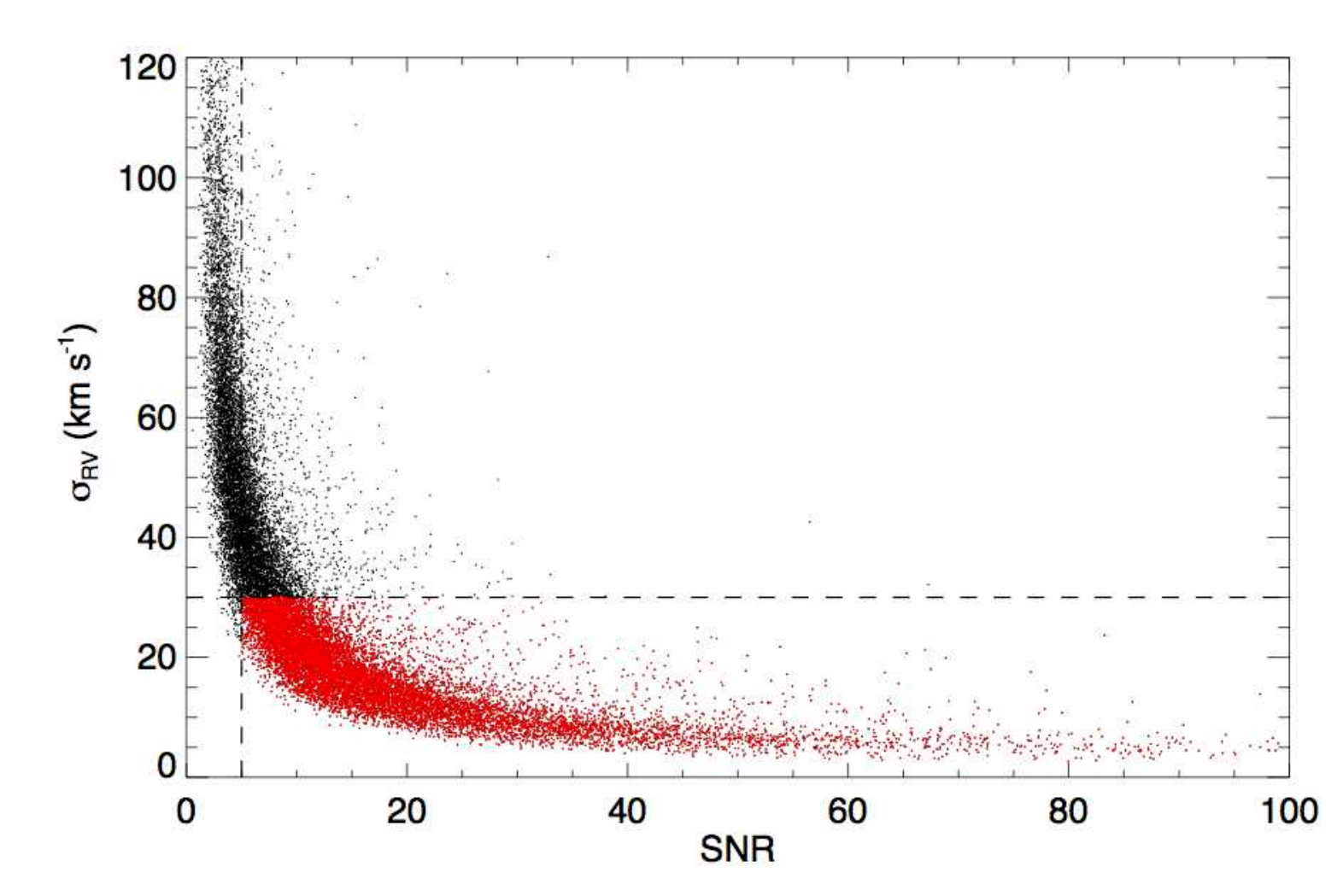}  
  \caption{Uncertainties  in RV  as  a  function of  SNR  of the  SDSS
    spectra  for the  our entire  sample, a  total of  20,247 objects.
    Objects with SNR$<$5 are associated to large uncertainties and are
    not considered in our analysis.   All the white dwarf spectra with
    SNR$>$20 have $\sigma_{RV}<20\,$km s$^{-1}$.}
\label{fig:sn_eRV}
\end{figure}

The low spectral  resolution and the low  SNR (see Fig.~\ref{fig:SNR})
for most of SDSS DA  white dwarf spectra together  with the broad
Balmer lines make  deriving precise RVs a challenging  task.  Thus, to
improve the RV determinations we implemented a Fourier filter. The aim
of this filter  is to suppress some of the  low-frequency power, making
the Balmer lines narrower.  We also designed the filter for an optimal
removal of  photon noise  at the high-frequency  end of  the spectrum.
Also, to  derive the  RV uncertainties  we used  the cross-correlation
coefficient  ($r$), where  $r$ is  the ratio  of the  correlation peak
height    to    the    amplitude   of    the    antisymmetric    noise
\citep{TonryDavis1979}.  Fig.~\ref{fig:sn_eRV} shows the uncertainties 
from the cross-correlation in RV as  a function of  the SNR  of the SDSS  
spectra for our  20,247 DA white dwarfs.   For stars  with SNR$<$5,  
$\sigma_{\rm RV}$  is always larger than $\sim  25$~km~s$^{-1}$. For 
spectra with SNR larger than 20 we found $\sigma_{\rm  RV}$$<$15~km~s$^{-1}$.

\begin{figure}
  \includegraphics[width=1.12\columnwidth]{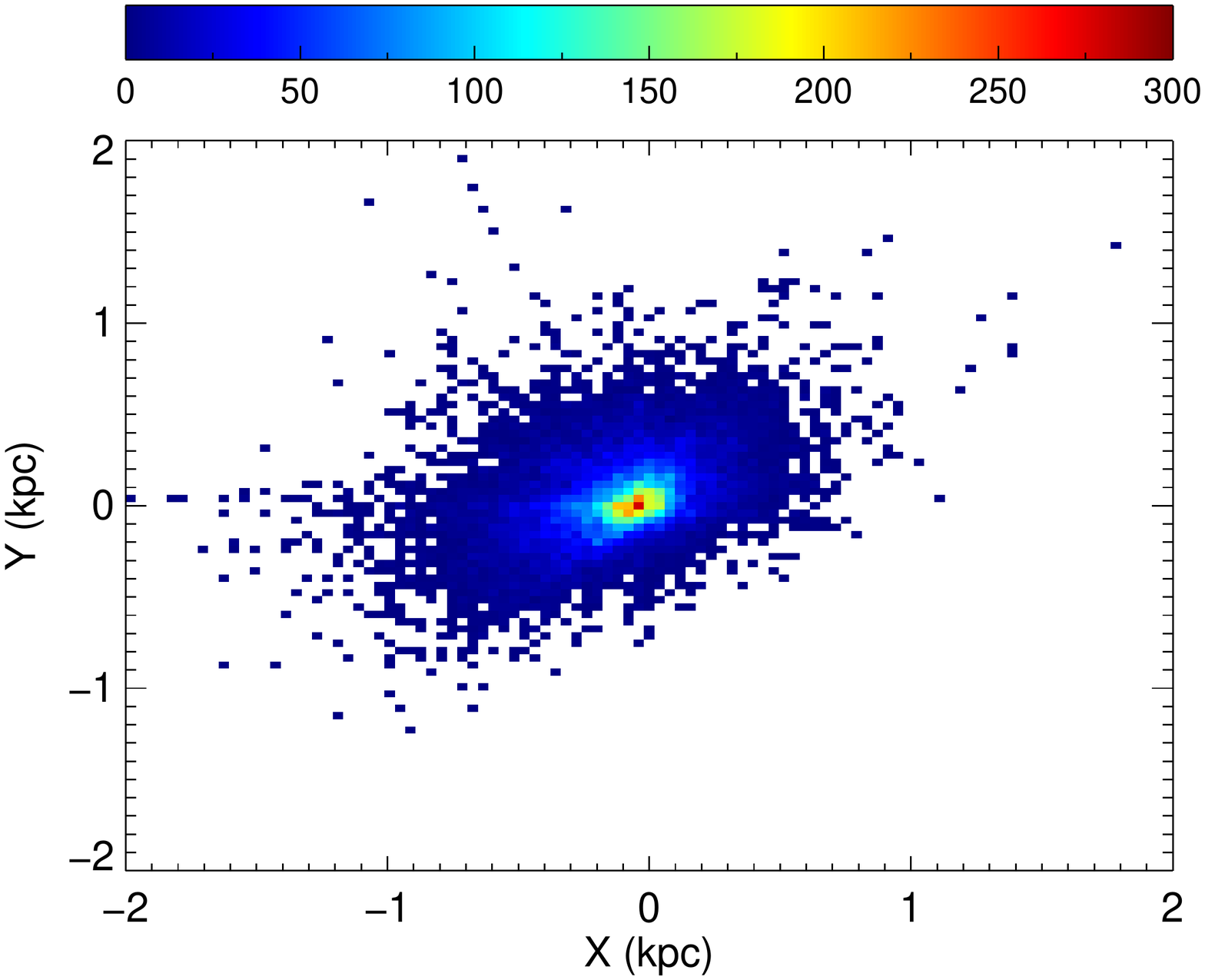}
  \includegraphics[width=1.12\columnwidth]{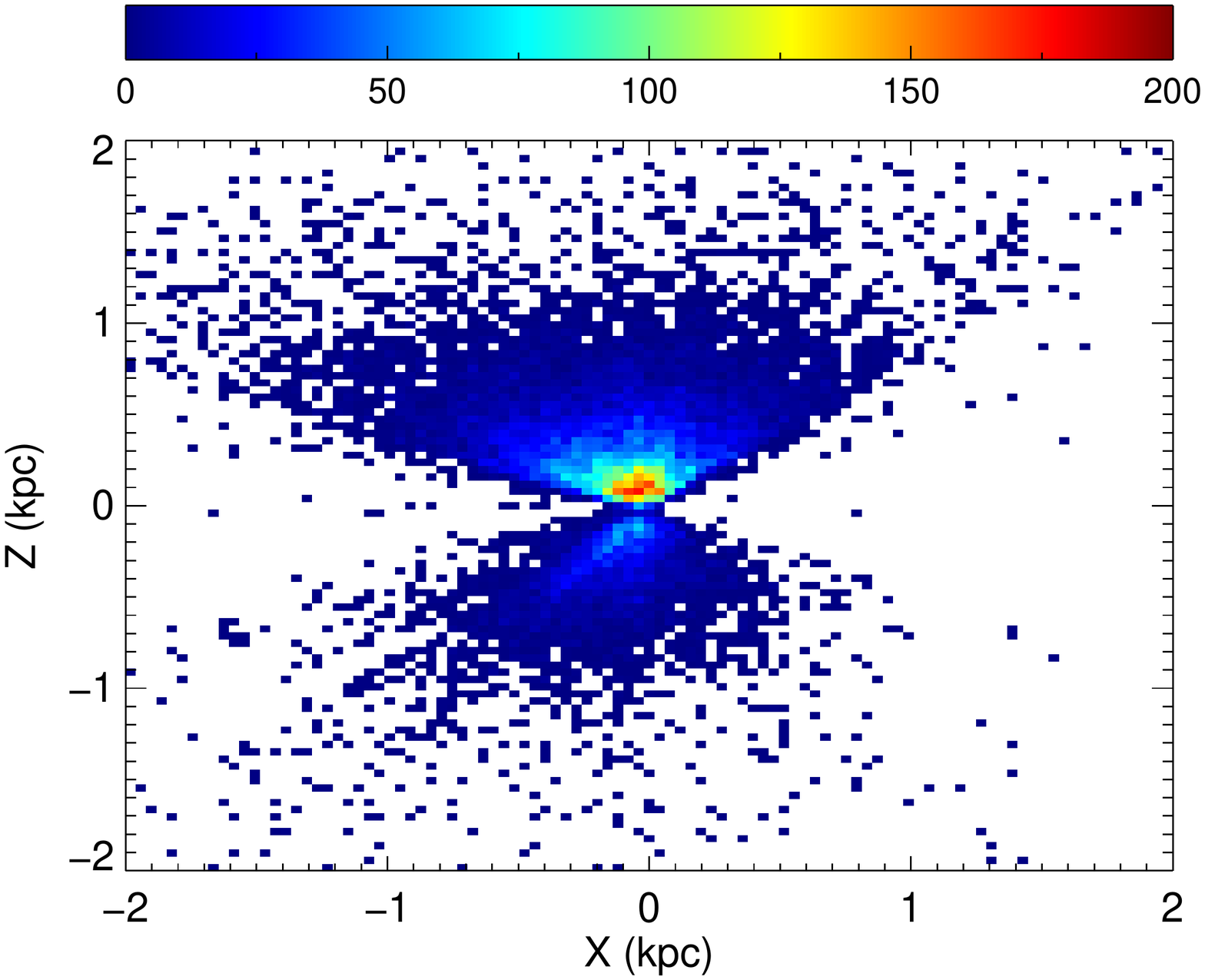}
  \includegraphics[width=1.12\columnwidth]{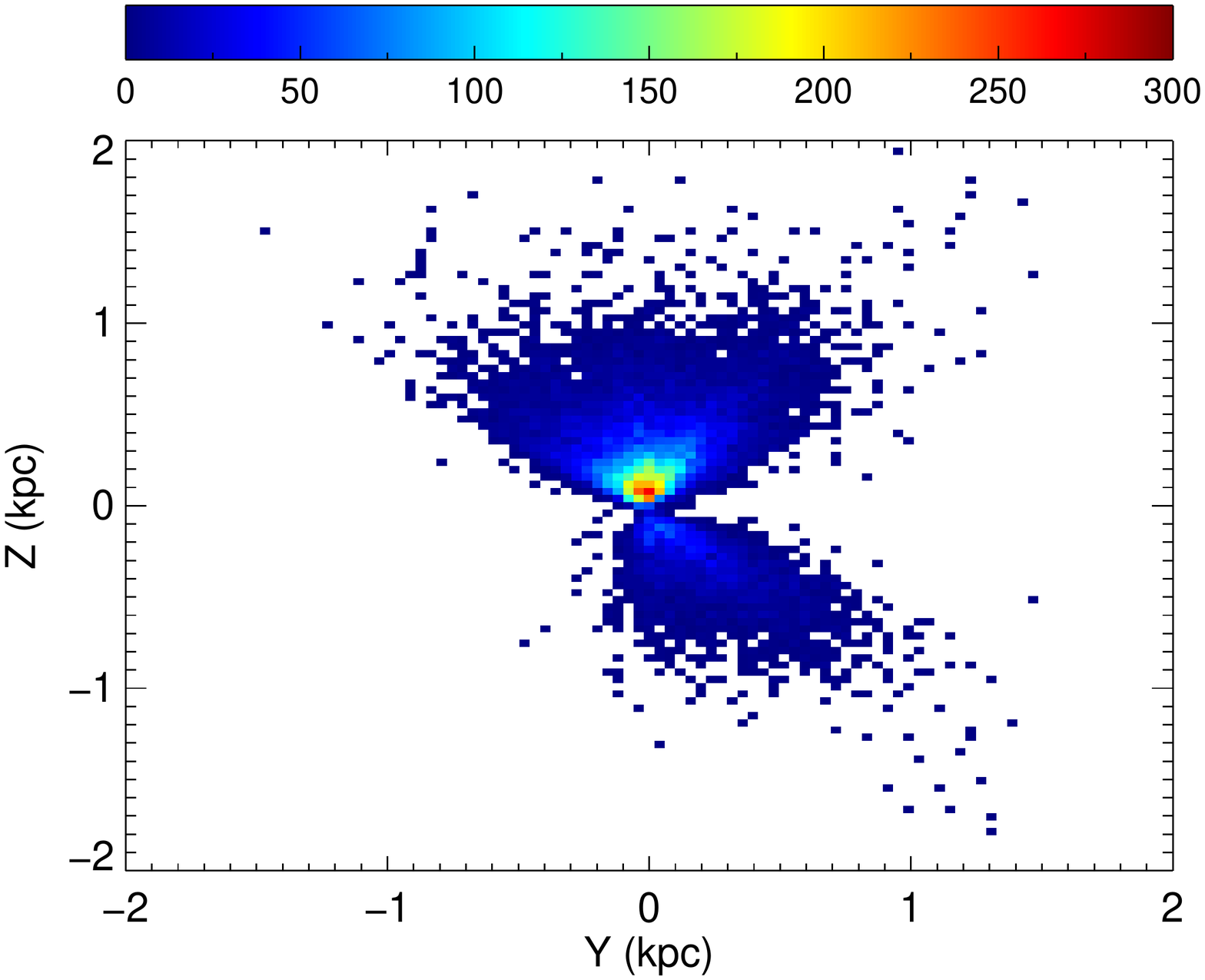}
  \caption{The $XYZ$ distribution of the  white dwarf sample color coded according to 
  the spatial density, see text for details.}
\label{fig:XYZ_space}
\end{figure}

The spectra of white  dwarfs are affected by gravitational redshifts 
due to  their high surface gravities. Thus, the velocities measured by
the cross-correlation  technique (hereafter RV$_{\rm cross}$)  are the
sum of two individual  components \citep{falconetal10-1},  an 
intrinsic radial velocity component (RV)  and the contribution of the
gravitational redshift RV$_{\rm grav}=GM/Rc$.   In this expression $G$
is the gravitational  constant, $c$ is the speed of  light and $M$ and
$R$ are the white dwarf mass  and radius respectively. Note that these
quantities are known for  each star (see Sect.~\ref{s-param}).  Hence,
the radial velocities of our SDSS DA white dwarfs are finally obtained
as RV=RV$_{\rm cross} - GM/Rc$. The error contribution from the 
gravitational redshift RV$_{\rm grav}$ has a typical value of 
1.5 km s$^{-1}$, and all the sample has $\sigma$ $<$ 4 km s$^{-1}$ 
when SNR$>$20. 

\subsection{Distances and spatial distribution}

We  derived the  distances  to all  SDSS DA  white  dwarfs from  their
distance moduli $M_\mathrm{g}-g-A_\mathrm{g}$, where $M_\mathrm{g}$ is
the SDSS  $g$-band absolute magnitude (see  Sect.\,\ref{s-param}), $g$
is  the SDSS  $g$-band apparent  magnitude and  $A_\mathrm{g}$ is  the
$g$-band  extinction.   $A_\mathrm{g}  =  E(B-V)  R_\mathrm{g}$,  with
$E(B-V)$ interpolated in the tables of \citet{schlafly+finkbeiner11-1}
at  the  specific  coordinates  of  each white  dwarf,  and  we  adopt
$R_\mathrm{g}=3.3$  \citep{yuanetal13-1}.    Because  the   values  of
$E(B-V)$  given  by  \citet{schlafly+finkbeiner11-1} are  for  sources
located at infinity, it is  likely that they have been over-estimated,
as white  dwarfs are  intrinsically faint and  are generally  found at
distances 0.5--1\,kpc \citep{rebassa-mansergasetal15-1}.  Hence,
we   computed  an   independent   estimate  of   $E(B-V)$  using   the
three-dimensional extinction model  of \citet{chenetal99-1}, where our
preliminary distance determination was used as input parameter.  If the
difference between  both values  of $E(B-V)$ was  larger than  0.01, we
adopted  the estimate  obtained using  the three-dimensional  extinction
map. We then computed again the extinction $A_\mathrm{g}$.  Afterwards,
we determine again the distance, which was then used to calculate a new
value of  $E(B-V)$ from the  three-dimensional map.  This  was repeated
for each white dwarf until the  difference between the adopted and the
calculated $E(B-V)$ became $\leq$0.01.

To check which regions  of the sky are probed by  our SDSS white dwarf
sample,  and  before examining  the  space  velocities, we  study  the
spatial     distribution     of      stars     in     our     catalog.
Fig.~\ref{fig:XYZ_space} shows  the spatial distribution of the white dwarfs sample.  
We use  a right-handed Cartesian  coordinate system
with $X$ increasing towards the  Galactic center, $Y$ in the direction
of  rotation  and  $Z$  positive   towards  the  North  Galactic  Pole
(NGP). The white dwarf sample probes a region between $0.02<d<2\,$kpc,
but the vast majority of white dwarfs are located at distances between
0.1 and  $0.6\,$kpc.  Note as  well that  our sample has  white dwarfs
with relatively  large altitudes from  the Galactic plane ---  see the
middle  and bottom  panels  of  Fig.~\ref{fig:XYZ_space}. Finally,  we
mention  that most  white dwarfs  have distance  uncertainties smaller
than     $0.1\,$kpc,      being     the    typical relative uncertainty in distance  
around 5$\%$. All the stars with a SNR$>$20 have a relative error better than 
20$\%$.

\begin{figure}
  \includegraphics[width=\columnwidth]{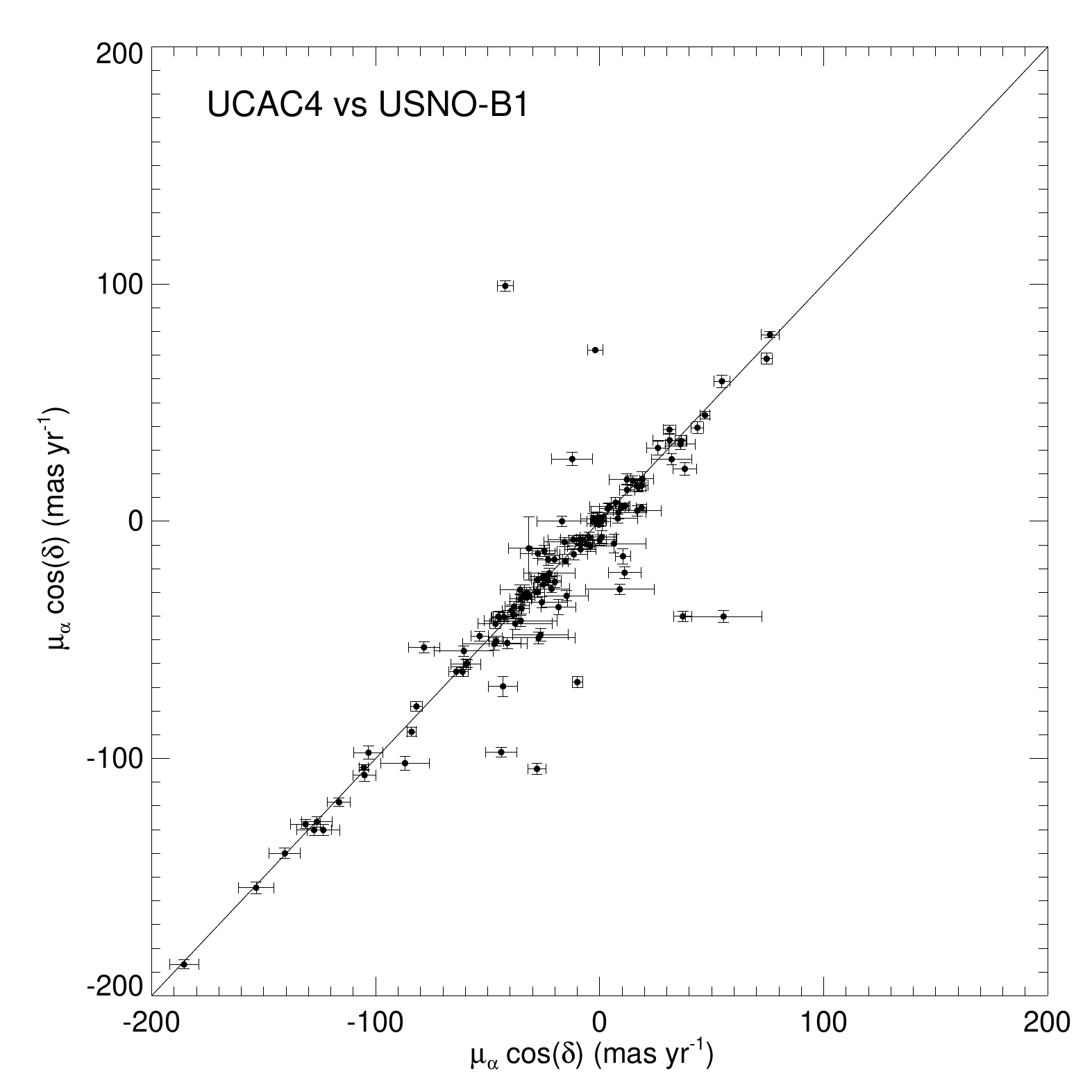}
  \includegraphics[width=\columnwidth]{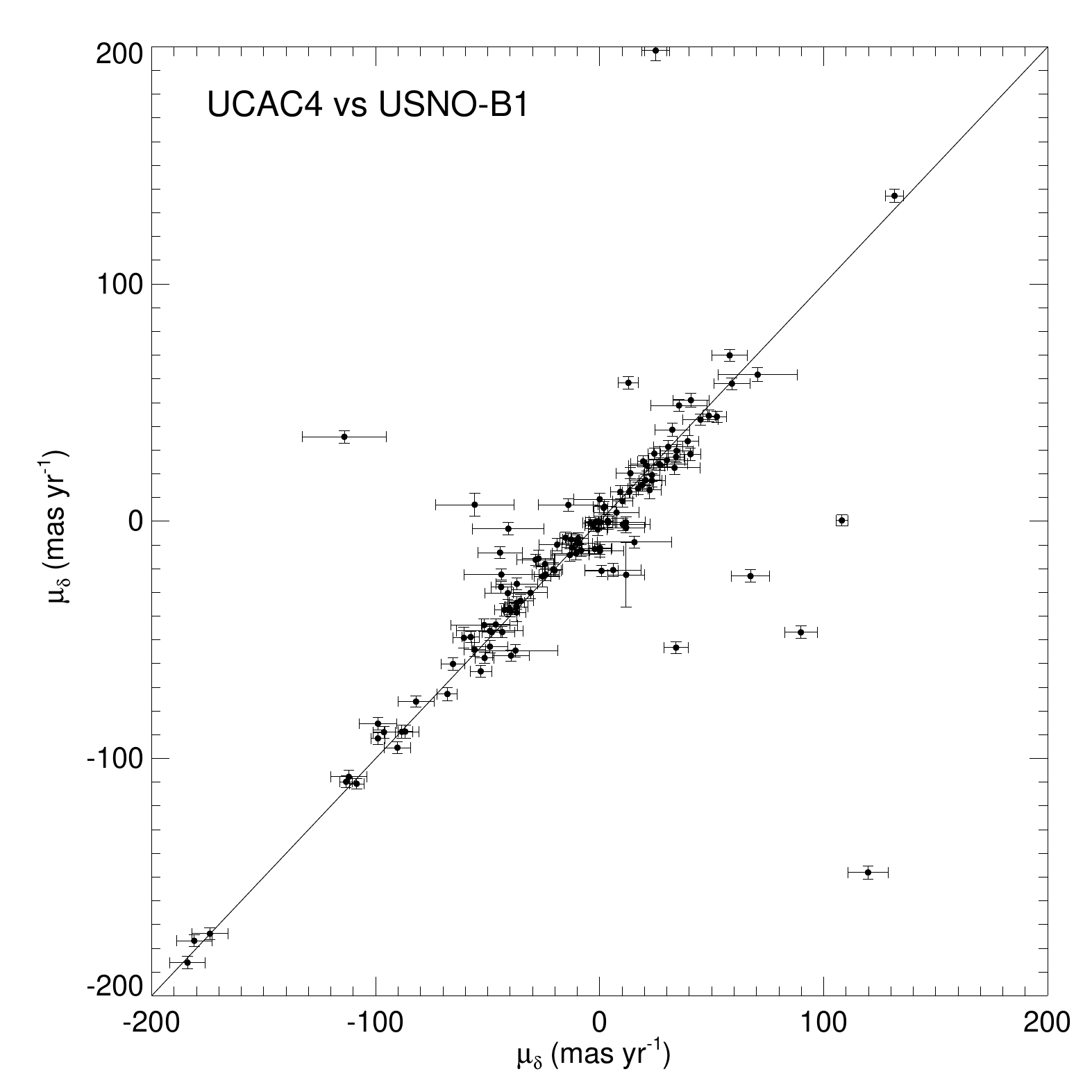}
  \caption{A comparison of  the proper motions of 125  white dwarfs in
    common  between the  USNO-B1 and  UCAC4 catalogs.   Note that,  in
    general, there is a good agreement between both sets of data.}
  \label{fig:PM}
\end{figure} 

\begin{figure}
  \includegraphics[width=1.05\columnwidth]{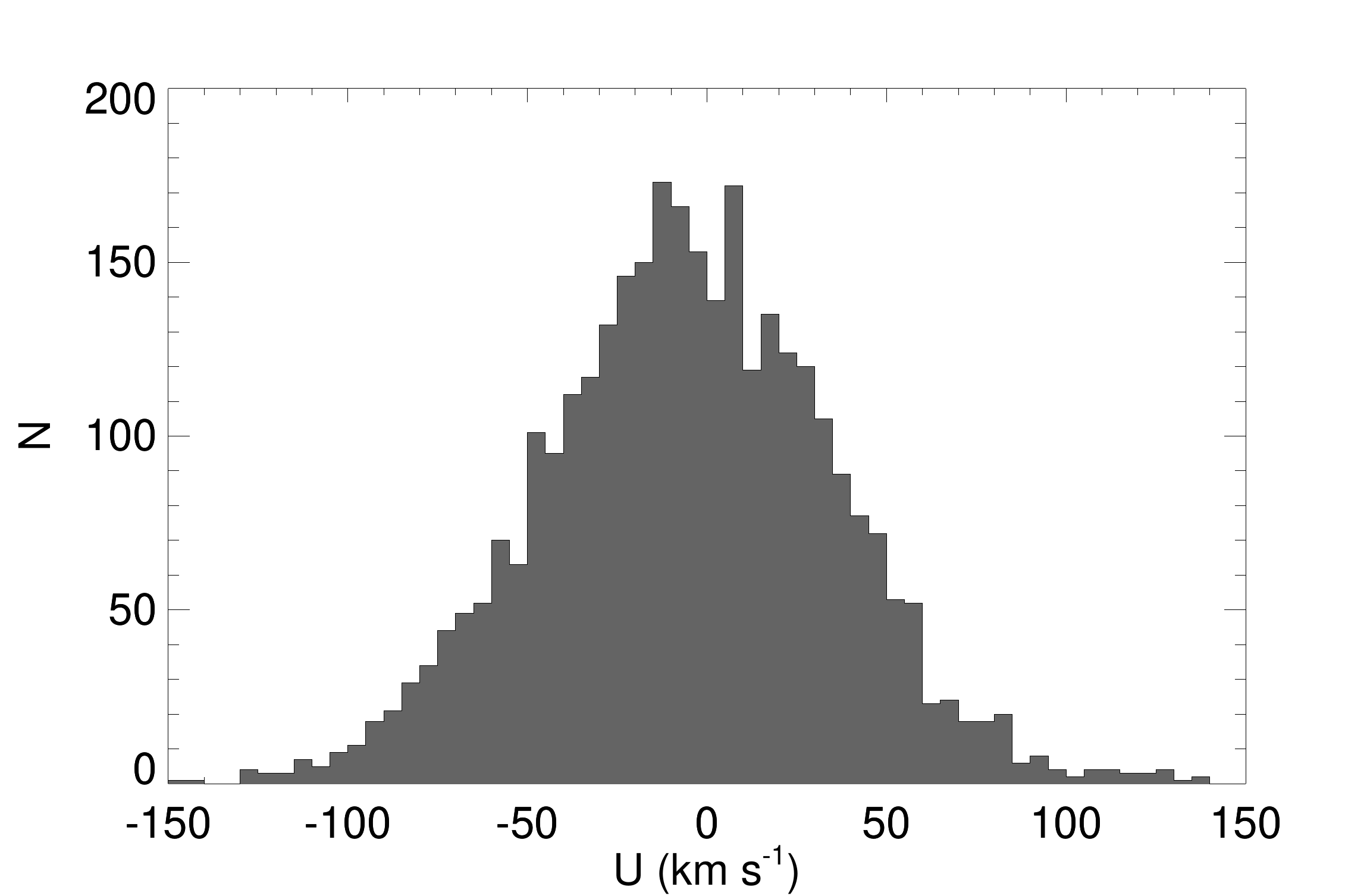}
  \includegraphics[width=1.05\columnwidth]{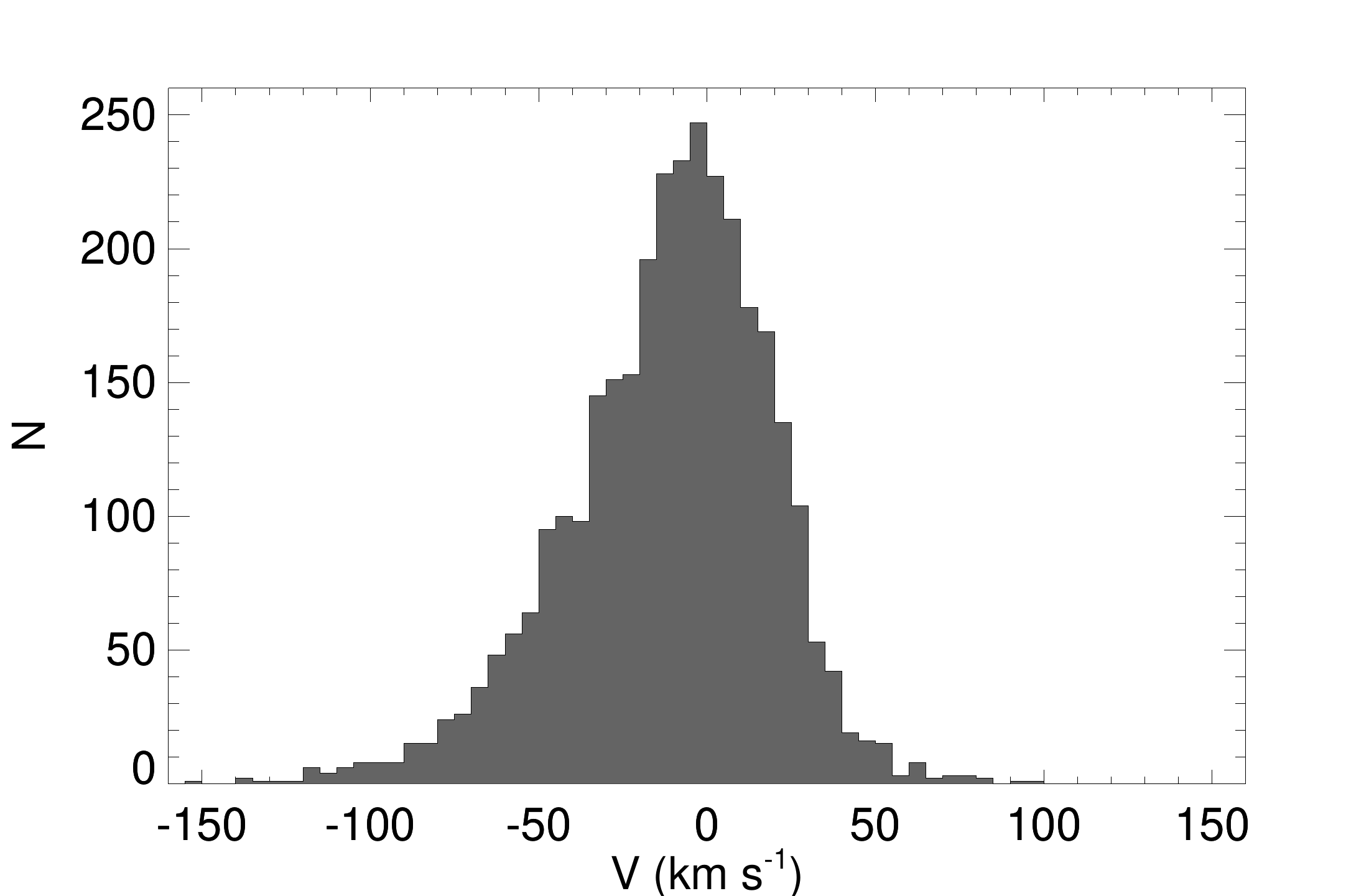}
  \includegraphics[width=1.05\columnwidth]{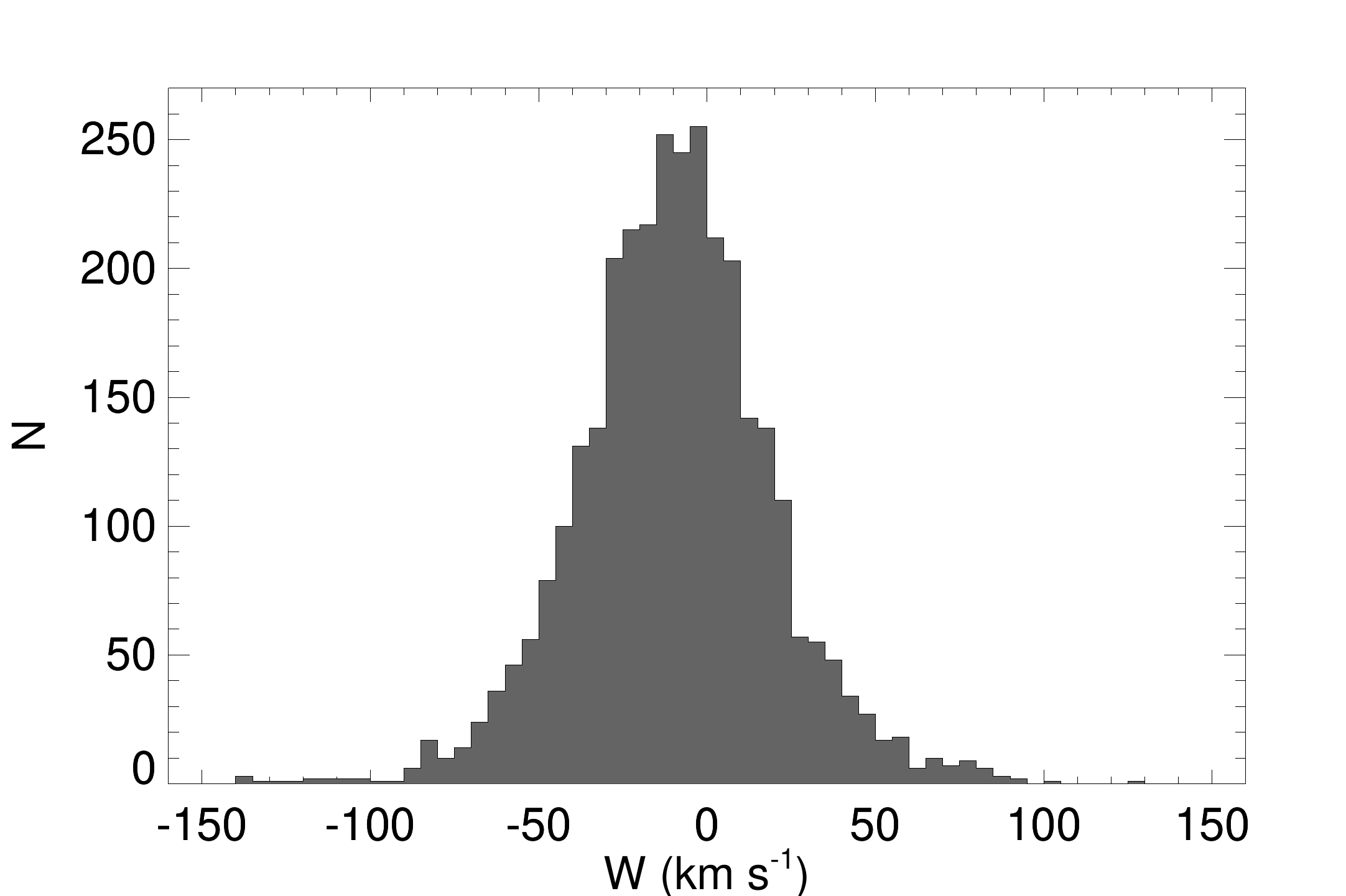}
  \caption{Distributions  of the  space  velocities $(U,\,V,\,W)$  for
    those white dwarfs with spectra with SNR$>$20, $M_{\rm WD}> 0.45\,
    M_{\sun}$    and    $\Delta U,\,\Delta V,\Delta W<35$~km~s$^{-1}$.     The
    distribution  of  the $V$  component  shows  a long  tail  towards
    negative  values, a  consequence  of the  asymmetric drift.  These
    stars lag behind the LSR.}
\label{fig:velo_cartesian}
\end{figure}

Finally, we also estimated the  Galactocentric distances $R_{\rm G}$ of
each white  dwarf in our  sample. This  was done using  their distances
$d$,  and  Galactic  coordinates $(l,b)$  (see  Fig.\ref{fig:aitoff}).
Adopting a Sun  Galactocentric distance, $R_{\sun}=8.34\pm0.16\,$kpc
\citep{Reid2014}, we  found that most  white dwarfs in this  study are
located between $7.8<R_{\rm G}<9.3\,$kpc.   In the following sections,
we will use $R_{\rm G}$ together  with $Z$ to investigate the velocity
distributions.

\subsection{Proper motions}

We obtained proper motions and  their associated uncertainties for the
objects       of      our       sample       using      the       {\tt
casjobs}\footnote{http://casjobs.sdss.org/casjobs/}          interface
\citep{li+thakar08-1},  which combines  SDSS and  re-calibrated USNO-B
astrometry \citep{munnetal04-1,  Munn2014}.  These proper  motions are
calculated  from the  USNO-B1.0  plate  positions re-calibrated  using
nearby galaxies  together with  the SDSS position  so that  the proper
motions  are more  accurate  and absolute.   By  measuring the  proper
motions  of  quasars,  \citet{munnetal04-1,  Munn2014}  estimate  that
1$\sigma$  error is  $\sim4\,$mas~yr$^{-1}$.  In  Fig.~\ref{fig:PM} we
compare the  measurement of the  proper motion, in right  ascension, and
declination for 125 objects  in common in our white dwarf
data-set  between  SDSS-USNO-B  \citep{Munn2014} and  UCAC4  catalogue
\citep{Zacharias2013}.   We  found  a   good  agreement  between  both
measurements for most of the common objects.  Unfortunately, for 4,360
DA white dwarfs in our sample there are no available proper motions.

\subsection{Space velocities}
\label{sec:phase_space}

\begin{figure}
  \includegraphics[width=1.05\columnwidth]{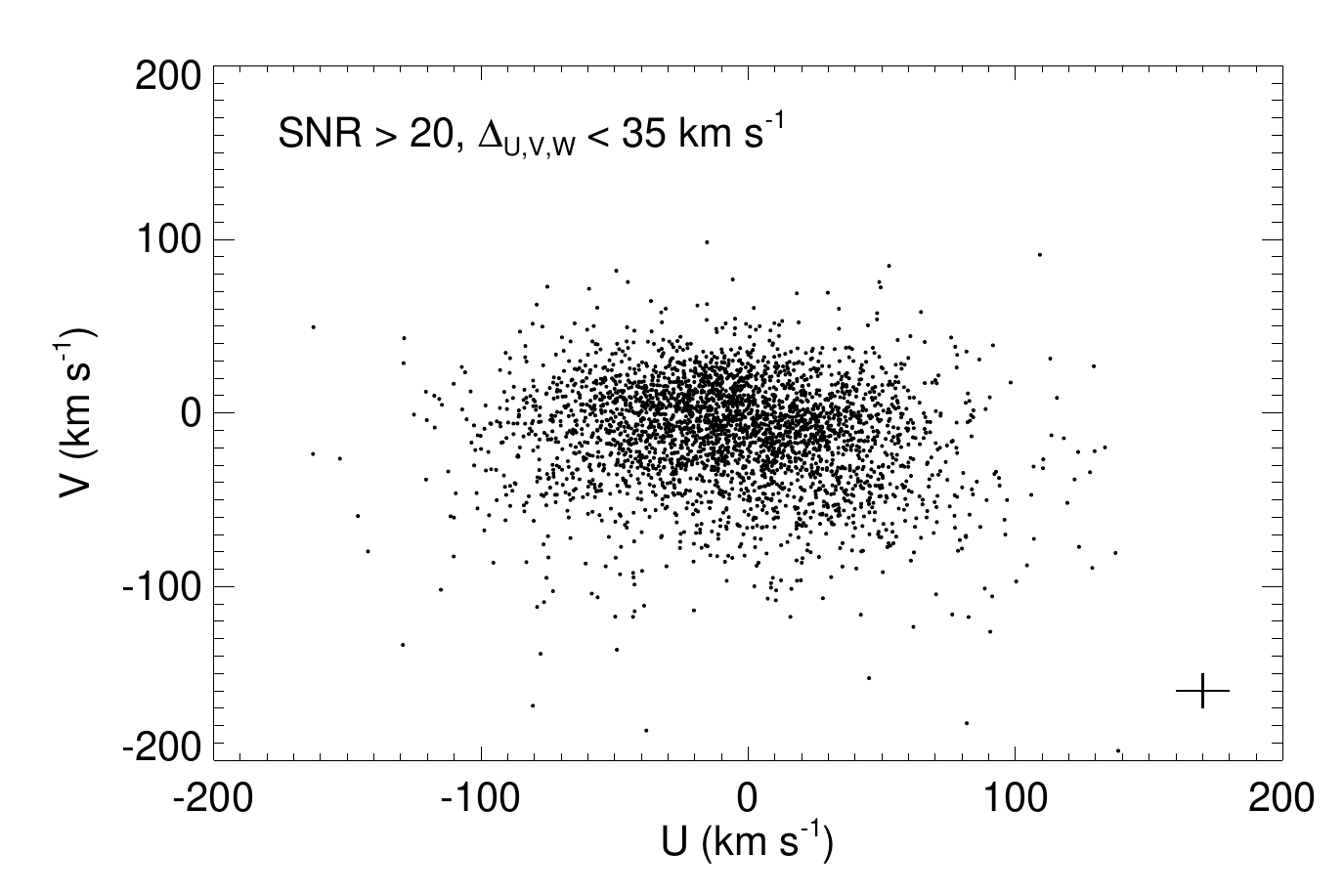}
  \includegraphics[width=1.05\columnwidth]{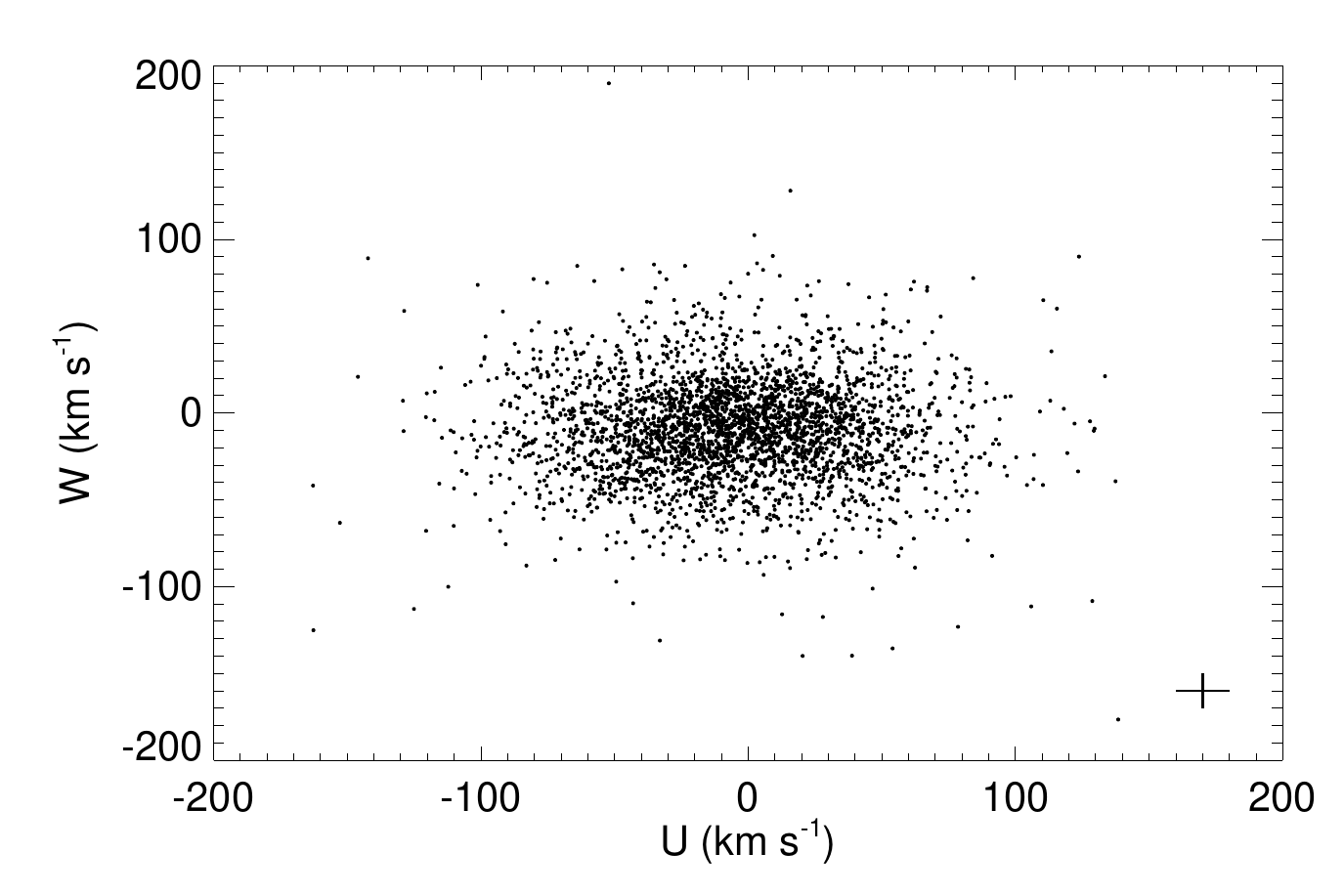}
  \includegraphics[width=1.05\columnwidth]{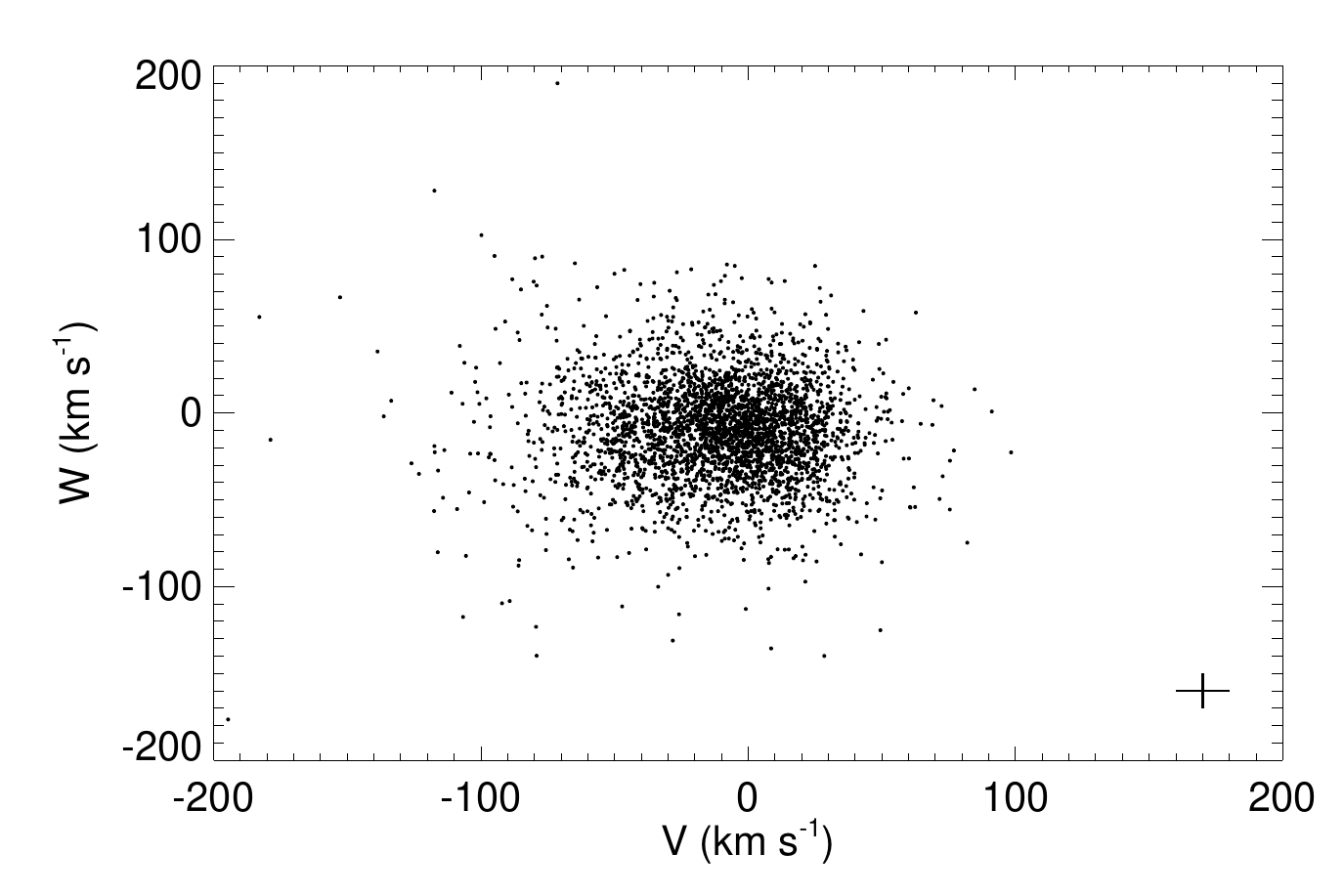}
  \caption{$U-V$, $U-W$, and $V-W$ diagrams for the selected sample of
    white dwarfs.}
\label{fig:phase_space}
\end{figure}

We computed  the velocities in  a cartesian Galactic  system following
the method developed by \citet{Johnson1987}.   That is, we derived the
space  velocity  components  $(U,\,V,\,W)$ from  the  observed  radial
velocities,   proper  motions   and   distances.    We  considered   a
right-handed Galactic reference system,  with $U$ positive towards the 
anticenter, $V$ in the direction of rotation,  and $W$ positive towards  the 
North  Galactic  Pole (NGP).   The  uncertainties in  the
velocity components $U$, $V$ and  $W$ were derived using the formalism
of  \citet{Johnson1987}. Within  this framework  the equation  for the
variance   of    a   function   of   several    variables   is   used.
\citet{Johnson1987}  assumed   that  the  matrix  used   to  transform
coordinates  into  velocities  introduces  no error  in  $U$,  $V$  or
$W$. Consequently,  the only sources  of error are the  distances, proper  
motions and  radial  velocities. Furthermore,  this method
assumes that the errors of the measured quantities are uncorrelated,
i.e.  that the covariances are zero.   We found that the typical error
is         ($\Delta          U,\,\Delta         V,\,\Delta         W$)
$\sim(6.5,\,8.3,\,10.5)\,$km~s$^{-1}$  and  that   nearly  the  entire
sample has $\Delta U,\,\Delta V,\Delta W<35\,$km~s$^{-1}$.

\begin{figure}
  \includegraphics[width=1.02\columnwidth]{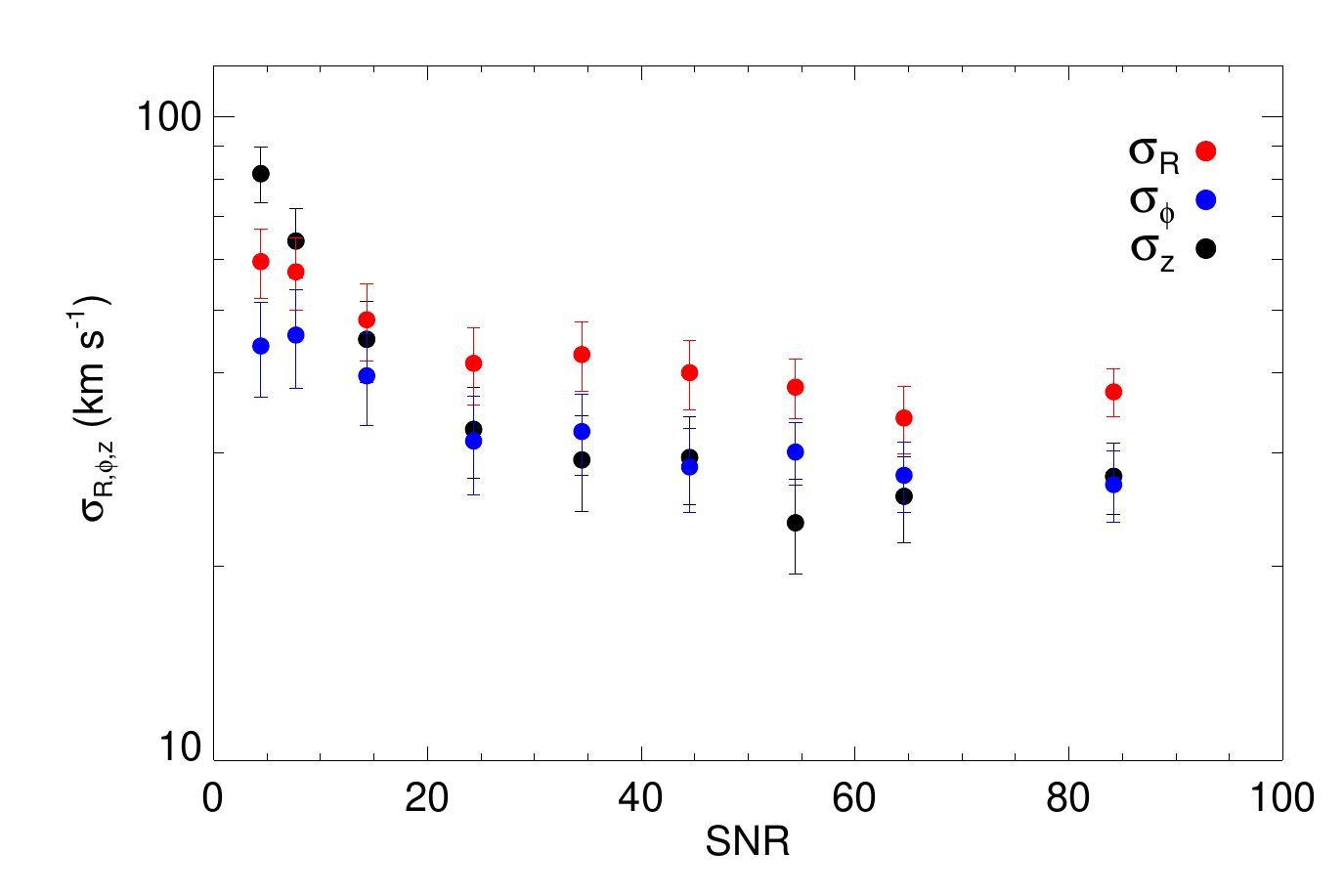}
  \caption{Distribution of  the velocity  dispersion as a  function of
    the SNR.   The velocity dispersion  for white dwarfs  with spectra
    with SNR$<$20 is  dominated by random errors while  for stars with
    spectra with SNR$>$20 the average value of the velocity dispersion
    of the three velocity components remains constant independently of
    the uncertainties, suggesting a physical origin.}
\label{fig:sig_SNR}
\end{figure}

Fig.~\ref{fig:velo_cartesian} shows the $U$, $V$ and $W$ distributions
for the  SDSS DA white  dwarfs.  Only  white dwarfs with $M_{\rm  WD}>0.45\, M_{\sun}$  --- to  avoid
contamination from close binaries, see Sect.\,\ref{sec:masses} --- and spectra with
SNR$>$20 ---  to ensure  a RV error  smaller than  20~km~$s^{-1}$, see
Sect.\,\ref{sec:rvs} --- and reasonable  uncertainties   $\Delta  U,\,   \Delta  V$,   and  $\Delta
W<35$~km~s$^{-1}$ were selected.   Also, in Fig.~\ref{fig:phase_space}
we show the phase-space diagrams for the same sample. These diagrams reveal 
interesting substructure in the Galactic disc, i.e. moving groups in the $U$-$V$ plane \citep{2004A&A...418..989N}. 
A detailed study of the overdensities and outliers objects seen in the phase-space using a population synthesis code \citep{Torres2002} is underway.

Since  white dwarfs  in our  sample  are located  at relatively  large
distances  from the  Sun, we  also use  cylindrical coordinates,  with
$V_{\rm R}$,  $V_{\rm   \phi}$  and  $V_{\rm z}$  defined   as  positive  with
increasing $R_{\rm  G}$, $\phi$ and  $Z$, with the latter  towards the
NGP.  We took the motion of the Sun with respect to the Local Standard
of     Rest    (LSR) \citep{Schonrich2010},     that    is
($U_{\sun},\,V_{\sun},\,W_{\sun})=(-11,\,   +12,\,  +7)\,$km~s$^{-1}$.
The LSR was assumed to be on a circular  orbit, with circular velocity
$\Theta = 240 \pm  8$~km~s$^{-1}$ \citep{Reid2014}.  With these values
we  computed  the  cylindrical  velocities  of  our  sample,  $V_{\rm R}$,
positive towards the Galactic center, in consonance with the usual $U$
velocity component, $V_{\rm \phi}$, towards  the direction of rotation and
$V_{\rm z}  =   W$,  positive   towards  the  NGP   --  see   appendix  in
\citet{Williams2013} for more details.

To  compute the  uncertainties of  the velocities  in the  cylindrical
coordinate  system   we  cannot   assume  that  the   observables  are
uncorrelated.  To  propagate the non-independent random  errors in the
velocities in  this coordinate frame  we used a Monte  Carlo technique
that takes  into account the  uncertainties in the  distances ($\Delta
X,\,\Delta Y,\,\Delta  Z$) and  in the  space velocities  in cartesian
coordinates  ($\Delta U,\,\Delta  V,\,\Delta  W$).   This Monte  Carlo
algorithm  for error  propagation  allows to  easily  track the  error
covariances.  In essence,  we generated  a distribution  of 1,000  test
particles around each input  value of $(XYZ,\,UVW)$, assuming Gaussian
errors  with standard  deviations given  by the  formal errors  in the
measurements for each white dwarf, and then we calculate the resulting
($V_{\rm R}$,  $V_{\rm \phi}$,  $V_{\rm z}$)  and  their  corresponding  1$\sigma$
associated uncertainties.

\section{Velocity maps and age gradients}
\label{Velo_maps}

In the  previous sections we have  introduced our SDSS DA  white dwarf
sample  and  we  have  explained  how  we  have derived  the  stellar
parameters, ages, distances, radial  velocities and proper motions for
each star.   With this information  at hands we analyze  the kinematic
properties of  the observed sample.   This is fully  justified because
our sample of  DA white dwarfs culled from the  SDSS DR\,12 catalog is
statistically significant,  independently of the  restrictions applied
to the full  sample to use only high-quality  data. These restrictions
are explained  below, and quite  naturally, reduce the total  number of
white dwarfs.  Finally,  we emphasize that a thorough  analysis of the
effects of the  selection procedures and observational  biases using a
detailed             population             synthesis             code
\citep{Torres1999,Torres2002,Torres2004} is underway, and we refer the
reader to a forthcoming publication.

\begin{table*}
\caption{Average  values  for   the  Galactic  cylindrical  velocities
  $(V_{\rm R},\,V_{\rm \phi},\,V_{\rm z})$ and their  corresponding dispersions as
  a function of the Galactocentric distance $R_{\rm G}$.}
\begin{center}
\begin{tabular}{ccccccc}
\hline
\hline
$R_{\rm G}$ (kpc) & $\langle V_{\rm R}\rangle$ (km s$^{-1}$)  & $\sigma_{\rm R}$ (km s$^{-1}$) & $\langle V_{\rm \phi} \rangle$ (km s$^{-1}$) & $\sigma_{\rm \phi}$ (km s$^{-1}$) & $\langle V_{\rm z}\rangle$ (km s$^{-1}$)  & $\sigma_{\rm z}$ (km s$^{-1}$) \\ 
\hline
$<$ 7.90     & +11.3 $\pm$ 5.2 & 31.1 $\pm$ 6.1   & +215.2 $\pm$ 5.3 & 25.7 $\pm$ 5.3 &  +8.9 $\pm$ 4.3 & 33.9 $\pm$ 6.5\\
7.90 -- 8.15 & +2.5 $\pm$ 1.0 & 35.9 $\pm$ 2.2   & +232.9 $\pm$ 0.9 & 26.5 $\pm$ 1.7 & --5.1 $\pm$ 0.9 & 24.5 $\pm$ 1.5\\
8.15 -- 8.40 & +2.1 $\pm$ 0.2 & 26.1 $\pm$ 0.6  & +224.6 $\pm$ 0.2 & 21.5 $\pm$ 0.5 & --10.2 $\pm$ 0.2 & 23.7 $\pm$ 0.4\\
8.40 -- 8.65 & +14.2 $\pm$ 0.3 & 33.7 $\pm$ 0.8 & +233.5 $\pm$ 0.3 & 22.7 $\pm$ 0.6 & --10.5 $\pm$ 0.2 & 27.1 $\pm$ 0.6\\
8.65 -- 8.85 & +5.4 $\pm$ 0.9 & 34.1 $\pm$ 2.1 & +236.4 $\pm$ 0.9 & 23.5 $\pm$ 1.4 & --2.9 $\pm$ 0.9 & 16.4 $\pm$ 1.2\\
$>$ 8.85     & +3.4 $\pm$ 2.8 & 36.0 $\pm$ 4.3  & +237.3 $\pm$ 2.7 & 22.1 $\pm$ 2.8 & --2.7 $\pm$ 2.9 & 21.5 $\pm$ 2.9\\
\hline
\end{tabular}
\end{center}
\label{tab_R}
\end{table*}

\begin{table*}
\caption{Average  values  for   the  Galactic  cylindrical  velocities
  $(V_{\rm R},\,V_{\rm \phi},\,V_{\rm z})$ and their  corresponding dispersions as
  a function of the Galactic height $Z$.}
\begin{center}
\begin{tabular}{ccccccc}
\hline
\hline
$Z$ (kpc) & $\langle V_{\rm R} \rangle$ (km s$^{-1}$)  & $\sigma_{\rm R}$ (km s$^{-1}$) & $\langle V_{\rm \phi}\rangle$ (km s$^{-1}$) & $\sigma_{\rm \phi}$ (km s$^{-1}$) & $\langle V_{\rm z}$ $\rangle$ (km s$^{-1}$)  & $\sigma_{\rm z}$ (km s$^{-1}$) \\ 
\hline
$< -0.5$           & +4.8 $\pm$ 4.8 & 42.0 $\pm$ 7.9 & +212.7 $\pm$ 4.8 & 23.9 $\pm$ 4.9 &  +13.0 $\pm$ 4.5 & 27.8 $\pm$ 5.5\\
$-0.50$ -- $-0.25$ & +17.0 $\pm$ 1.2 & 37.1 $\pm$ 2.8 & +226.9 $\pm$ 1.1 & 31.9 $\pm$ 2.3 &   +7.3 $\pm$ 1.3 & 23.4 $\pm$ 1.9\\
$-0.25$ -- 0.00    & +11.1 $\pm$ 0.3 & 32.4 $\pm$ 1.2 & +226.6 $\pm$ 0.3 & 24.7 $\pm$ 0.9 &  +10.6 $\pm$ 0.3 & 20.8 $\pm$ 0.9\\
0.00 -- +0.25      & +2.1 $\pm$ 0.1 & 27.2 $\pm$ 0.6 & +225.7 $\pm$ 0.1 & 21.6 $\pm$ 0.4 & --14.0 $\pm$ 0.2 & 21.9 $\pm$ 0.4\\
+0.25 -- +0.50     & +5.4 $\pm$ 0.4 & 34.4 $\pm$ 1.0 & +235.3 $\pm$ 0.4 & 23.7 $\pm$ 0.7 &  --9.5 $\pm$ 0.4 & 24.4 $\pm$ 0.7\\
$>$ +0.50          & +4.9 $\pm$ 1.8 & 37.6 $\pm$ 2.9 & +237.8 $\pm$ 1.9 & 24.7 $\pm$ 2.1 &  --2.6 $\pm$ 1.8 & 29.5 $\pm$ 2.4\\
\hline
\end{tabular}
\end{center}
\label{tab_Z}
\end{table*}

As mentioned, the  catalog used in this work contains  20,247 DA white
dwarfs culled  from the  SDSS DR\,12.   To study  the behavior  of the
velocity  components  with  respect  to  the  Galactocentric  distance
($R_{\rm G}$)  and to  the vertical distance  ($Z$) we  restricted our
sample to employ  only stars with data of the  highest quality.  First
of all, we  explored the effects of selecting stars  with spectra with
qualities above a given SNR  threshold.  We found that both the 
resulting velocity dispersions and the average velocities are
sensitive to  this choice.  Hence,  a natural question  arises, namely
what  is the  optimal SNR  cut to  achieve our  science goals  without
discarding an excessive number of  stars.  To address this question in
Fig.~\ref{fig:sig_SNR}  we  examine  the   behavior  of  the  velocity
dispersion  as a  function of  the SNR  cut.  The  error bars  are the
standard  deviation of  the uncertainties  in the  velocities for  the
corresponding SNR bin.   Note that for white dwarfs  with spectra with
SNR$<$20  random errors  dominate the  velocity dispersion,  while for
stars  with  spectra with  SNR$>$20  the  dispersion remains  constant
within the error bars, suggesting that a physical dispersion, which is
not due  to the uncertainties  in the observables, exists.   Hence, we
only selected  stars with spectra  having SNR$>$20.  We  also excluded
from  this  sub-sample all  white  dwarfs  with masses  below  $0.45\,
M_{\sun}$.  In this way we avoid an undesirable contamination by close
binaries   ---   see   the  discussion   in   Sect.\,\ref{sec:masses}.
Furthermore,  we   only  considered   white  dwarfs   with  reasonable
determinations, and consequently we excluded those stars for which the
velocity errors  were large  --- see  Sect.\,\ref{sec:phase_space}. In
particular,   we  adopted   a   cut   $\Delta_{U,V,W}<35\,$km~s$^{-1}$. 
We also  did not consider those  white dwarfs with velocities 
$-600\,<\,V_{\rm R},V_{\rm \phi},V_{\rm z}\,<\, 600$~km~s$^{-1}$ to
remove outliers. This resulted  in a sub-sample of  3,415 DA
white dwarfs.  Since the aim of  this section is to gain a preliminary
insight  on  the  general  behavior of  the  velocities  and  velocity
dispersions, we do  not distinguish between halo, thick  and thin disc
white dwarfs.

\begin{figure}
  \centering
  \includegraphics[width=1.03\columnwidth]{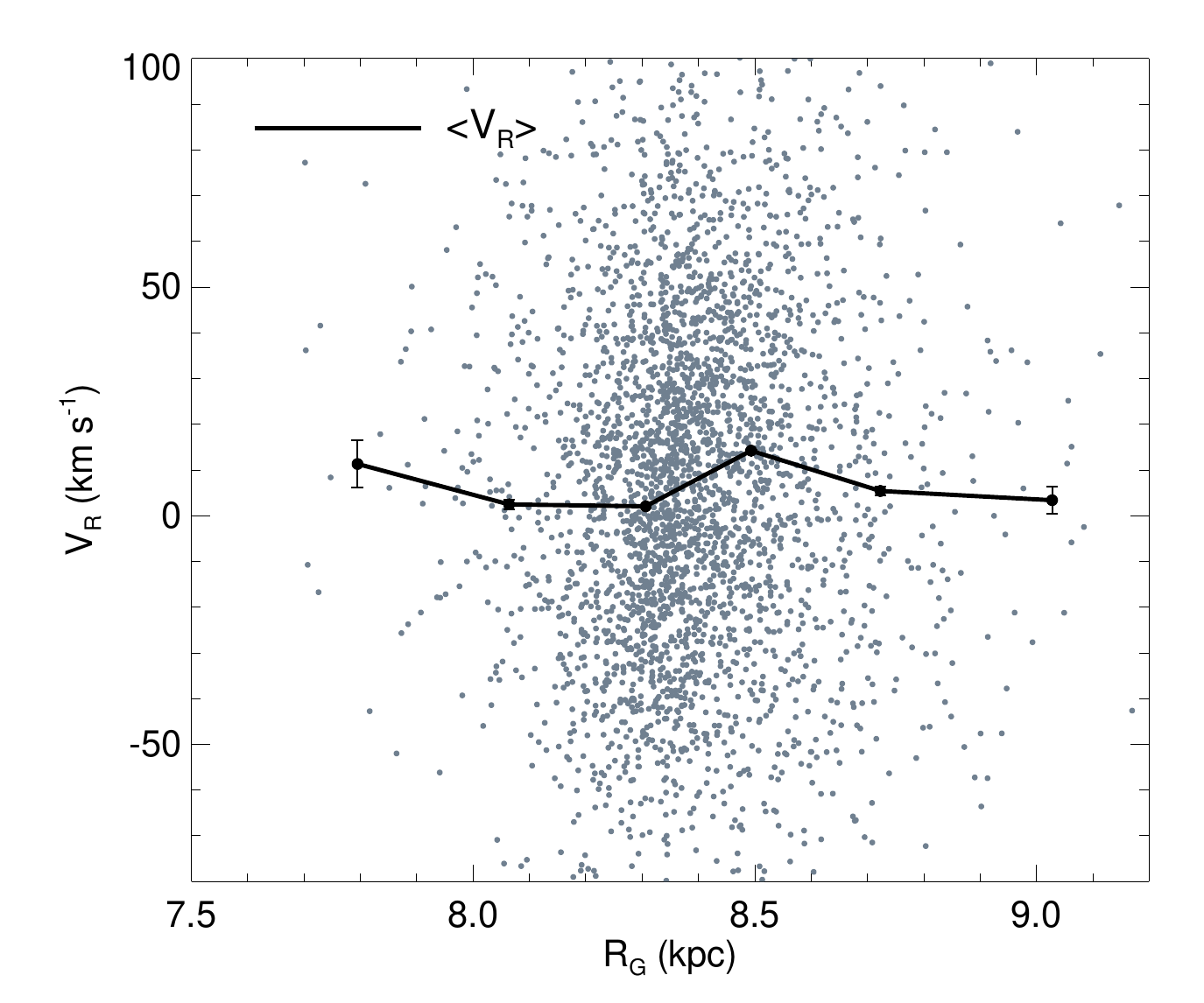}
  \includegraphics[width=1.03\columnwidth]{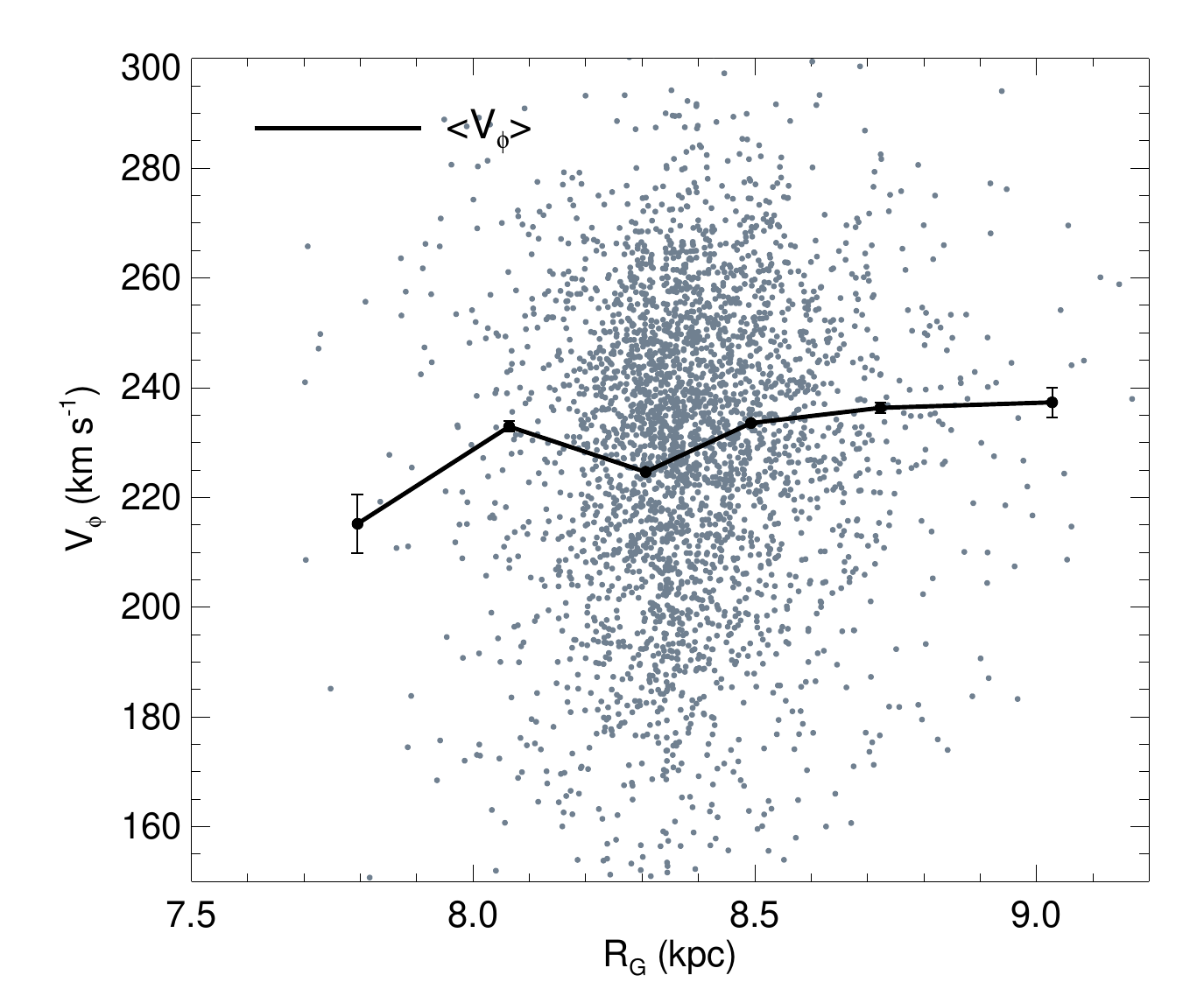}
  \includegraphics[width=1.03\columnwidth]{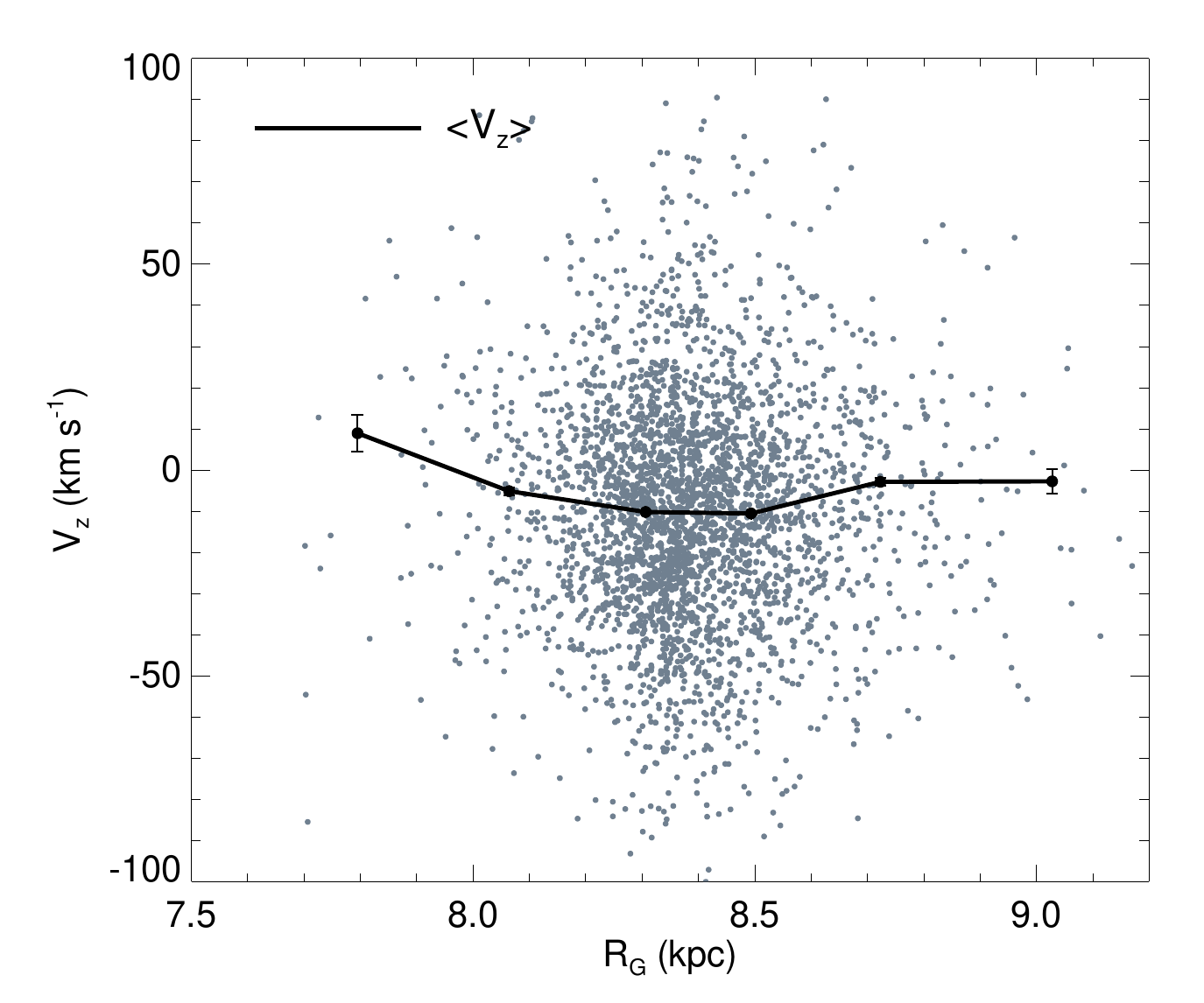}
  \caption{Average velocities as a function of $R_{\rm G}$. See text for more details.}
\label{fig:R_velo}
\end{figure}

\begin{figure}
  \centering
  \includegraphics[width=1.03\columnwidth]{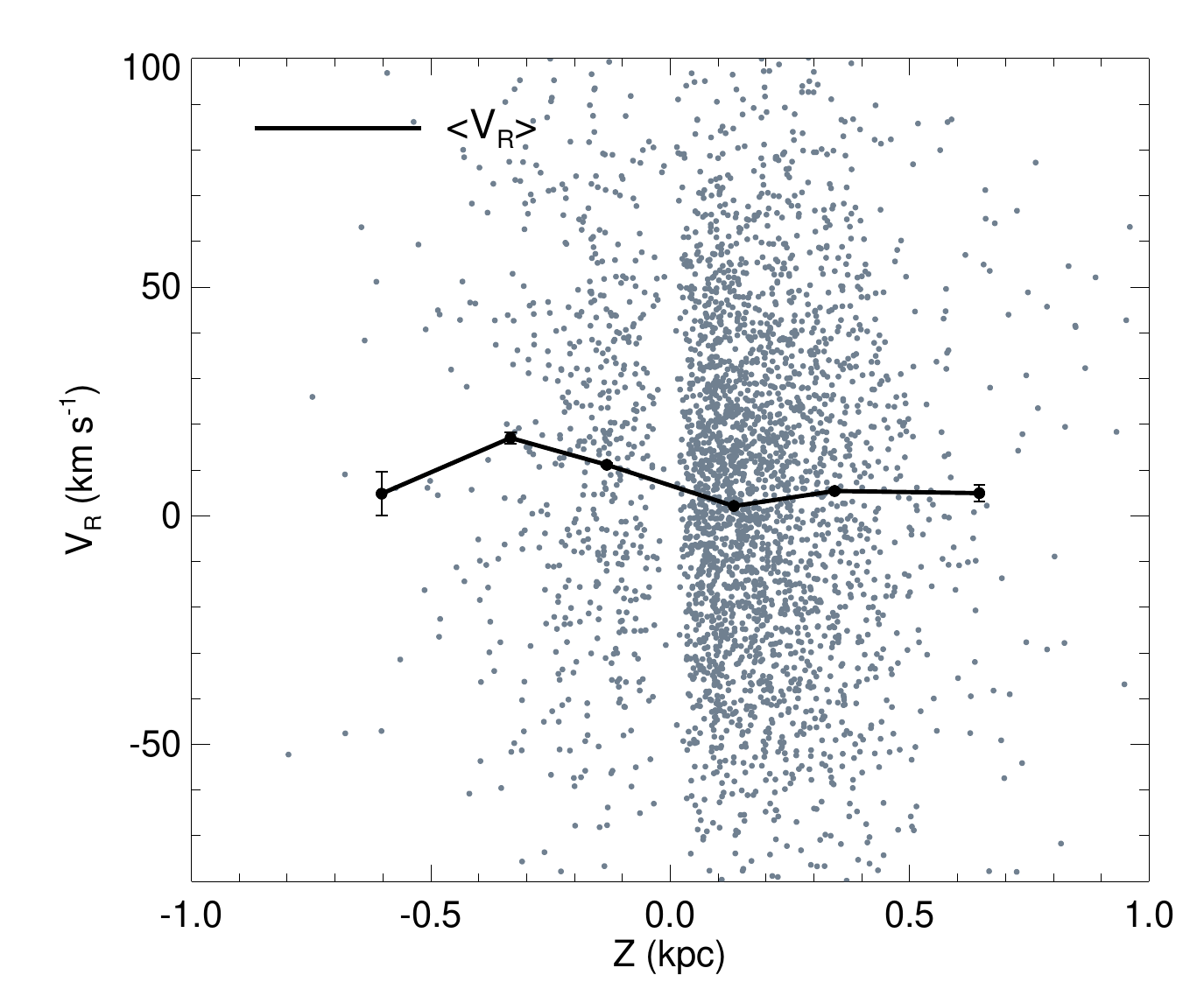}
  \includegraphics[width=1.03\columnwidth]{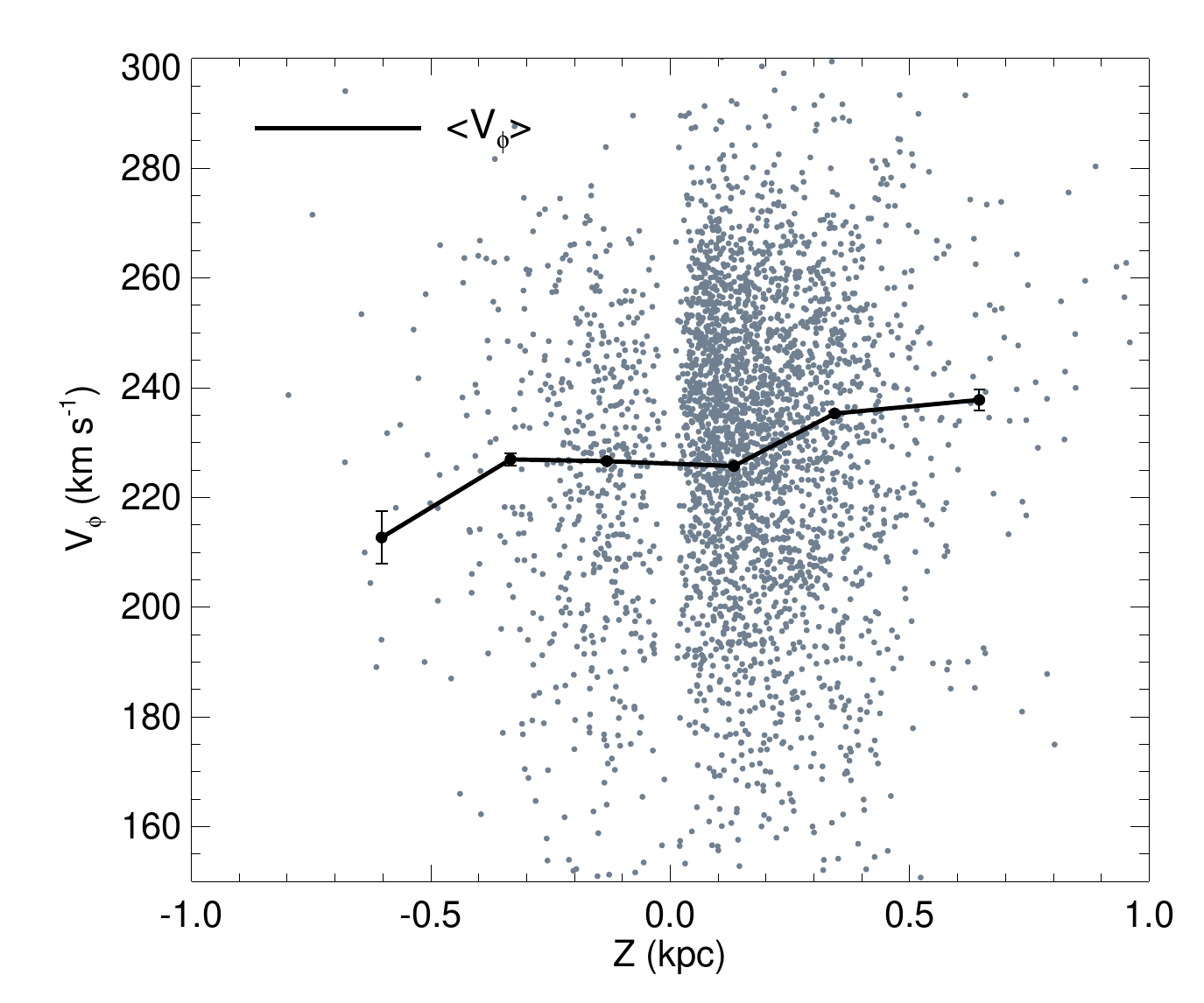}
  \includegraphics[width=1.03\columnwidth]{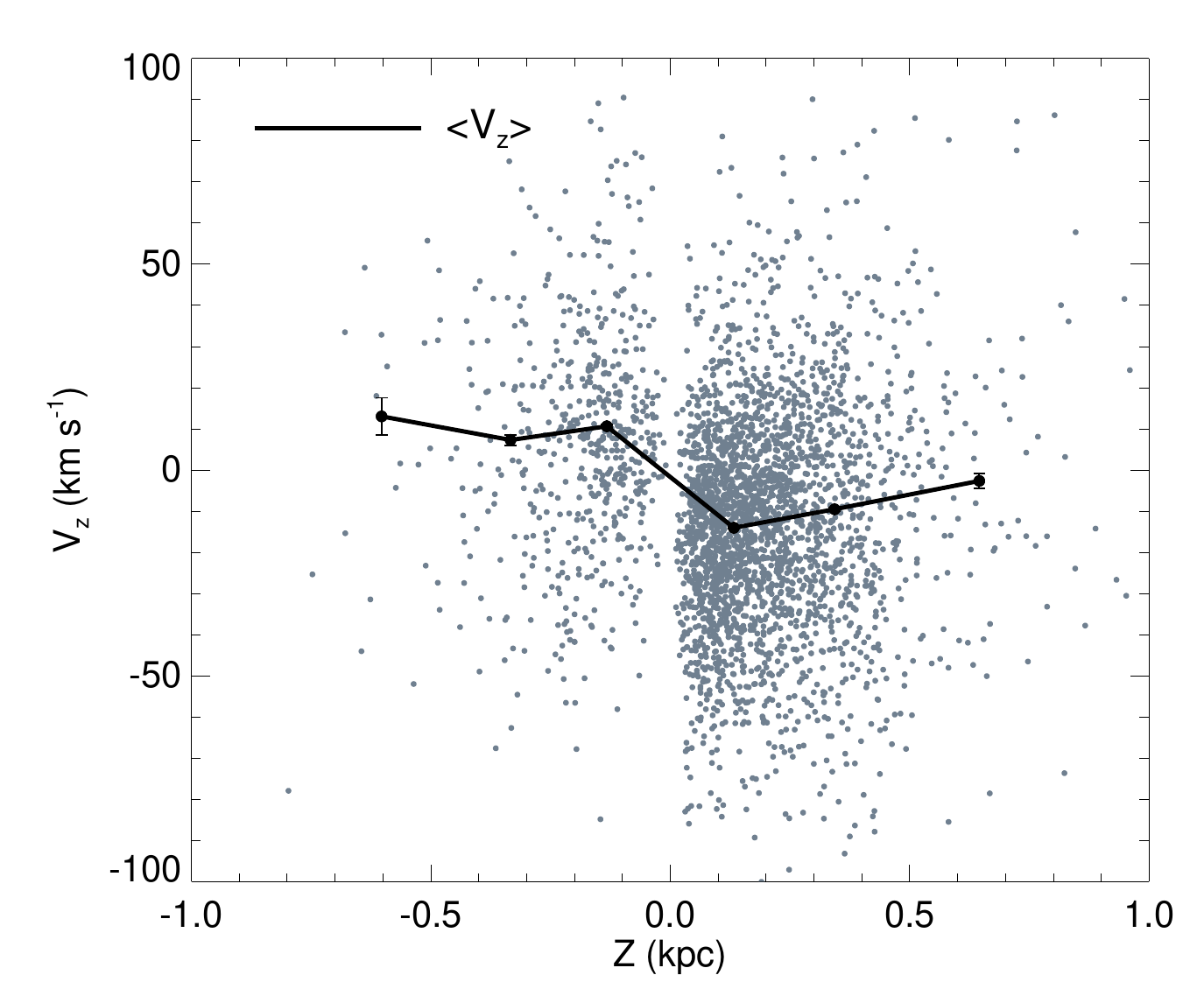}
  \caption{Average velocities as a function of $Z$. See text for more details.}
\label{fig:Z_velo}
\end{figure}

This sub-sample allowed us to study  the behavior of the components of
the velocity  as a  function of  the Galactocentric  distance ($R_{\rm
G}$)  and the  vertical distance  ($Z$).  We used  weighted means  and
weighted velocity dispersions. The expression for the weighted mean of
any (say $\omega$) of the three components of the velocities is:
\begin{equation}
\langle V_{\omega}\rangle = \frac{\displaystyle\sum_{i=1}^{N}
  \eta_{i}  V_{\omega_i}} {\displaystyle\sum_{i=1}^{N}
  \eta_{i}}
\end{equation}
where $V_{\omega_i}$  is the  $\omega$ component  of the  velocity for
each star.   In this expression the  weights are given by  $\eta_{i} =
1/\sigma^{2}_{\omega_i}$,   being    $\sigma_{\omega_i}$   the   error
associated to the  three components of the  velocity ($\Delta V_{\rm R},\,
\Delta V_{\rm \phi},\, \Delta V_{\rm z}$)  of each individual white dwarf.
The variance of $\langle V_{\omega}\rangle$ is given by:
\begin{equation}
\sigma_{\omega}^2 = \frac{1}{\displaystyle\sum_{i=1}^{N}
  1/\sigma_{\omega_i}^2}
\end{equation}
To compute the velocity dispersion  of the stellar velocity components
we used a weighted variance:
\begin{equation}
\sigma_{V_{\omega}}^2 = \frac{\displaystyle\sum_{i=1}^{N}
  \eta_{i}(V_{\omega_i} - \langle
  V_{\omega}\rangle)^{2}}{k\displaystyle\sum_{i=1}^{N} \eta_{i}}
\end{equation}
where $k=(N'-1)/N'$, and $N'$ is  the number of non-zero weights.  The
error bars  of the  velocity dispersion in  Figs.~\ref{fig:R_velo} and
\ref{fig:Z_velo} weere computed employing the expression
\begin{equation}
\Delta{V_{\omega}} = (2N)^{-1/2} \sigma_{V_{\omega}}
\end{equation}
where $N$ is the number of white dwarfs in each bin of $R_{\rm G}$ and
$Z$ respectively  (see also Table~\ref{tab_R}  and Table~\ref{tab_Z}),
used to calculate $\sigma_{V_{\omega}}$.

\begin{figure*}
  \centering
  \includegraphics[width=1.04\columnwidth]{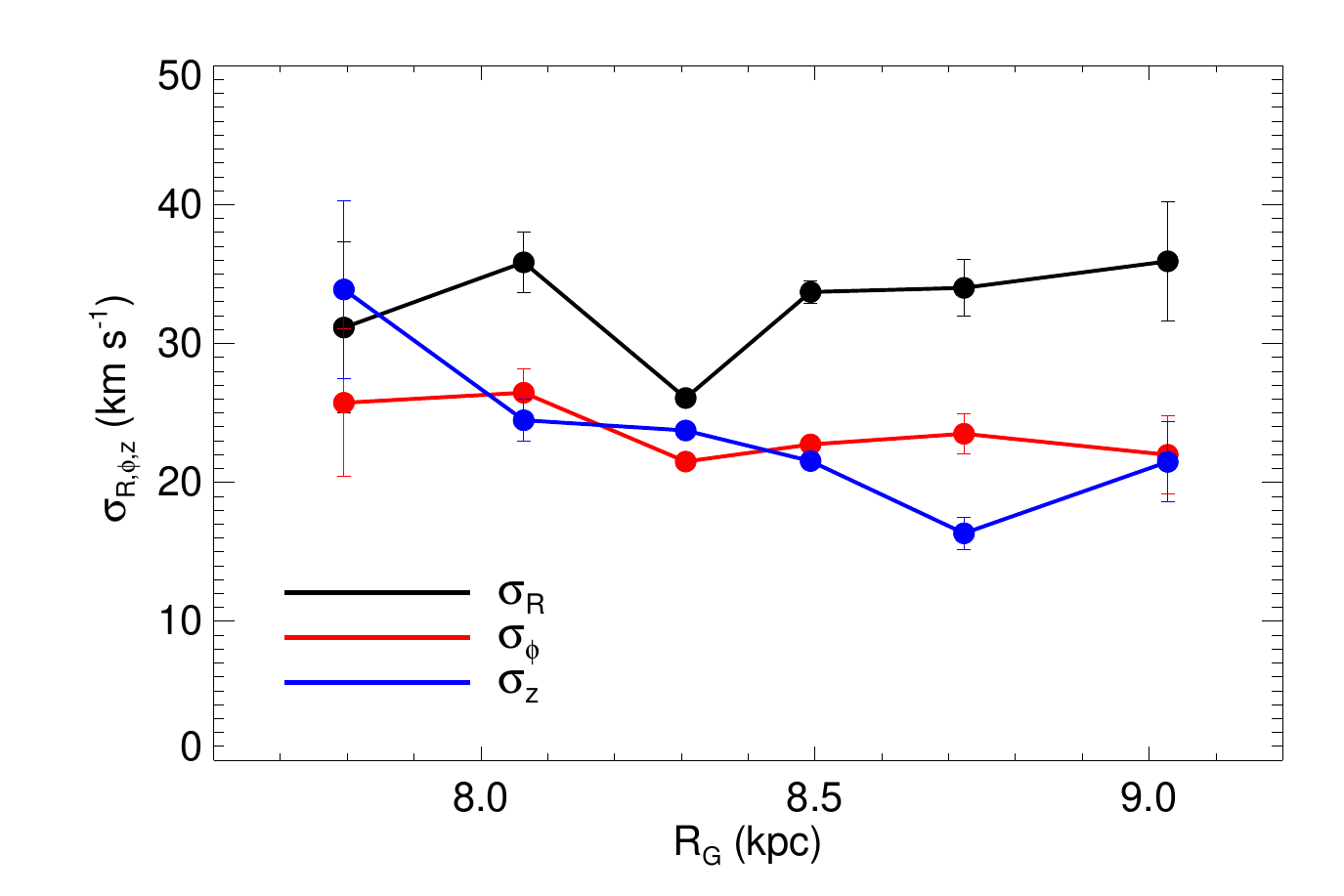}
  \includegraphics[width=1.04\columnwidth]{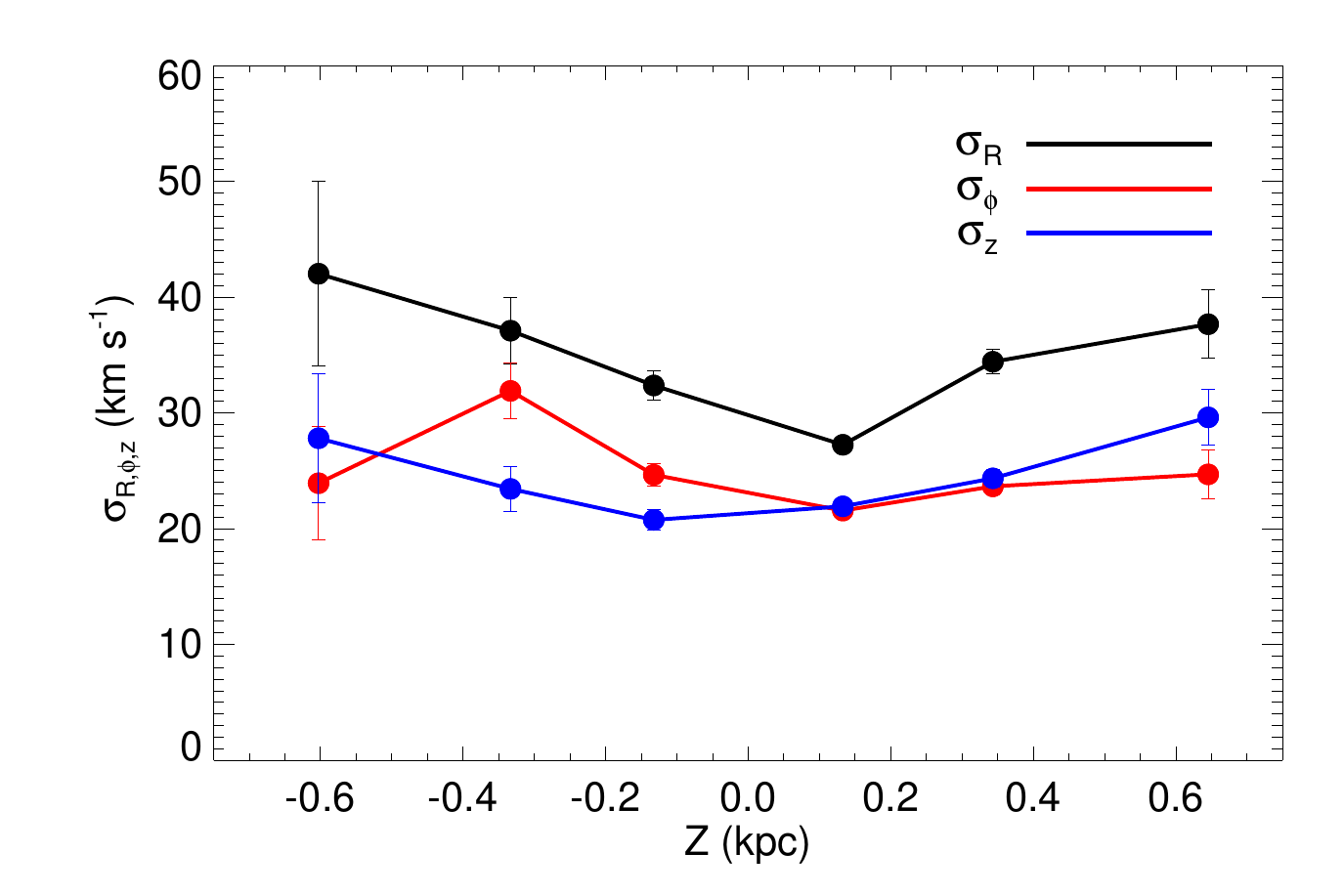}
  \caption{Left panel: variation of the velocity dispersion for the three velocity components 
  with respect to the Galactocentric distance. Right panel: variation of the velocity dispersion 
  with respect to the vertical Galactic distance $Z$.}
\label{fig:sigma_RZ}
\end{figure*}

\subsection{Space velocities in the local volume}

In  Figs.~\ref{fig:R_velo}  and  \ref{fig:Z_velo} we  show  the  three
components  of the  space velocity  $(V_{\rm R},\,V_{\rm \phi},\, V_{\rm z})$  for
each white dwarf  in this sub-sample as a function  of $R_{\rm G}$ and
$Z$, respectively.   We also  show the average  (black lines). In Fig.~\ref{fig:sigma_RZ} 
we show the velocity dispersion as a function  of $R_{\rm G}$ and
$Z$. The velocity dispersion components are corrected for the measuring errors using the 
standard deviation of the error distribution per each bin. Most white
dwarfs in  this sub-sample  with precise kinematics  are close  to the
Galactic  plane,  with  altitudes   from  the  Galactic  plane  within
$\pm0.5\,$kpc.  Also, the vast majority  of these stars are located at
Galactic radii ranging from $R_{\rm G}  = 8.0$~kpc to 8.8~kpc.  In the
local volume covered by our sample  we found that the average value of
the radial  velocity $V_{\rm R}$  first slightly increases  for increasing
distances from  the center  of the  Galaxy, and  then decreases  up to
$R_{\rm  G} \simeq  8.5$~kpc.  From  this point  outwards the  average
velocity  remains nearly  constant.   The  radial velocity  dispersion
$\sigma_{\rm R}$ slightly increases as $R_{\rm G}$ increases (see Table~\ref{tab_R}). 
For white dwarfs around $R_{\rm G}\sim 8.0\,$kpc
we  found  $V_{\rm R}\sim  +2.5\pm1.0\,$km~s$^{-1}$,   while  for  $R_{\rm  G}\sim
9.0\,$kpc  we  have  $V_{\rm R}\sim+3.4\pm2.8\,$km~s$^{-1}$.  However,  at  larger
distances, for example, $R_{\rm G}<7.9$~kpc and $Z<-0.5\,$kpc, Poisson
noise  contributes  significantly. This  noise  is  beyond the  formal
measurement errors of the measured  kinematic properties as these bins
contain only  a few white  dwarfs (note  that Poisson noise  scales as
$N^{-1/2}$, where $N$ is the number of objects in the bin).

\citet{Siebert2011} and  \citet{Williams2013}, using  a sample  of red
clump stars from the RAVE survey that covers a  volume larger than that
probed by our sample of white  dwarfs; also reported a radial gradient
in the  mean Galactocentric radial velocity,  $V_{\rm R}$.  In particular,
if we restrict ourselves to  stars with altitudes $-0.5<Z<0.5\,$kpc we
find  a small negative gradient  $\partial\langle  V_{\rm R}\rangle/\partial  R_{\rm  G}
=-3\pm5\,$km~s$^{-1}$~kpc$^{-1}$ while  the gradient of  the velocity
dispersion      is      $\partial      \sigma_{\rm R}/\partial      R_{\rm
G}=+3\pm4\,$~km~s$^{-1}$~kpc$^{-1}$. Interestingly, $V_{\rm R}$  shows a
small   gradient   in   the   vertical   direction,   $\partial\langle
V_{\rm R}\rangle/\partial Z=-5\pm6\,$km~s$^{-1}$~kpc$^{-1}$.  However, the
errors are too large to draw any robust conclusion.  For the variation
of  $\sigma_{\rm R}$  with  respect  to the  vertical  distance  we  found the 
following, $\partial \sigma_{\rm R}/\partial  Z =-23\pm5\,$km~s$^{-1}$~kpc$^{-1}$ for for Z $<$ 0,
suggesting a substantial gradient while $\partial \sigma_{\rm R}/\partial  Z =+19\pm3\,$km~s$^{-1}$~kpc$^{-1}$ for Z $>$ 0. 
$\sigma_{\rm R}$ clearly increases when moving away from the Galactic plane (see Table.~\ref{tab_Z}).

As  well known  from the  Jeans equation  $V_{\rm \phi}$ decreases  as the
asymmetric  drift  increases,  and   hence  when  $\sigma_{\rm  \phi}$
increases  \citep{BinneyTremaine2008}. We  found  a positive  gradient
with    respect    to    $R_{\rm  G}$,    $\partial\langle    V_{\rm
\phi}\rangle/\partial  R_{\rm   G}=+15\pm  5\,$km~s$^{-1}$~kpc$^{-1}$.
Moving inwards,  $V_{\phi}$ decreases. For the  velocity dispersion we
found  a  negative  gradient  in   terms  of  $R_{\rm  G}$,  $\partial
\sigma_{\rm \phi}/\partial  R_{\rm G}=-3\pm1\,$km~s$^{-1}$~kpc$^{-1}$, 
the  velocity dispersion increases when  moving inwards. Also,
it  is  interesting to  investigate  the  profile of  $V_{\rm \phi}$  with
respect  to the  vertical  height.  Our  data allows  us  to detect  a
significant  gradient,  $\partial\langle V_{\rm  \phi}\rangle/\partial
Z=+17\pm4\,$km~s$^{-1}$~kpc$^{-1}$, suggesting  that the  outer region
of the South Galactic Pole, rotates  slower than those regions close to
the North.  However, as mentioned above,  the number of
white dwarfs  with $Z<-0.5\,$kpc  is small,  and Poisson  noise, among
other selection  biases, may  play a  role.  The  velocity dispersion,
$\sigma_{\rm   \phi}$  also   shows   a   small  gradient. For Z $<$ 0, we have  $\partial
\sigma_{\rm \phi}/\partial Z=+3\pm18\,$km~s$^{-1}$~kpc$^{-1}$ while for Z $>$ 0, $\partial
\sigma_{\rm \phi}/\partial Z=+6\pm2\,$km~s$^{-1}$~kpc$^{-1}$.  We
also    found   that    $\sigma_{\rm    \phi}$   increases    ($\Delta
\sigma_{\rm \phi}\sim$~4~km~s$^{-1}$) when $\lvert Z\rvert$ increases from
zero    to     $0.6\,$kpc,    reaching    a    minimum     value    of
$\sigma_{\rm \phi}=+21.6\pm0.4\,$km~s$^{-1}$ at $0<Z<+0.25\,$kpc.

Figs.~\ref{fig:R_velo}, \ref{fig:Z_velo}, \ref{fig:sigma_RZ} also shows how the mean value of $V_{\rm z}$ and
$\sigma_{\rm z}$  behave with  respect to  $R_{\rm  G}$ and  $Z$. A  small
gradient     is     found,     $\partial     V_{\rm z}/\partial     R_{\rm
G}=-6\pm7\,$km~s$^{-1}$~kpc$^{-1}$, suggesting that  for this range of
values of  $Z$ the mean  vertical velocity increases when  $R_{\rm G}$
decreases. The vertical velocity dispersion, $\sigma_{\rm z}$, clearly
decreases  when moving  away from  the Galactic  center, the  gradient
being          $\partial           \sigma_{z}/\partial          R_{\rm
G}=-10\pm4\,$km~s$^{-1}$~kpc$^{-1}$.  The profile of the mean value of
$V_{z}$ as  a function of  the vertical  distance, $Z$, also  shows an
interesting trend. Specifically, while  for white dwarfs with positive
$Z$  (in the  North Galactic  hemisphere) we  have negative values in
$\langle V_{\rm z} \rangle$, for white  dwarfs with $Z<0$ (in the South 
Galactic hemisphere), $\langle V_{\rm z} \rangle$ is positive. Our dataset
shows        a         gradient        $\partial        V_{\rm z}/\partial
Z=-18\pm8\,$km~s$^{-1}$~kpc$^{-1}$.  We  also found that  the velocity
dispersions increase when moving away from the Galactic plane, $\Delta
\sigma_{\rm z}\sim10\,$km~s$^{-1}$ from  the Galactic plane  to $0.6\,$kpc
with  a significant  gradient,    $\partial     \sigma_{\rm z}/\partial
Z=-15\pm1\,$km~s$^{-1}$~kpc$^{-1}$ for Z $<$ 0 and $\partial     \sigma_{\rm z}/\partial
Z=+15\pm2\,$km~s$^{-1}$~kpc$^{-1}$ for Z $>$ 0. We  note  that the  variation  of
$\sigma_{\rm z}$ with $Z$ has received  significant attention as it can be
used to trace the vertical potential of the disc \citep{Smith2012}.

For the local sample of white dwarfs  used in this study we found that
the  inner  part  of  the  Galaxy  is  hotter  than  the  outer  part,
$\sigma_{\rm R,\phi,z}(R_{\rm G}<R_{\sun})>\sigma_{\rm R,\phi,z}(R_{\rm
G}>R_{\sun})$. The radial gradient in the velocity dispersion is well 
known and also detected in other galaxies, e.g. \citet{Kruit_Freeman2011}. We  
also found that the  velocity dispersions increase when  moving away  
from  the Galactic  plane, $\sigma_{\rm  R,\phi,z}$ ($\lvert Z\rvert = 0)<\sigma_{\rm R,\phi,Z}$ ($\lvert Z\rvert$ $>$ 0). 
Figs.~\ref{fig:R_velo}, \ref{fig:Z_velo}, \ref{fig:sigma_RZ} and  Tables~\ref{tab_R} 
and \ref{tab_Z} summarize the results discussed here.

\subsection{Radial and vertical age gradients}

\begin{figure*}
  \centering  
  \includegraphics[width=\columnwidth]{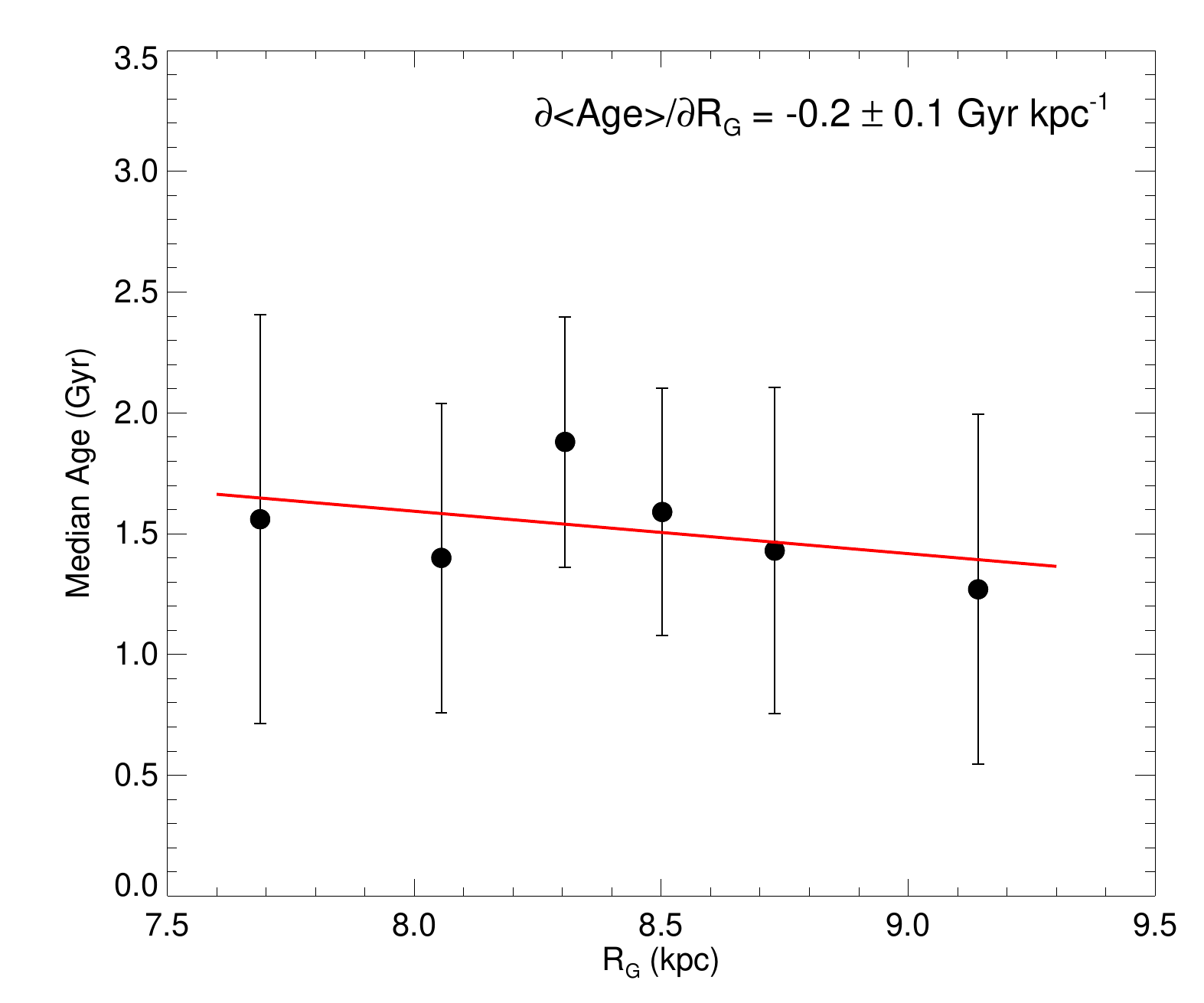}
  \includegraphics[width=\columnwidth]{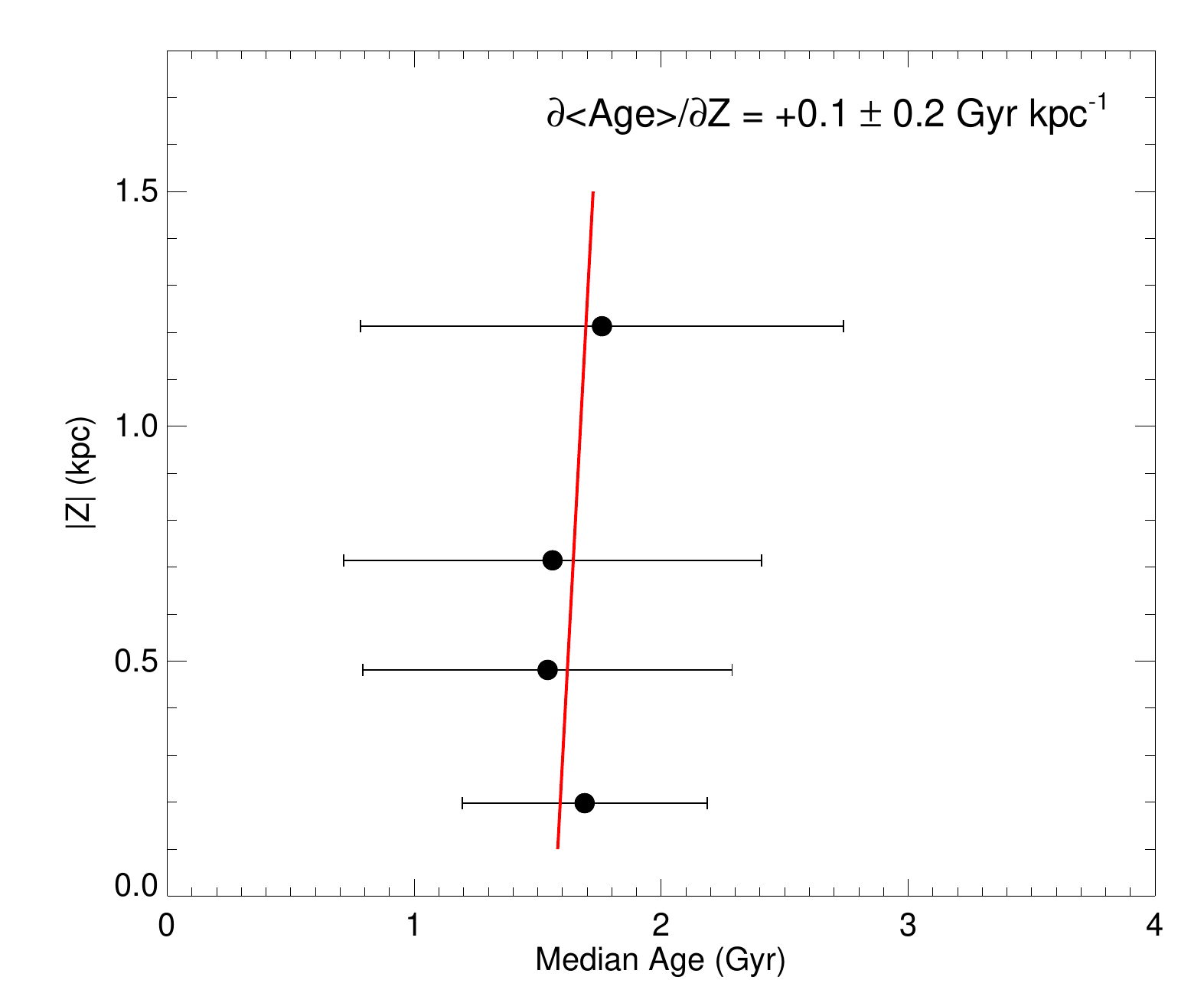}
  \caption{Median  age as  a function  of the  Galactocentric distance
    $R_{\rm G}$ (left panel) and vertical distance $Z$ (right panel).}
  \label{fig:R_Z_age}
\end{figure*}

In this section we study the  radial and vertical age gradients on the
Milky Way disc  using the derived white  dwarf ages.  \citet{West2011}
reported  variations in  the fraction  of active  M dwarfs of  similar
spectral type at increasing Galactic latitudes as an indirect evidence
for the  existence of a vertical  age gradient in the  Milky Way disc.
Recently,  \citet{Casagrande2016} using  red giants and seismic  ages by  
{\sl Kepler} mission reported  that old  red giants  dominate at
increasing Galactic  heights.  They also reported  a vertical gradient
of approximately 4~Gyr~kpc$^{-1}$, although with a large dispersion of
ages  at  all  heights  and  considerably  large  uncertainties,  that
prevented them to derive meaningful conclusions.

Note that,  as mentioned  before, our  data-set contains  white dwarfs
with  ages $<4.5$~Gyr.  Hence, intermediate-old  stars are  missing in
this study and this may introduce a bias in our results.  By selecting
intervals in $R_{\rm  G}$ and $\lvert Z\rvert$ we  estimated the median
age. For  stars within  this bin  we also computed  the mean  values of
$R_{\rm G}$ and $\lvert Z\rvert$.  Then, we used a linear regression to
compute  the  age gradients  as  well  as their  correspond  $1\sigma$
uncertainties.  In Fig.~\ref{fig:R_Z_age} the  median stellar age as a
function  of  Galactocentric distance  $R_{\rm  G}$  (left panel)  and
height $\lvert Z\rvert$ (right panel) for $7.6<R_{\rm G}<9.2\,$kpc and
$0.1<\lvert  Z\rvert<1.3\,$kpc   are  displayed,  together   with  the
resulting linear fit (red line).  The error bars in any given bin were
computed as $\Delta_{\tau}=(2N)^{-1/2}\sigma_{\tau}$, where $N$ is the
number of  white dwarfs in  each bin. We  found a very  small negative
radial   and  positive   vertical   age  gradients,   $\partial\langle
\tau\rangle/\partial    R_{\rm   G}=-0.2\pm0.1\,$Gyr~kpc$^{-1}$    and
$\partial\langle  \tau\rangle/\partial  Z=+0.1\pm0.2\,$Gyr~kpc$^{-1}$,
respectively.  However,  these results together with  their associated
uncertainties  are  also  compatible  with zero  radial  and  vertical
gradients for the volume probed in our study.

\begin{figure}
  \centering 
  \includegraphics[width=\columnwidth]{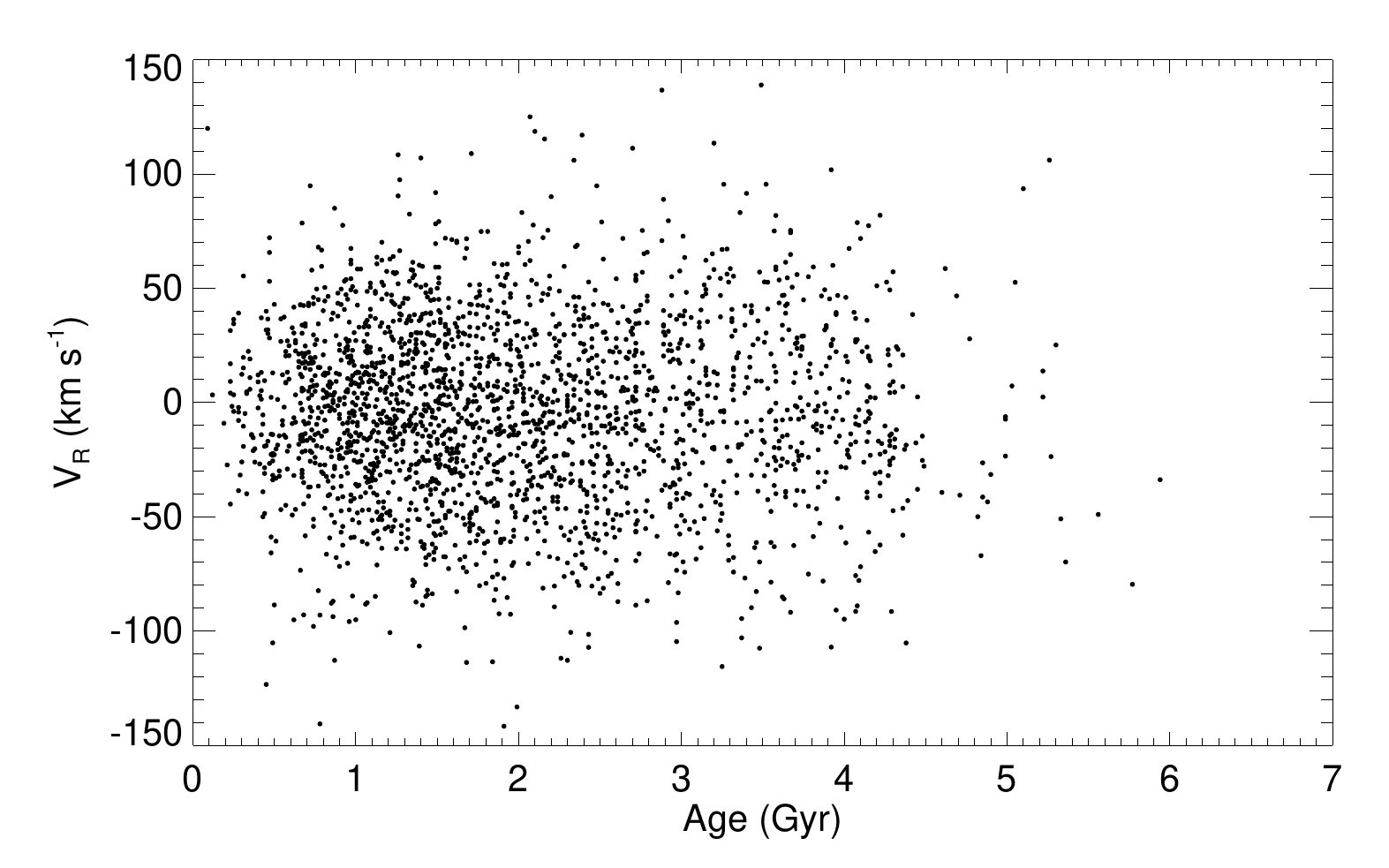}
  \includegraphics[width=\columnwidth]{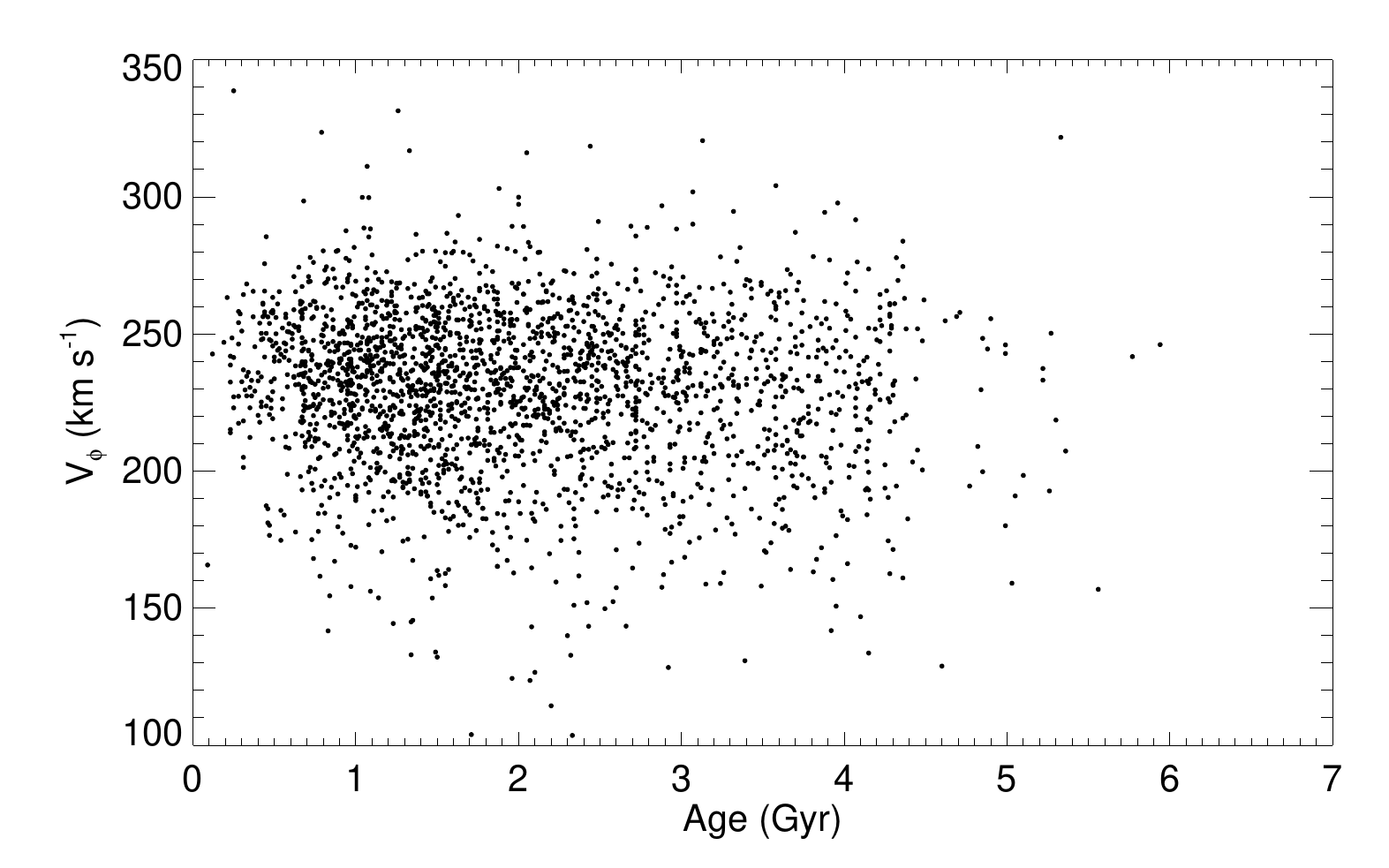}
  \includegraphics[width=\columnwidth]{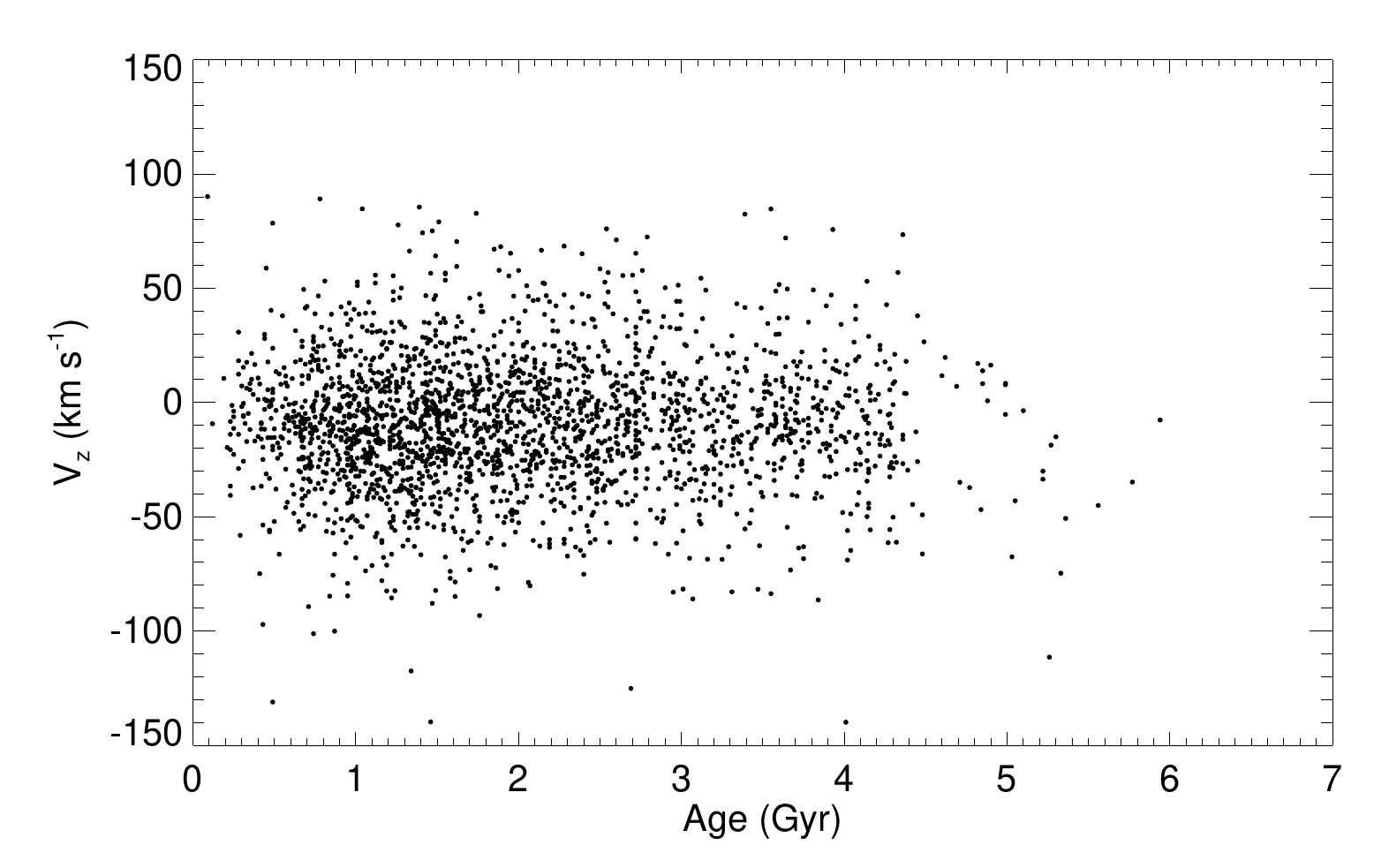}
  \caption{Age-velocity  relation  for  the three  components  of  the
    stellar velocity, $V_{\rm R}$, $V_{\rm \phi}$, and $V_{\rm z}$, the IFMR
    of  \citet{Catalan2008} is used  to  estimate  the  ages   of  the  white
    dwarfs. Our sample is limited to stars younger than 4.5~Gyr.}
  \label{fig:age_velo}
\end{figure} 

\section{The age-velocity relation}
\label{AVR}

In  this section  we  explore the  age-velocity  and the  age-velocity
dispersion  relationships  of  the  Galactic disc.  We restricted our sample, as 
discussed in Sect.~\ref{Velo_maps}, to those stars with the highest quality  
data.  Our final sample contains 3,415 DA  white  dwarfs  for  which  the typical  age  uncertainty  is  $\le
0.5$~Gyr.  Fig.~\ref{fig:age_velo}  shows the three components  of the
velocity in cylindrical  coordinates ($V_{\rm R},\,V_{\rm \phi},\,V_{\rm z}$) as a
function  of  age  when  the  IFMR  of  \citet{Catalan2008}  is  used.
Unfortunately,  when this  IFMR is  used our  sample does  not contain
stars older than  4.0~Gyr, and the same occurs  when the relationships
of  \citet{ferrarioetal05-1}  and  \citet{Gesicki2014}  are  employed.
This  is due  to the  tight quality  requirements we  have introduced.
Hence,  we are  restricted  to the  study  of the  latest  few Gyr  of
evolution from the formation of the Milky Way disc.

\begin{figure}
  \centering  
  \includegraphics[width=\columnwidth]{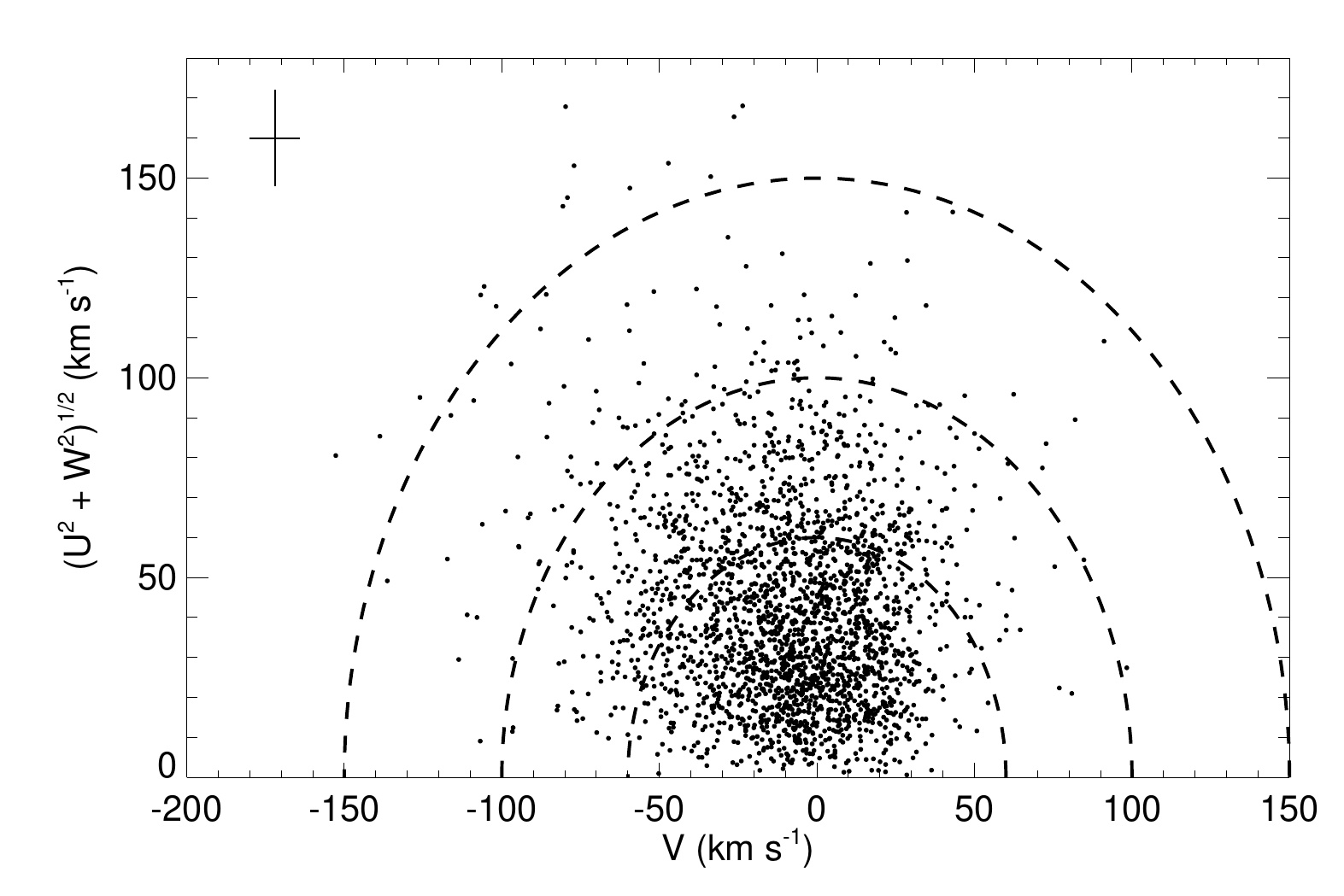}
  \caption{Toomre  diagram  for  our  white  dwarf  sample.   The  $V$
    component of  the stellar velocity represents  the stellar angular
    momentum  and  has  an  asymmetric distribution  as  discussed  in
    Fig.~\ref{fig:velo_cartesian},  while the  combination of  $U$ and
    $W$ is a measure of the  orbital energy.  Dotted lines show curves
    of constant speed, i.e.,  the curves for which $U^{2}+V^{2}+W^{2}$
    remains constant.}
  \label{fig:Toom_WD}
\end{figure} 

\subsection{Thin, thick disc and halo sample}
\label{thin_thick}

There is observational  evidence that a sizable fraction  of the thick
disc is chemically  different from the dominant thin  disc. Studies of
abundance populations based on  high resolution spectroscopy find that
the   abundance   distribution   is    bimodal   ---   see   ,   e.g.,
\citet{Navarro2011},  \citet{Fuhrmann2011},   and  \citet{Bensby2014}.
This strongly suggests  that the thin and thick disc  have a different
physical origin.  Moreover, most of the thick Galactic disc population
is    kinematically   hotter    than   that    of   the    thin   disc
\citep{Freeman2012}. However, the strong  gravities of white dwarfs do
not allow to use the metallicity to classify white dwarfs, because all
metals  are  diffused  inwards   in  short  timescales,  resulting  in
atmospheres made of  pristine hydrogen. Thus, to  study the membership
of white dwarfs in our sample to the thick or to the thin disc we rely
exclusively on their kinematical properties. In view of this, a Toomre
diagram is useful to understand  the characteristics of final DA white
dwarf  sample.   This diagram  combines  vertical  and radial  kinetic
energies  as a  function of  the rotational  energy.  Several  studies
based in  the analysis of  the abundances of the  Galactic populations
--- e.g., \citet{Nissen2004} and \citet{Bensby2014} --- have concluded
that to a first approximation  low-velocity stars (that is, those with
$V_{\rm tot}<70\,$km~s$^{-1}$) mainly belong to the thin disc, whereas
stars with  $V_{\rm tot}>70\,$km~s$^{-1}$ but with  velocities smaller
than $\sim$~200~kms$^{-1}$,  are likely to  belong to the  thick disc. 
However, the velocity cutoff selection population strongly depends on the 
size of the sample. Moreover, a pure kinematical selection of thin/thick 
disc stars can introduce a severe bias in our studies, as observed by different authors where subpopulations 
were selected using individual chemical abundances \citep{2013A&A...554A..44A,Bensby2014}.

According  to this,  from Fig.~\ref{fig:Toom_WD}  it follows that our
sample is dominated by thin disc white dwarfs, although certainly some
stars in the  sample exhibit thick disc kinematics. Moreover, Fig. 17 shows that the $\sigma_{\rm z}$ velocity dispersion rises with $\lvert Z \rvert$, which could be at least partly due to a component hotter than the thin disc. In order to study the fraction of thin, thick and halo stars in our WD sample, we performed a gaussian mixture model on their V$_{\rm z}$ distribution, to see if the velocity distribution shows multiple components. The model uses maximum likelihood to get the parameters for gaussian components, and takes account of the measuring errors for the velocities of individual stars.  We find that the z-velocity distribution is dominated by a single colder component, with a velocity dispersion of about 23 $\pm$ 0.5 km s$^{-1}$ and showing no significant change with age.  A weak hotter component is also seen, containing about 5$\%$ of the total sample (about  100 stars), with a velocity dispersion of 72 $\pm$ 6 km s$^{-1}$, again showing no significant change with age. 

We can compare the dispersion of the colder component of the white dwarfs (23 km s$^{-1}$) with the AVR from the \citet{2011A&A...530A.138C} sample of Geneva-Copenhagen Survey (GCS) stars. This is probably the best source of age-velocity data available for nearby stars.  The GCS stars have isochrone ages and are mostly turnoff stars. Fig.~\ref{fig:sigmaZ_Casa} shows the AVR for GCS stars with  [Fe/H] $>$ -0.3; this cut removes most of the thick disc stars from the sample. The GCS stars show a well defined AVR, with the $\sigma_{\rm z}$ velocity dispersion rising with age from about 10 km s$^{-1}$ for the youngest stars, and levelling out at about 23 km s$^{-1}$ for stars older than about 5 Gyr. We note that the 23 km s$^{-1}$ dispersion of the colder component of the white dwarfs (ages mostly $<$ 4 Gyr) is in excellent agreement with the dispersion of the older thin disc stars in Fig 23.  The velocity dispersion of the weak hotter component from the mixture model is about 72 km s$^{-1}$. This is much larger than the typical z-velocity dispersion of thick disc stars near the sun (about 40 km s$^{-1}$), and suggests that there may also be a few halo WDs in the sample. However, in this we will not attempt to  fully analyze  and characterize  each kinematic  component of  our sample.  Thus, we  do not apply any {\sl a  priori} kinematical cut to distinguish between thin and thick stars to study the AVR.

\subsection{The age-velocity dispersion relation}

The  age-velocity dispersion  relation  is a  fundamental relation  to
understand local Galactic dynamics.  It  reflects the slow increase of
the  random velocities  with age  due to  the heating  of the  stellar
disc. We study this relation using  the ages estimated using the three
IFMRs discussed in Sect.~\ref{ages_WD}.  To  this end, we first binned
the  data in  intervals of  1~Gyr, and  then we  computed the  stellar
velocity  dispersions  and  their   associated  errors  following  the
expressions discussed in Sect.~\ref{Velo_maps}. The velocity dispersion 
components were also corrected for the measuring errors using the 
standard deviation of the error distribution for each age bin.

\begin{table}
\caption{Velocity dispersions as  a function of the age  for the three
  IFMRs studied in this work.}
\begin{center}
\begin{tabular}{cccc}
\hline
\hline
Age (Gyr) & $\sigma_{\rm R}$ (km s$^{-1}$) & $\sigma_{\rm \phi}$ (km s$^{-1}$) & $\sigma_{\rm z}$ (km s$^{-1}$) \\ 
\hline
&\multicolumn{3}{c}{\citet{Catalan2008}}\\
\cline{2-4}
0 -- 1 & 24.6 $\pm$ 0.8 & 17.7 $\pm$ 0.6  & 19.8 $\pm$ 0.6 \\
1 -- 2 & 24.6 $\pm$ 0.6 & 19.2 $\pm$ 0.4  & 22.6 $\pm$ 0.5 \\
2 -- 3 & 28.9 $\pm$ 0.9 & 21.8 $\pm$ 0.7  & 23.5 $\pm$ 0.7 \\
3 -- 4 & 30.3 $\pm$ 1.2 & 22.5 $\pm$ 0.9  & 21.4 $\pm$ 0.8\\
4 -- 5 & 31.1 $\pm$ 2.0 & 26.9 $\pm$ 1.7  & 21.8 $\pm$ 1.4\\
\hline
&\multicolumn{3}{c}{\citet{ferrarioetal05-1}}\\
\cline{2-4}
0 -- 1 & 24.5 $\pm$ 0.6 & 17.9 $\pm$ 0.5  & 19.5 $\pm$ 0.5 \\
1 -- 2 & 27.0 $\pm$ 0.6 & 20.5 $\pm$ 0.4  & 22.1 $\pm$ 0.5 \\
2 -- 3 & 27.1 $\pm$ 1.0 & 21.8 $\pm$ 0.8  & 25.8 $\pm$ 1.0 \\
3 -- 4 & 32.3 $\pm$ 2.0 & 25.9 $\pm$ 1.6  & 22.9 $\pm$ 1.4\\
\hline
&\multicolumn{3}{c}{\citet{Gesicki2014}}\\
\cline{2-4}
0 -- 1 &  24.0 $\pm$ 0.7 & 17.5 $\pm$ 0.5 & 19.4 $\pm$ 0.6  \\
1 -- 2 &  27.2 $\pm$ 0.5 & 20.5 $\pm$ 0.4 & 21.9 $\pm$ 0.4 \\
2 -- 3 &  26.9 $\pm$ 1.0 & 22.8 $\pm$ 0.8 & 25.1 $\pm$ 1.0 \\
3 -- 4 &  37.3 $\pm$ 3.6 & 24.2 $\pm$ 2.3 & 24.8 $\pm$ 2.4   \\
\hline
\end{tabular}
\end{center}
\label{tab_AVR}
\end{table}

\begin{figure}
  \centering  
  \includegraphics[width=1.03\columnwidth]{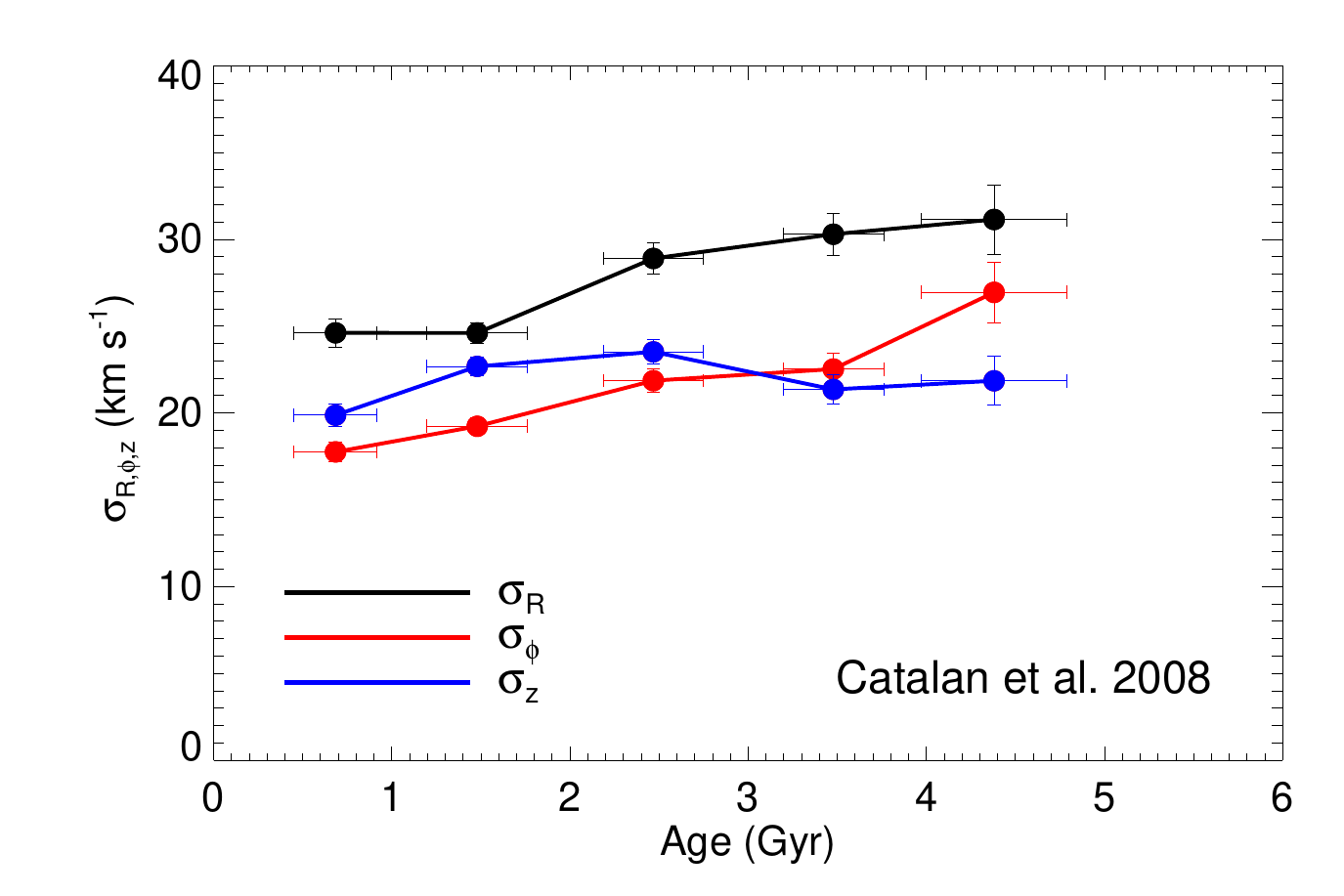}
  \includegraphics[width=1.03\columnwidth]{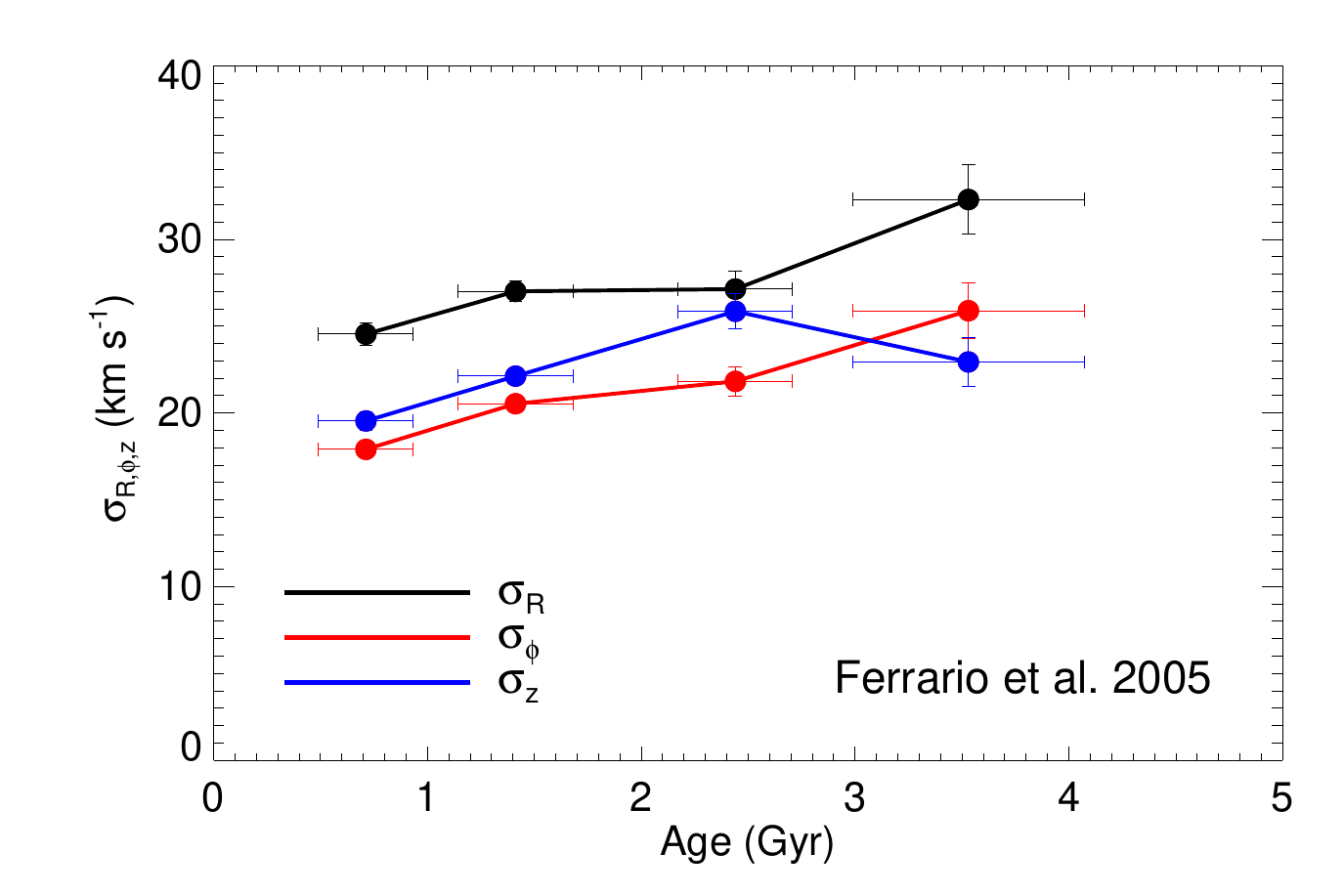}
  \includegraphics[width=1.03\columnwidth]{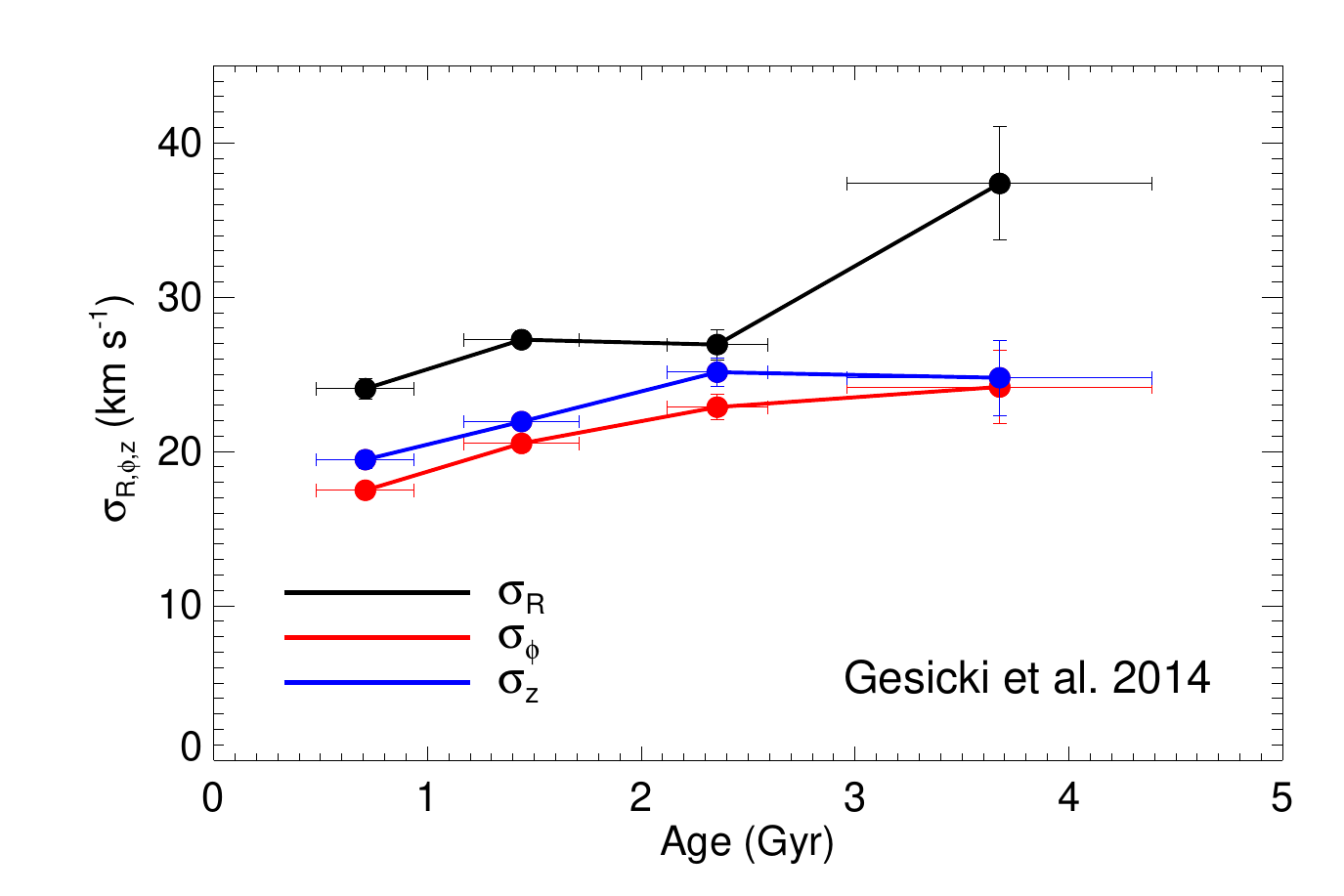}
  \caption{Age-velocity dispersion  relation for  a total of  2,345 DA
    white dwarfs  binned in 1~Gyr  intervals. The top panel  shows the
    results   using   the   ages    obtained   using   the   IFMR   of
    \citet{Catalan2008}, the middle panel shows the ages derived using
    the IFMR of \citet{ferrarioetal05-1}, whereas for the bottom panel
    the IFMR  of \citet{Gesicki2014}  was adopted.   The corresponding
    dispersions,  $\sigma_{\rm R}$, $\sigma_{\rm  \phi}$ and  $\sigma_{\rm z}$
    are shown in red, blue and black, respectively.}
  \label{fig:AVdR}
\end{figure}

Our results indicate that the three components of the stellar velocity
dispersion  increase  with time,  independently of  the adopted IFMR.  
This is illustrated in Fig.~\ref{fig:AVdR} --- see also
Table~\ref{tab_AVR}.  When    we    use    the   IFMR    of
\citet{Catalan2008}  for  stars  older  than  2.5~Gyr  $\sigma_{\rm \phi}$
increases with time, while $\sigma_{\rm R}$ remains nearly constant within
the errors, and $\sigma_{\rm Z}$ remains also constant.  However, it should be
taken into  account that  the number  of stars  older than  3.5~Gyr is
small for  all three IFMRs,  although the  number of stars  older than
3.5~Gyr    is     significantly    larger    for    the     IFMR    of
\citet{Catalan2008}.  Thus,   only  in  this  case   the  age-velocity
dispersion relation for ages longer than 4~Gyr, but still shorter than
5.5~Gyr can  be explored.  For the  rest of the IFMRs  the results are
less significant. Nevertheless, it is  interesting to realize that, as
it occurs for  the IFMR of \citet{Catalan2008}, for both  the IFMRs of
\citet{ferrarioetal05-1} and \citet{Gesicki2014}  all three components
of  the stellar  velocity dispersion  increase for  ages smaller  than
$\sim  2.5\,$Gyr,  while  for   ages  larger  than  this  $\sigma_{\rm R}$ 
and $\sigma_{\rm \phi}$ increases both cases, while $\sigma_{\rm z}$ saturates. 

\begin{figure}
  \centering  
  \includegraphics[width=1.05\columnwidth]{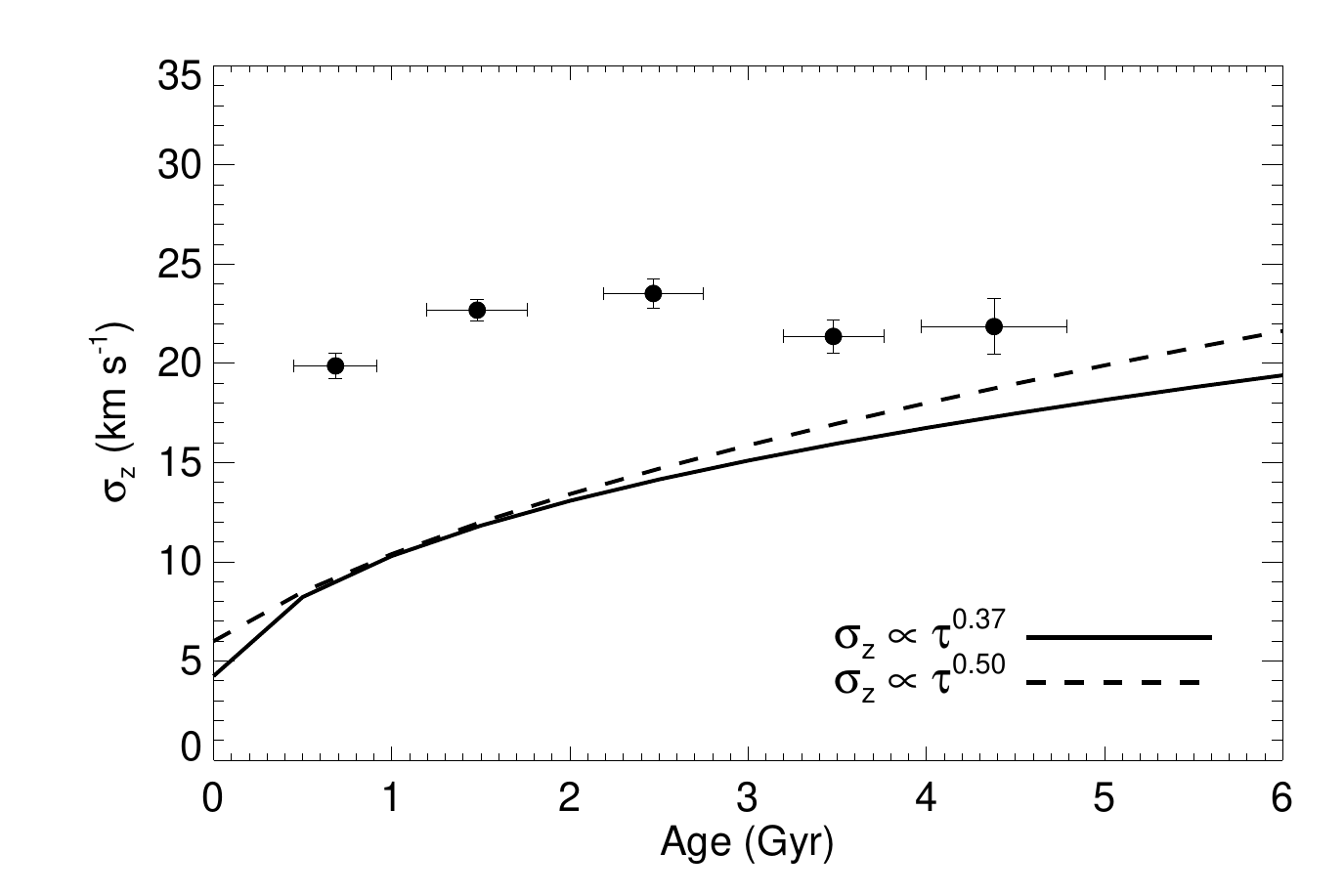}
  \caption{Age-velocity dispersion relation for the vertical component
    of  the  velocity  using  the  ages  derived  using  the  IFMR  of
    \citet{Catalan2008}.   Dynamical  heating  functions  following  a
    power  law,  $\sigma_{\rm z}\propto\tau^{\alpha}$, with  $\alpha=0.37$
    (solid line) and 0.50 (dashed line) are also represented.}
  \label{fig:sigmaZ}
\end{figure}

Age-velocity dispersion  relations are  suitable to  constrain heating
mechanisms taking place in the  Galactic disc during the first epochs.
Since  dynamical streams  are  well mixed  in  the vertical  direction
\citep{Seabroke2007} we  will use the age-$\sigma_{\rm z}$ relation.  The
vertical velocity dispersion obtained  when the ages derived employing
the  semi-empirical IFMR  of  \citet{Catalan2008},  together with  the
associated uncertainties, is  displayed in Fig.~\ref{fig:sigmaZ} using
black  dots.   In this  figure  we  also  show two  dynamical  heating
functions following  a power law  as a  function of the  age ($\tau$),
$\sigma_{\rm z}\propto\tau^{\alpha}$, with $\alpha=0.35$  (solid line) and
0.50     (dashed    line)     as    labelled     in    this     figure
\citep{Wielen1977,Hanninen2002,Aumer2009}.  Clearly,  our results fall
well above  the predictions  of these power  laws for the younger ages.  
In  particular, we found  that $\sigma_{\rm z}\,\sim\,22\pm0.5\,$km  s$^{-1}$ for  white dwarfs
with  ages between  $0.5$ and  2.0~Gyr.  Newly born  stars, with  ages
$\sim\,300\,$Myr, typically  have $\sigma_{\rm z}\,\sim\,5.0\,$km~s$^{-1}$
\citep{Torra2000}.  For  open clusters with ages  around $1.5\,$Gyr it
is     found     that    $\sigma_{\rm z}\,\sim\,15.0\pm5\,$km     s$^{-1}$
\citep{VandePutte2010}.  These  values are significantly  smaller than
these obtained using the population of Galactic white dwarfs.

\begin{figure}
  \centering  
  \includegraphics[width=1.05\columnwidth]{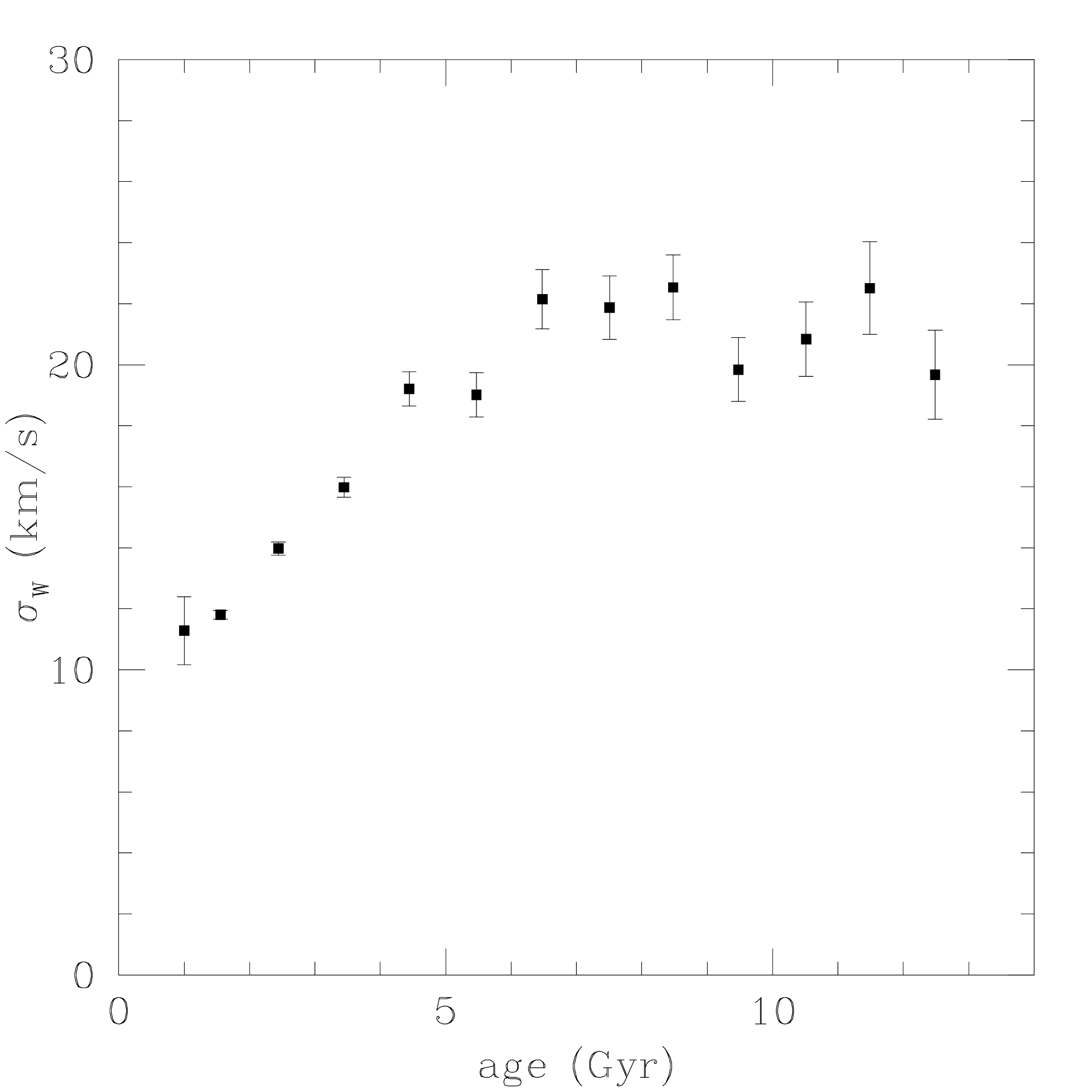}
  \caption{Vertical velocity dispersion $\sigma_W$ {\it vs} stellar age for 
  the solar neighborhood, for stars with [Fe/H] $> -0.3$; data from \citet{2011A&A...530A.138C}.}
  \label{fig:sigmaZ_Casa}
\end{figure}

A highly  effective mechanism for scattering  stars, especially during
the  first $\sim  3$~Gyr \citep{Lacey1984,Aumer2016},  is interactions
with Giant Molecular Clouds  (GMCs).  In particular, using simulations
of the  orbits of  tracer stars  embedded in  the local  Galactic disc
\citet{Hanninen2002}  derived  an   age-velocity  dispersion  relation
$\sigma\propto\tau^{0.2}$  for  the  heating caused  by  GMCs.   Other
widely accepted mechanism  is heating by transient spiral  arms in the
disc  \citep{DeSimone2004,   Minchev2006}.   Combining   both  heating
mechanisms   leads   to   a    power   law   with   larger   exponent,
$\sigma_{\rm z}\propto\tau^{0.4\pm0.1}$                \citep{Jenkins1990,
BinneyTremaine2008}.   

The age-velocity dispersion relation derived from the present WD sample indicates that the Galactic population of white dwarfs may have experienced an additional source of heating, which adds to the secular evolution of the Galactic disc seen in Fig.~\ref{fig:sigmaZ_Casa}. The GCS stars in the age range of our WDs ($<$ 4 Gyr) have velocity dispersions that are clearly less than the 23 km s$^{-1}$ found for the colder component of the WDs. The  origin of  this heating mechanism remains  unclear. One  possibility is  that some  thick disc-halo 
stars may have  contaminated the younger bins of the  AVR (see Sect.~\ref{thin_thick} where 
we reported the existence of a hot component in our sample), 
thus making the age-velocity dispersion hotter.  Another possibility is that there
is an intrinsic  dispersion for these stars.  Given  that white dwarfs
are  the  final products  of  the  evolution  of  stars with  low  and
intermediate masses  they might  have experienced  a velocity  kick of
$\sim  10$~km~s$^{-1}$  during the  final  phases  of their  evolution
\citep{Davis2008}. It is also important to keep in mind that, although
we have excluded in our analysis  all white dwarfs with masses smaller
than $0.45\,  M_{\sun}$, because  they are expected  to be  members of
close binaries,  a possible  contamination by close  pairs ---  of the
order  of  $\sim$10 per  cent  \citep{Maoz2016,  Badenes2012} ---  may
influence the  age-velocity dispersion  relation. Finally, there  is a
last  possible explanation.   Our sample  of  DA white  dwarfs is  not
homogeneously distributed  on the sky,  but instead is drawn  from the
SDSS.   The   geometry   of   the  SDSS   is   complicated   ---   see
Fig.~\ref{fig:aitoff} --- and there is  a large concentration of stars
around the NGP. Even  more, the fields of the SDSS are
not equally  distributed in  the polar cap.  This means  that, possibly,
important  selection  effects  could  affect the  results.  All  these
alternatives  need   to  be  carefully  explored   employing  detailed
population synthesis  models must be  employed that take  into account
the   sample  biases   and   selection  procedures  to  check   their
verisimilitude. We postpone this study for a future publication.

\section{Conclusions}
\label{conclu}

We have used the largest  catalog of white dwarfs with hydrogen-rich 
(DA) atmospheres currently available  (20,247 stars) obtained from the
SDSS  DR12   to  compute   effective  surface   temperatures,  surface
gravities, masses,  ages, photometric distances and  radial velocities
as well as the components of  their velocities when proper motions are
available.   For  the  first  time   we  investigated  how  the  space
velocities $V_{\rm R},\,V_{\rm \phi},\,V_{\rm z}$ depend on the Galactocentric
radial distance $R_{\rm G}$ and on  the vertical height $Z$ in a large
volume around the Sun using a sample of DA white dwarfs.
Our understanding of the chemical and dynamical evolution of the Milky
Way  disc has  been  hampered  over the  years  by  the difficulty  of
measuring accurate ages of stars  in our Galaxy. However, white dwarfs
are natural  cosmochronometers, and this  motivated us to use  them to
study  the  kinematical  evolution  of our  Galaxy.   Accordingly,  we
derived ages  for each individual  white dwarf  in our catalog  and we
computed  averaged ages  as well.  We did  this using  three different
initial-to-final mass relationships.  In a  second step we studied the
sensitivity  of  the  individual  and averaged  ages  to  the  adopted
initial-to-final mass relationship, finding  that average ages are not
affected by the  choice of this relationship,  although individual
ages can be significantly different.  Additionally, we found that when
our  preferred  choice,  the  initial-to-final  mass  relationship  of
\citet{Catalan2008},  is  employed  the  number of  stars  older  than
2.5~Gyr is  larger than  that predicted  by the  initial-to-final mass
relationships of \citet{ferrarioetal05-1} and \citet{Gesicki2014}.  In
all  three  cases, nevertheless,  we  found  a  paucity of  old  white
dwarfs. Naturally, this  arises because most of the white dwarfs  older than 6~Gyr
have small effective surface  temperatures, $T_{\rm eff}<7000\,$K, and
luminosities.  Consequently, they fall out  the magnitude limit of the
SDSS.
In a  subsequent effort  we studied the  age-velocity relation  of the
Galactic  disc during  the last  few  Gyr. To  do this  we selected  a
sub-sample  of stars  for  which proper  motions  and accurate  radial
velocities  were  available.  For  this  sub-sample  of stars  precise
kinematics  were derived  (see  Sect.~\ref{Velo_maps}).  White  dwarfs
belonging  to this  sub-sample  are within  $\pm0.5\,$kpc  in $Z$  and
between 7.8  and $\sim$ 9.0~kpc in $R_{\rm  G}$, and allowed us  to study how
their kinematical properties depend on  $R_{\rm G}$ and $Z$. Our results can
be summarized as follows.
We found  that the mean value  of the radial velocity,  $\langle V_{\rm R}
\rangle$ increases when moving from the inner regions of the Galaxy to
distances  beyond   the  Solar  circle,  while   the  radial  velocity
dispersion $\sigma_{\rm R}$  slightly increases as $R_{\rm  G}$ increases.
Similar  radial  gradients have  been  reported  for the  RAVE  survey
\citep{Siebert2011,  Williams2013}, which  uses  red  clump stars  and
probes a  volume larger than  the one studied here.   Additionally, we
found that $V_{\rm R}$  shows a small gradient in  the vertical direction.
However, the uncertainties are still  large, preventing us to draw any
robust conclusion.  Finally, we also  found that $\sigma_{\rm R}$ clearly increases 
when moving away from the Galactic plane.
We  also found  that $V_{\rm  \phi}$ decreases  for decreasing  $R_G$,
while  $\sigma_{\rm  \phi}$  increases,  and  that  $V_{\phi}$  has  a
significant  gradient in  the $Z$  direction, $\partial\langle  V_{\rm
\phi}\rangle/\partial  Z=+17\pm4\,$km~s$^{-1}$~kpc$^{-1}$,  suggesting
that the  regions of  South Galactic hemisphere  probed by  our sample
rotate slower than the those of the North Galactic hemisphere. We also
found  that  $\sigma_{\rm  \phi}$   increases  when  $\lvert  Z\rvert$
increases from zero to 0.6~kpc.
We  found that  the mean  vertical velocity  increases as  $R_{\rm G}$
decreases,  while  the  vertical  velocity  dispersion,  $\sigma_{\rm z}$,
clearly  decreases   when  moving  away  from   the  Galactic  center.
Interestingly, while for white dwarfs  with positive $Z$ (belonging to
the North  Galactic hemisphere) we  found negative values  of $\langle
V_{\rm z} \rangle$, for stars with  $Z<0$ (belonging to the South Galactic
hemisphere), $\langle V_{\rm z} \rangle$ is  positive.  We also found that
the velocity dispersions  increase when moving away  from the Galactic
plane. 
The age-velocity dispersion  relation was also studied.  We found that
the age-velocity dispersion relation  derived using the present sample
of  DA white  dwarfs may have experienced  an additional  heating, in addition to the  secular evolution of the 
Galactic  disc. However, the origin of this heating mechanism  remains unclear. 
We advanced several hypothesis in the previous section which may explain it, but we defer a 
thorough evaluation of these  alternatives  for  a  future publication where we will employ population  
synthesis techniques to carefully  reproduce the selection procedures and observational biases.
To conclude,  we demonstrated  that white  dwarfs can be  used to
study the  dynamical evolution of  our Galaxy. This has  been possible
because now  we have a  large database  of white dwarfs  with accurate
measurements of  their stellar  parameters, a  byproduct of  the SDSS.
However,  the catalog  of  white dwarfs  presented  here has  inherent
limitations, due to the selection procedures and observational biases.
Thus, modeling  the kinematical  properties derived from  this sample
requires  significant   theoretical  efforts,  and   deserves  further
studies.   These  studies are  currently  underway,  and will  be  the
subject  of future  publications.  Nevertheless, the  results, will  be
rewarding, as we will have independent determinations of the dynamical
properties of our Galaxy.


\section*{Acknowledgments}

BA gratefully acknowledge the financial support of the Australian Research Council through Super Science Fellowship FS110200035. BA also thanks Maurizio Salaris (Liverpool John Moores University) and Daniel Zucker (Macquarie University/AAO) for lively discussions. TZ thanks the Slovenian Research Agency (research core funding No. P1-0188). This   research   has   been   partially  funded   by   MINECO   grant AYA2014-59084-P and by the AGAUR.

\bibliographystyle{mnras} 
\bibliography{AVR_bib.bib}
\label{lastpage}

\end{document}